\documentclass [11pt]{book}
\usepackage[pagebackref,colorlinks=true,linkcolor=blue, citecolor=blue]{hyperref}
\usepackage[nosumlimits,nointlimits,nonamelimits]{amsmath}
\usepackage{amssymb,amsthm,amsfonts}

\newcommand\N{\mbox{$\mathbb N$}}
\newcommand\R{\mbox{$\mathbb R$}}

\def\B{\{0,1\}}
\def\cond{\mid}
\def\D{{\cal D}}

\providecommand\floor[1]{\lfloor#1\rfloor}
\providecommand\ceil[1]{\lceil#1\rceil}

\providecommand\abs[1]{\lvert#1\rvert}
\providecommand\bigabs[1]{\bigl\lvert#1\bigr\rvert}

\def\poly{\mbox{\rm poly}}

\def\epsilon{\varepsilon}
\def\phi{\varphi}

\def\ared{\leq_{\rm AvgP}}

\def\avg{{\rm Avg}}
\def\heur{{\rm Heur}}
\def\bptime{{\rm BPTIME}}
\def\zptime{{\rm ZPTIME}}
\def\dtime{{\rm DTIME}}
\def\size{{\rm SIZE}}

\def\co{{\rm co}}
\def\np{{\rm NP}}
\def\p{{\rm P}}
\def\zpp{{\rm ZPP}}
\def\bpp{{\rm BPP}}
\def\pspace{{\rm PSPACE}}
\def\exptime{{\rm EXP}}

\def\am{{\rm AM}}

\def\pcomp{\mbox{\sc PComp}}              % computable ensembles
\def\psamp{\mbox{\sc PSamp}}              % samplable ensembles
\def\uniform{\ensuremath{\mathcal{U}}}    % the uniform ensemble

\def\bh{{\rm BH}}                         % bounded halting
\def\ubh{U^{\mathrm{BH}}}       % uniform bounded halting
\def\calubh{\mathcal{U}^{\mathrm{BH}}}
\def\sat{{\rm SAT}}                       % satisfiability
\def\trisat{{\rm 3SAT}}                     % 3sat
\def\svp{{\rm SVP}}                     % shortest vector problem
\def\usvp{{\rm uSVP}}                    % unique shortest vector problem

\newtheorem{theorem}{Theorem}
\newtheorem{lemma}[theorem]{Lemma}
\newtheorem{corollary}[theorem]{Corollary}
\newtheorem{proposition}[theorem]{Proposition}
\newtheorem{claim}[theorem]{Claim}

\newtheorem{definition}[theorem]{Definition}

\newenvironment{proof*}[1][Proof]{\par\trivlist
   \item[\hskip \labelsep{\itshape {#1}.}]}
   {\hspace*{0pt plus1fill}\fboxsep2.5pt\fboxrule.5pt\raise3pt\hbox{\fbox{}}\endtrivlist}

\DeclareMathOperator\E{\mathbf{E}}

\DeclareMathOperator\pr{\mathbf{Pr}}
\DeclareMathOperator\supp{Supp}

\def\xor{\oplus}

\def\calD{\mathcal{D}}
\def\calU{\mathcal{U}}

\def\hypref#1#2{#1~\ref{#2}}
\def\secref#1{\hypref{Section}{#1}}
\def\thmref#1{\hypref{Theorem}{#1}}

\def\propref#1{\hypref{Proposition}{#1}}
\def\dfnref#1{\hypref{Definition}{#1}}

\newenvironment{change}{}{}

\title {\bf\Huge Average-Case Complexity}

\author{{\large\sc Andrej Bogdanov}\thanks{{\tt
andrejb@cse.cuhk.edu.hk}. Chinese University of Hong Kong. Work  done at DIMACS and at
the Institute for Advanced Study, Scool of Mathematics, and
partially supported by the National Science Foundation grant CCR 0324906.}  \and
{\large\sc Luca Trevisan}\thanks{{\tt L.Trevisan@unibocconi.it}. Bocconi University. Work done at
U.C. Berkeley, supported by  US-Israel Binational Science
Foundation Grant 2002246 and by NSF Grant and by the National Science Foundation under grant CCF 0515231.}}

\date{October 2006, Revised August 2021}

\begin{document}

\pagenumbering{roman}

\maketitle

\newpage

\section*{Abstract}

We survey the average-case complexity of problems in NP.

We discuss various notions of good-on-average algorithms, and
present completeness results due to Impagliazzo and Levin.
Such completeness results establish the fact that if a certain
specific (but somewhat artificial) NP problem is easy-on-average
with respect to the uniform distribution, then all problems in NP
are easy-on-average with respect to all samplable distributions.
Applying the theory to natural distributional problems remain an
outstanding open question. We review some natural distributional
problems whose average-case complexity is of particular interest and
that do not yet fit into this theory.

A major open question whether the existence of hard-on-average
problems in NP can be based on the P$\neq$NP assumption or on
related worst-case assumptions. We review negative results showing
that certain proof techniques cannot prove such a result. While the
relation between worst-case and average-case complexity for general
NP problems remains open, there has been progress in understanding
the relation between different ``degrees'' of average-case
complexity.  We discuss some of these ``hardness amplification''
results.

\tableofcontents

\newpage

\setcounter{page}{0}
\pagenumbering{arabic}

\chapter{Introduction}

The study of the average-case complexity of intractable problems
began in the 1970s motivated by two distinct applications: the
developments of the foundations of cryptography and the search
for methods to ``cope'' with the intractability of $\np$-hard
problems.

All definitions of security for cryptographic problems require that
any efficient algorithm that tries to ``break'' the protocol
``succeeds'' only with very small probability. The formalizations of
{\em breaking} and {\em succeeding} depend on the specific
application, but it has been known since the 1980s that there is a
unifying concept: no cryptographic task (for example, electronic
signature, or data encryption) is possible unless {\em one-way
functions} exist.\footnote{The realizability of many cryptographic
tasks, in fact, is {\em equivalent} to the assumptions that one-way
functions exist.} Informally, a one-way function is an efficiently
computable function $f: \B^* \to \B^*$ that maps $\B^n$ to $\B^n$
and such that, if we are given $f(x)$ for a random $x\in \B^n$, it
is intractable (in time polynomial in $n$) to find a pre-image $x'$
such that $f(x')=f(x)$. In particular, the existence of one-way
functions implies that there is a search problem in $\np$ (given $y\in
\B^n$, find $x\in \B^n$ such that $f(x)=y$) that is intractable to
solve on random inputs sampled from a simple distribution (the
distribution $f(x)$ where $x$ is chosen randomly from $\B^n$). The
fact that all of cryptography is predicated on the existence of
average-case intractable problems in $\np$ is a main motivation for the
study of the theory we describe in this survey.

In particular, a long-standing open question is whether it is
possible to {\em base the existence of one-way functions on the
$\p \neq \np$ assumption}, or related ones (such as NP-complete problems
not allowing polynomial size circuits).

The second motivation for the study of the average-case complexity
of problems in $\np$ comes from the analysis of heuristic algorithms.
Unless $\p = \np$, we cannot hope for efficient algorithms that solve
$\np$-complete problem exactly on all inputs. We may hope, however, for
algorithms that are ``typically efficient'' on inputs sampled from
distributions that occur in practice. In order to understand the
limitations of such an approach, it would be desirable to have an
``average-case analog'' of the theory of $\np$-completeness. Such a
theory would enable us to prove that for certain problems, with
respect to certain distributions, it is impossible to have
algorithms that perform well on ``typical'' inputs, unless an entire
class of presumably intractable problems can be efficiently solved.

The basic foundations of such a theory have been laid out.
Surprisingly subtle difficulties arise even when just developing the
analogs of trivial elements of the theory of $\np$-completeness, such
as the definition of {\em computational problem}, the definition of
{\em efficient algorithm}, the definitions of {\em reduction} and
{\em completeness}, and the equivalent complexity of {\em decision
versus search} for $\np$-complete problems. In this survey we will
discuss these difficulties, and show how they were resolved. We will
see a number of results, insights, and proof techniques whose
usefulness goes beyond the study of average-case complexity.

The  right techniques to apply such a theory to natural problems and
distributions have not been discovered yet. From this point of view,
the current state of the theory of average-case complexity in $\np$ is
similar to the state of the theory of inapproximability of $\np$
optimization problems before the PCP Theorem.

Finding ways of {\em applying this theory to natural problems} is
another outstanding open question in this area.

\section{Roadmap}

In this section we give an overview of the content of this survey.

\subsection{Definitions of Tractability.} The first difficulty in
developing a theory of average-case intractability is to come up
with a formal definition of what it means for a problem to be
``intractable on average'' or, equivalently, what it means to be
``average-case tractable.'' A natural definition would be to
consider an algorithm efficient-on-average if it runs in {\em
expected polynomial time}. Such a definition has various
shortcomings (related to the fact that it is too restrictive). For
example, if an algorithm $A$ runs in time $t(x)$ on input $x$, and
its simulation $B$ (on a different model of computation) runs in
time $t^2(x)$ on input  $x$, it is natural that we would like our
definition to be such that $A$ is efficient-on-average if and only
if $B$ is. Suppose, however, that our inputs come from the uniform
distribution, and that $A$ runs in time $n^2$ on all inputs of
length $n$, except on one input on which $A$ takes time $2^n$. Then
the expected running time of $A$ is polynomial but the expected
running time of $B$ is exponential. Looking at the {\em median}
running time of an algorithm gives us a more robust measure of
complexity, but still a very unsatisfactory one: if an algorithm
runs in polynomial time on $70 \%$ of the inputs, and in exponential
time on $30 \%$ of the inputs, it seems absurd to consider it an
efficient-on-average algorithm. The right way to capture the notion
of ``efficient on typical instances'' should be that it is fine for
an algorithm to take a large amount of time on certain inputs,
provided that such inputs do not occur with high probability: that
is, inputs requiring larger and larger running times should have
proportionally smaller and smaller probability. This is the idea of
Levin's definition of average-case complexity. In (an equivalent
formulation of) Levin's definition \cite{L86}, an algorithm is
polynomial-time-on-average if there is a constant $c>0$ such that
the probability, over inputs of length $n$, that the algorithm takes
more than time $T$ is at most $\poly(n)/T^c$. As usual in complexity
theory, various choices can be made in the definition: we may look
at deterministic algorithms, randomized algorithms, or
non-uniform families of circuits. An additional choice is whether we
require our algorithm to always be correct, but possibly run in
superpolynomial time on some inputs, versus requiring the algorithm
to always run in polynomial time, but to give an incorrect answer to
some inputs. This will lead to several possible definitions, each
meaningful in some applications. (See Section \ref{sec:defavg}.) The
important thing will be that almost all the other results we discuss
in this survey are based on reductions that preserve tractability
under all of these definitions. Hence, the treatment of
completeness, reductions, families of distributions and decision
versus search is independent of the specific notion of tractability
that one is interested in.

\subsection{Reductions Between Distributional Problems} Let $L$ be a
decision problem and $\D$ be a distribution over
inputs,\footnote{Additional difficulties arise in defining how to
specify $\D$.} we call the pair $(L,\D)$ a {\em distributional
problem}. All the definitions of average-case tractability have a
characteristic in common: than an algorithm $A$ is efficient for
$(L,\D)$ if a certain set of ``bad'' inputs has low probability
under $\D$. (Depending on the cases, the bad inputs could be the
ones where the algorithm $A$ takes very long time, or those on which
$A$ outputs an incorrect answer, and so on.) This motivates the
following definition of reduction \cite{L86}: we say that $(L,\D)$
reduces to $(L',\D')$ if there is a polynomial time computable
function $f$ such that $x\in L$ if and only if $f(x)\in L'$ and, in
addition, for every input  $y$, the probability of generating $y$ by
picking $x$ at random according to $\D$ and then computing $f(x)$ is
at most $\poly(\abs{x})$ larger than the probability of sampling $y$ at
random from $\D'$.\footnote{When the second condition holds, we say
that $\D'$ {\em dominates} $\D$.} The motivation for this definition
is the following. Suppose that $A'$ is a good algorithm for
$(L',\D')$, so that the set $B'$ of inputs that are bad for $A'$ has
small probability according to $\D'$. Consider the following
algorithm for $(L,\D)$: on input $x$, output $A'(f(x))$. Now, the
bad inputs for this algorithm are the inputs $x$ such that $f(x)\in
B'$. The probability of sampling such an $x$, according to $\D$,
however, is upper bounded by $\poly(\abs{x})$ times the probability of
sampling an element of $B'$ according to $\D'$, which we had assumed
to be small. Hence, we have a good algorithm for $(L,\D)$, and the
definition of reduction preserves average-case tractability. Note
that, in this argument, we used nothing about the definition of
tractability except the notion of ``bad'' input. (See also Section
\ref{sec:complete}.)

\subsection{A Completeness Result} Having given the definition of
computational problem and of reduction, we will present a {\em
completeness result} \cite{L86}. We consider the {\em bounded
halting} problem \bh, where on input $(M,x,1^t)$ we have to
determine whether the non-deterministic Turing machine $M$ accepts
input $x$ within $t$ steps. This problem is readily seen to be
NP-complete. We show that for every distributional problem $(L,\D)$,
where $L$ is in NP and $\D$ is a {\em polynomial-time computable}
distribution there is a reduction from $(L,\D)$ to $(\bh,\calubh)$,
where $\calubh$ is a reasonable formalization of the notion of a
``uniformly chosen'' random input for \bh. Informally, the reduction
maps an input $x$ into the triple $(M',C(x),1^t)$ where $C$ is a
(carefully chosen) injective polynomial time computable encoding
function; $M'$ is a non-deterministic machine that first recovers
$x$ from $C(x)$ and then simulates the non-deterministic polynomial
time Turing machine that decides whether $x\in L$ (recall that $L$
is in NP); and $t$ is a polynomial upper bound to the running time
of $M'$. The main claim in the analysis of the reduction is that,
for $x$ selected from $\D$, $C(x)$ is ``approximately'' uniformly
distributed. Technically, we show that the distribution of $C(x)$ is
dominated by the uniform distribution. This will follow from a
choice of $C$ as an information-theoretic optimal compression
scheme. 

The completeness result implies that if $(\bh,\calubh)$ has a
good-on-average algorithm (according to one of the possible
definitions) then all problems $(L,\D)$ where $L$ is in $\np$ and $\D$
is polynomial time computable also have good-on-average algorithms.

The proof uses the fact that all {\em polynomial time computable}
distributions $\D$  allow polynomial time computable optimal
compression schemes. Many natural distributions are polynomial-time
computable, but there are a number of important exceptions. The
output of a pseudorandom generator, for example, defines a
distribution that is not  optimally compressible in polynomial time
and, hence, is not polynomial time computable.

\subsection{Decision versus Search} The second result that we
present, due to Ben-David et al. \cite{BCGL92}, shows that if
$(\bh,\calubh)$ has a good-on-average algorithm, then for all $\np$
relations $R$ and all polynomial time computable distributions $\D$,
there is an efficient algorithm that, given $x$ sampled from $\D$,
almost always finds a $y$ such that $R(x,y)$, provided that such a
$y$ exists. This shows that the question of whether there are
intractable-on-average search problems in NP (with respect to
polynomial-time computable distributions) is equivalent to the
question of whether there are intractable-on-average decision
problems in NP (with respect to such distributions). Both questions
are equivalent to the specific decision problem $(\bh,\calubh)$ being
intractable. 

\subsection{Computable, Samplable, and Arbitrary Distributions} 
The restriction of the completeness result to
samplable distributions is quite undesirable, because it rules out
reasonably natural distributions that can occur in certain
applications. Ideally, it would be desirable that the theory put no
restriction whatsoever on the distributions, and that we could prove
results of the form ``if there is a good-on-average algorithm for
$(\bh,\calubh)$, then for every $L$ in $\np$ and every distribution $\D$
there is a good-on-average algorithm for $(L,\D)$.'' The conclusion,
however, is equivalent to $\p = \np$.\footnote{This was first proved by
Levin. In \secref{sec:nonsamplable} we present a later proof by Li and
Vit\'anyi \cite{LV92}.} More specifically, there is a distribution
$\D$ such that, for every language $L$ in $\np$, if there is a
good-on-average algorithm for $(L,\D)$ then there is a
good-on-worst-case algorithm for $L$. As we discuss below, there are
difficulties in relating the worst-case complexity to the
average-case complexity of all problems in $\np$, and so it seems
unlikely that the theory can be generalized to handle completely
arbitrary distributions. An important intermediate case between
polynomial-time computable distributions and arbitrary distributions
is the class of {\em polynomial time samplable distributions}. Such
class includes some natural distributions that are not polynomial
time computable (for example, the output of a pseudorandom
generator), and an argument can be made that any distribution that
occurs ``in nature'' should be samplable. Impagliazzo and Levin
\cite{IL90} show that the completeness result can be extended to all
samplable distributions. That is, if $(\bh,\calubh)$ admits a
good-on-average algorithm, then for every problem $L$ in $\np$ and
every samplable distribution $\D$, the problem $(L,\D)$ has a
good-on-average algorithm. In Sections \ref{sec:il} and
\ref{sec:invert} we present two proofs of this result. A simpler
one, appearing in the paper of Impagliazzo and Levin, which applies
only to some (but not all) definitions of ``good-on-average,'' and a
second proof, also due to Impagliazzo and Levin, but unpublished,
that is more complex but that applies to all definitions. The first
proof is similar to the proof of the completeness result for
polynomial-time computable distributions, but using a randomized
encoding scheme. An input $x$ for $L$ is mapped into an input
$(M',(r,C(r,x)),1^t)$ for $\bh$, where $r$ is randomly chosen. The
desired properties of the randomized encoding $C$ are: (i) over the
choices of $r$, the encoding $x \rightarrow (r,C(x,r))$ is
``approximately injective,'' and (ii) the distribution $(r,C(x,r))$
is ``approximately uniform'' when $r$ is uniformly chosen and $x$ is
sampled from $\D$. Some additional difficulties arise: in order to
compute the randomized encoding one needs some extra information
about $x$, and the reduction just ``guesses'' all possible values
for this extra information, and, for technical reasons, this forces
us to work with the {\em search} rather than the {\em decision}
version of $L$. This is done without loss of generality given the
reduction of Ben-David et al. \cite{BCGL92}. The idea for the second
proof is that, if $S$ is the sampling algorithm for $L$, and $L$ is
hard-on-average over the outputs of $S$, then the problem ``on input
$r$, is it true that $S(r)\in L$?'' should be hard-on-average with
respect to the uniform distribution. This intuition is quite
difficult to translate into a proof, especially in the case in which
the computation of the sampler $S$ is a one-way function.

\subsection{Worst Case versus Average Case} In order to unify the
theory of average-case complexity with the rest of complexity
theory, it would be highly desirable to prove a theorem of the form
``if $\p \neq \np$ then there is a hard-on-average problem $(L,\D)$
where $L$ is in $\np$ and $\D$ is samplable.'' In order to prove such a
result via a reduction, we would need to find an oracle algorithm
$R$ (the reduction) such that if $A$ is a good-on-average algorithm
for $(L,\D)$ then $R^A$ is a good-on-worst-case algorithm for, say,
3SAT. Feigenbaum and Fortnow \cite{FF93} show that (under standard
assumptions) such a result cannot be proved via a {\em non-adaptive
random self-reduction}, that is, via an algorithm $R$ that makes
non-adaptive queries and such that each query has the distribution
$\D$ (regardless of the input of $R$). Bogdanov and Trevisan
\cite{BT03} show that the same impossibility result holds even if
$R$ is allowed to make arbitrary non-adaptive queries, provided that
$R$ works for arbitrary oracles. It remains possible that a
worst-case-to-average-case reduction in $\np$ exists which makes {\em
adaptive} access to the oracle, or that uses the {\em code} of the
algorithm $A$ (and, hence, does not work for arbitrary oracles).
Guttfreund and Ta-Shma \cite{GT06} make some progress in the latter
direction. An even more ambitious goal is to show, via reductions
that ``if $\p \neq \np$ then one-way functions exist.'' The result of
Bogdanov and Trevisan rules out the possibility of proving such a
result via oracle non-adaptive reductions; Akavia et al.
\cite{AGGM06} present a simpler proof in the setting of one-way
function (which, unlike the Bogdanov-Trevisan proof, works also in
the uniform setting) and are also able, for a restricted class of
one-way functions, to rule out non-adaptive reductions. See Section
\ref{sec:ff}.

\subsection{Degrees of Average-Case Intractability} If a problem $L$
is worst-case intractable, then every efficient algorithm  makes an
infinite number of  mistakes; if a problem $(L,\D)$ is average-case
intractable, then every efficient algorithm makes
mistakes\footnote{Or {\em fails}, depending on the definition of
average-case tractability that we are using.} on a set of inputs
that has noticeably large probability according to $\D$. Given the
difficulties in relating these two settings, it is interesting to
ask what happens if we consider different quantitative formulations
of ``noticeably large.'' O'Donnell \cite{OD02} shows that any
quantification between $1/2 - 1/n^{.33}$ and $1/\poly(n)$ leads 
essentially to an equivalent intractability
assumption. O'Donnell's argument, presented in Section
\ref{sec:amp}, gives a far-reaching generalization of Yao's XOR
Lemma \cite{Y82}.

\subsection{Specific Problems} Eventually, we would like the theory
to talk about the complexity of specific natural problems with
specific natural distributions.  It follows from Cook's reduction
that if there is a hard-on-average problem $(L,\D)$ where $L$ is in
NP and $\D$ is samplable, then every $\np$-hard problem is hard on
average with respect to some samplable distribution, albeit a very
unnatural one. On the other hand, Levin's completeness result shows
(under the same assumption) that there are hard-on average problems
$(L,\D)$ where $\D$ is uniform, but $L$ is quite artificial. Yet the
theory of average-case completeness has little to say about specific
cases of interest where both $L$ and $\D$ are natural: for instance
the hardness of \trisat\ or maximum independent set with respect to
natural distributions on inputs.

A specific problem whose average-case behavior has been widely
investigated is random $k$SAT with respect to the following
distribution of instances:  Choose at random $m_k(n)$ out of the $2^k\binom{n}{k}$ possible clauses of $k$SAT independently.  The 
tractability of this problem appears to depend
heavily on the number of clauses $m_k(n)$.  While it is believed
that random $k$SAT is hard for certain choices of $m_k(n)$, no
hardness result supporting this intuition is known.  However,
Feige~\cite{F02} shows the following surprising connection between
hardness of random \trisat\ and hardness of approximation:  Assuming
that random \trisat\ is hard for certain values of $m_3(n)$, it is
{\em worst-case hard} to approximate certain problems in $\np$
(e.g., maximum bipartite clique within $n^{-\epsilon}$ for some
$\epsilon > 0$.) 

For certain {\em lattice problems} we know an equivalence between 
worst-case and average-case complexity \cite{A96,M04,MR04,Regev05}. 
If such equivalences could be proved for
$\np$-complete lattice problems we would have a positive solution to
the question of whether the existence of hard-on-average problems in
$\np$ can be based on the worst-case assumptions on $\np$-complete
problems.

\section{A Historical Overview}

In this section we review the historical progression towards the
results described in the previous section.

\subsection {One-Way Functions and Cryptography}

The average-case performance of algorithms  on random inputs has
been studied since the beginning of the modern theory of efficient
algorithms in the 1950s and 1960s. Such work was often focused on
problems for which worst-case polynomial time algorithms were also
known. Volume 3 of the Art of Computer Programming \cite{TAOCP3}
(published in 1973) extensively surveys  average-case analyses of
algorithms for problems such as sorting and median-finding.

The study of the average-case of (conjectured) intractable problem
began in the 1970s motivated by the development of the foundations
of cryptography and by interest in heuristic approaches to
NP-complete problems.

When Diffie and Hellman \cite{DH76} introduced the notion of public-key
cryptography, they speculated that one could base a trapdoor permutation on the difficulty
of an NP-complete problem.\footnote{Indeed, Diffie and Hellman give two main
justifications for their claim that ``we stand on the brink of a revolution
in cryptography:'' the availability of cheap and efficient computers (in the 1970s!) and the development of $\np$-completeness.}
Even, Yacobi and Lempel \cite{EY80,L79} devised a public key cryptosystem
such  that an efficient adversary that
breaks the system
{\em for every key} implies an efficient algorithm for an \np-complete problem.
An efficient adversary that breaks the system {\em on almost all keys}, however,
is also discussed.

Shamir \cite{S79} discusses the difficulty in formulating a definition of
intractability for cryptographic applications. Worst-case complexity is
immediately seen as inadequate. Furthermore, Shamir emphasizes that a cryptographic system
cannot be considered secure if there is an attack that takes expected polynomial time.
In fact, Shamir adds, it is not even enough to rule out
expected polynomial time attacks. Consider for example a system that can be
broken by an attacker  whose {\em expected} running time
is very large but whose {\em median} running time
is efficient. This is possible if the attacker takes a very long time, say, on
one third of the keys but is efficient otherwise. Even though the expected
running time of the adversary is large, such a system cannot be considered secure.

The median running time of an adversary is thus a better complexity measure
of the expected running time, Shamir notes, but one needs
to go beyond, and consider the running time of, say, the 1\% fraction of inputs on which
the algorithm is fastest. This short discussion anticipates the formal definition
of one-way function and the difficulties in defining a robust
notion of ``average-case tractability'' in Levin's theory of average-case complexity.

The work of Blum, Goldwasser, Micali and Yao \cite{GM84,BM84,Y82}
put cryptography on solid foundational grounds, and introduced the
modern definitions of one-way functions, trapdoor permutation,
pseudorandom generator, and secure encryption. In their definition,
an efficiently computable function $f$ is one-way if there is no
polynomial time algorithm that finds a preimage of $f(x)$ with  more
than inverse polynomial probability over the choice of $x$. This
means that if $f$ is a one-way function then the computational
problem ``given $y=f(x)$ find a pre-image of $y$,'' has no algorithm
of expected polynomial time, no algorithm of median polynomial time,
no algorithm that runs in polynomial time on the easiest $1\%$
fraction of inputs, and so on.

\subsection{Levin's Theory of Average-Case Intractability}

The development of the theory of $\np$-completeness gave
evidence that a large number of important computational problems do not
admit worst-case efficient algorithms and motivated the design of
good-on-average algorithms as a way to ``cope'' with intractability.

Following this approach, the goal
is to analyse worst-case super-polynomial time algorithms for $\np$-complete
problems and to show that on ``typical'' instances they are efficient.
A celebrated example is  Karp's algorithm for TSP in the plane \cite{K77}.
An annotated bibliography by Karp et al.~\cite{KLMR85} written in 1985
reports several results on average-case tractability of $\np$-complete problems on natural
distributions.

The initial success in the design of good-on-average algorithms led to
the question of the limitations of such an approach. Are there $\np$-complete
problems that, with respect to natural distributions, do not even have
good-on-average algorithms? Are there general techniques, analogous to the
theory of $\np$-completeness, to prove average-case intractability?
\footnote{Interestingly, around the same time (mid 1970s), another approach was
studied to ``cope'' with the intractability of $\np$-complete optimization problems, namely,
to design provably efficient {\em approximate} algorithm that deliver near-optimal solution, and
the question was asked of when not even such algorithms exist.
In the 1990s, the theory of probabilistically checkable proofs gave
a powerful tool to prove intractability of approximation problems.
A satisfactory and general  theory to
prove average-case intractability, unfortunately, does not exist yet.}

Levin \cite{L86} laid the foundations for a theory of the
average-case tractability of problems in $\np$. Levin introduced the
definition of average-case tractability and of reduction outlined
above, and proved the first completeness result, for the class
$(\np,\pcomp)$ of problems $(L,\D)$ such that $L$ is in $\np$ and $\D$
is polynomial-time computable.

Levin's paper, both in the one-page conference version  \cite{L84} and in the two-page
full version \cite{L86}, gives few details about the intuition behind the
definitions and the possibility of generalized or alternative definitions.
Several researchers contributed to
the goal of explaining and generalizing Levin's ideas, including Johnson \cite{J84:levin}, Gurevich and McCauley \cite{GM87}, and Goldreich  \cite{G88}.

Ben-David et al.~\cite{BCGL92} consider two issues not addressed in
Levin's paper.  One issue is the class of distributions to consider.
Levin restricts his attention to the class of ``polynomial time
computable distributions,'' that includes several natural
distributions but that excludes, for example, the output of a
pseudorandom generator and other natural distributions. Ben David et
al.~observe that the more general class of ``efficiently samplable''
distributions is a better formalization of the notion of natural
distribution and formulate the question of whether Levin's
completeness result can be extended to the corresponding class
$(\np,\psamp)$ of distributional problems $(L,\D)$ such that $L$ is
in NP and $\D$ is samplable. Another issue studied in \cite{BCGL92}
is the average-case complexity of {\em decision} versus {\em search}
problems, and their main result shows that if every decision problem
in $\np$ can be solved efficiently with respect to the uniform
distribution, then every search problem in $\np$ can also be solved
efficiently with respect to the uniform distribution. Impagliazzo
and Levin~\cite{IL90}, solving the main open question formulated in
\cite{BCGL92}, prove that there is a problem that is complete for
$(\np,\psamp)$. Notions of reducibility between distributional problems, and what notions suffice to prove the Impagliazzo-Levin theorem \cite{IL90} that we present in Section 5, are
studied in \cite{BG93}.

\subsection{Average-Case Intractability and Derandomization}

Yao \cite{Y82} proves that the existence of pseudorandom generators
implies the possibility of derandomizing probabilistic algorithms, and
that pseudorandom generators can be constructed using one-way permutations.
(H\aa stad et al. \cite{HILL99} later proved that the existence of one-way functions is sufficient.) The existence of a one-way permutation $f$ can be stated as the average-case intractability of the {\em distributional search problem} of inverting $f$ on a random input, so Yao's result proves
that a specific average-case assumption (for certain search problems within \np)
implies derandomization of probabilistic algorithms.
The connection between average-case complexity and
derandomization became more direct, simpler, and more general in
the work of Nisan and Wigderson \cite{NW94}. Their work requires the
existence of hard-on-average distributional {\em decision} problems in EXP.
The work of Nisan and Wigderson raised the

question of whether derandomization could be based on
{\em worst-case} assumptions about problems in $\exptime$ instead of average-case assumptions. The question led to the
study of worst-case versus average-case complexity in $\exptime$, and to such tools as {\em random-self-reduction} \cite{BFNW93},
{\em amplification of hardness} \cite{I95,IW97}, and
{\em error-correcting codes} \cite{STV99}. As a result of this decade-long
investigation we now know that worst-case and average-case are equivalent in
complexity classes such as $\exptime$ and $\pspace$. The interested reader
can find an account of such results in a survey paper by Trevisan \cite{T04:codes} (see, in particular, Section 4) and in a survey paper by Kabanets \cite{K02:survey}.

\subsection{Worst-Case versus Average Case within $\np$}

The proofs of the worst-case and average-case equivalence for complete problems in $\exptime$, $\pspace$ and other classes raise the question whether a similar worst-case and average-case equivalence also holds for intractable problems within $\np$.  This is related to fundamental questions in the foundations of cryptography:  Is it possible to base one-way functions on $\np$-completeness?
If so, what about one-way permutations, or public key encryption?

It is easy to see that one-way permutations cannot be based on $\np$-completeness, unless $\np = \co\np$ (or $\am = \co\am$ if one allows randomized reductions, or $\np/\poly = \co\np / \poly$ if one allows non-uniform reductions). Not even the intractability of {\em worst case} inversion can be based on $\np$-completeness (see \secref{sec:permutations}).

On the other hand  it is possible to define ``one-way functions'' that are
computable in polynomial time and that cannot have a ``worst-case inverter''
(that is, a polynomial time inverter that works on all inputs) unless $\p = \np$.  For this reason, when we ask whether the existence of one-way functions
(under the standard, average-case, definition) can be based on $\np$-completeness, we are asking a question about the {\em average-case complexity} of inverters.

To clarify before we continue:  The existence of one-way permutations implies the existence of one-way functions, which implies the existence of hard-on-average distributional problems in $(\np,\psamp)$\footnote{This implication is non-trivial; see \secref{sec:searchowf}.} which implies that $\p$ is different from $\np$. We do not know how to prove the inverse of any of those implications, even though we believe that all the statements are true, and so they all imply each other vacously.

We can ask, however, whether reverse implications can be proved via {\em reductions}, that is, for example, whether there is a distributional problem $(L,\D)$ in $(\np,\psamp)$ and a reduction $R$
such that, for every algorithm $A$ that solves $(L,\D)$ well on average, the reduction $R$
plus the algorithm $A$ give a worst-case algorithm for \trisat.%\footnote{We are being intentionally vague, so far, about what is a ``reduction,'' and what it means to ``solve well on average.''}

Feigenbaum and Fortnow \cite{FF93} study a special case of the above
question. They consider the case in which $R$ is a ``non-adaptive random self-reduction.'' They show that the existence of such a reduction implies the collapse of the polynomial hierarchy (which contradicts standard conjectures.) The result of Feigenbaum and Fortnow rules out a certain way of proving equivalence of worst-case and average-case for $\np$-complete problems,
including the way used in the work on $\exptime$ and $\pspace$ \cite{BFNW93,I95,IW97,STV99} (see \secref{sec:owfhard}).

In a celebrated breakthrough, Ajtai \cite{A96}, describes a distributional
problem in $(\np,\pcomp)$ whose average-case complexity is at least as high as the worst-case complexity of a related (promise) problem in $\np$ --- a version of the shortest vector problem for lattices in $\R^n$.  Ajtai also proves the existence of one-way functions that are based on the worst-case complexity of problems in $\np$. Ajtai and Dwork \cite{AD97} present
a public-key cryptosystem based on a worst-case assumption, and Micciancio and Regev \cite{M04,MR04,Regev05} present various improvements.
%Connection to "learning parity with errors".

The security of the cryptosystems of Ajtai, Dwork, Micciancio and Regev
relies on the worst-case complexity of problems that are not known to be $\np$-complete and, in fact, are in $\np \cap \co\np$. It remains an open question
whether these techniques can be refined and improved to the point where
cryptography primitives can be constructed that rely on the worst-case
complexity of an $\np$-complete problem.

Bogdanov and Trevisan \cite{BT03} prove that no non-adaptive
worst-case to average-case reduction exist for $\np$-complete problems unless $\np/\poly = \co\np/\poly$. Akavia et al. \cite{AGGM06} prove that one-way functions cannot be based on $\np$-complete problems via non-adaptive reductions unless $\am = \co\am$ (see \secref{sec:owfhard}).

It seems likely that reductions cannot relate worst case
and average case hardness in $\np$. What about different degrees of average-case
intractability?  For instance, if there exist distributional problems in $\np$ that are hard on some non-negligible fraction of instances, does it follow that there are distributional problems in $\np$ that are hard on almost all instances?  These questions have been answered in the affirmative by O'Donnell \cite{OD02} and Healy, Vadhan, and Viola \cite{HVV04} in the non-uniform setting and by Trevisan \cite{T03,T05} in the uniform setting (see \secref{sec:amp}.)

\chapter{Definitions of ``Efficient on Average''}
\label{sec:defavg}

A {\em distributional decision problem} is a pair $(L,\D)$ where $L$ is
a language and $\D$ describes how inputs are distributed. There are various possible formalizations of how $\D$ is specified, of what constitutes a ``natural'' subset of distribution of inputs to restrict to, and of what it means for a distributional problem to have  a good-on-average algorithm. We discuss the various definitions, and the relations among them, in this section, which closely follows the treatment given by Impagliazzo in \cite{I95:average}.

\section{Distribution over Inputs}
\label{sec:distributions}

There are at least two common conventions on  how to specify $\D$. The convention introduced by Levin \cite{L86} is that $\D$ is a probability distribution over the set $\B^*$ of all possible bit strings. This convention is convenient in many applications, and, for example, it leads to a simple
definition of reduction preserving average-case algorithms. Sometimes, however,
the single-distribution convention leads to counter-intuitive
definitions: in the {\em uniform distribution} over $\B^*$, as defined by Levin,
each binary string of length $n$ has probability $\Theta(n^{-2} 2^{-n} )$.
In the single-distribution setting it is also harder to quantify average-case
hardness and to give definitions of circuit complexity, and both of these notions are important for applications to derandomization.

The other possibility is to define for each $n$ a finite distribution $D_n$,
with the intuition that $D_n$ is a distribution over inputs of ``size'' $n$,
and to let $\D$ be the {\em ensemble} $\D = \{ D_n \}_{n > 0}$. This
convention is common in cryptography and derandomization. In cryptography, it
is common to call $n$ the {\em security parameter} of the distribution $D_n$.

In this paper we adopt the second convention, where $\D$ is an ensemble of
distributions.  When discussing average-case complexity with respect to
samplable ensembles, the two definitions are essentially equivalent, as we
discuss in \secref{sec:sampl}.

In \secref{sec:complete} we discuss an average-case analog of the notion of
NP-completeness. Intuitively, we would like a definition of ``average-case
$\np$-hard'' distributional problem $(L,\D)$ such that if $(L,\D)$ is
average-case tractable (a notion that has several possible formalizations,
more later on this) then for every problem $L'$ in $\np$ and every ensemble
$\D'$, the distributional problem $(L',\D')$ is also average-case tractable.
Unfortunately, such an approach is unlikely to work:
\begin{itemize}
\item  As we show in Section \ref{sec:nonsamplable}
below, a conclusion of the form ``for every problem $L'$ in $\np$ and every $\D'$, the
distributional problem $(L',\D')$ is average-case tractable'' implies $\p = \np$, even if we
allow very weak notions of average-case tractability;
\item As we  show in
Section \ref{sec:ff}, it is unlikely that we can use reductions to prove
statements of the form ''if $(L,\D)$ is average-case tractable then $\p = \np$,'' where $L$ is in $\np$
and $\D$ is, say, the uniform ensemble.
\end{itemize}
Together, these two results imply that an average-case analog of the theory of NP-completeness
cannot refer to the class of all distributional problems $(L,\D)$ with $L$ in NP, and that
it is necessary to put some restriction to the class of distributions to be considered.

The most natural restriction is to consider {\em samplable} ensembles, that is, ensembles of distributions
that can be realized as outputs of a polynomial time sampling algorithm. There are, in turn, several
possible formalizations of the notion of samplable distributions: among other choices, we may
require the sampling algorithm to {\em always} run in polynomial time (in which case
the sampler is said to run in {\em strict polynomial time}) or to run in {\em expected}
polynomial time (the latter notion itself has various possible formalizations), and we may
require the output of the sampler to be a {\em perfect}, {\em statistical} or {\em computational}
simulation of the true distribution. The distinction between these various notions of efficient
samplability is important in the study of {\em zero-knowledge} protocols, and we refer the
reader to the chapter on Zero Knowledge in Oded Goldreich's book \cite{G01:book}. For our
purposes, it will convenient to just consider the simplest definition,
corresponding to perfect sampling with strict polynomial running time.\footnote{We stress,
however, that the results that we prove about samplable ensembles remain true even if
we adopt more relaxed definitions of samplability.}

\begin{definition}[Samplable Ensemble]
An ensemble $\D = \{ D_n \}$ is {\em polynomial time samplable} if there is a randomized algorithm $A$
that, on input a number $n$, outputs  a string in $\B^*$ and:
\begin{itemize}
\item There is a polynomial $p$ such that, on input $n$, $A$  runs in time at most $p(n)$, regardless
of its internal coin tosses;
\item For every $n$ and for every $x\in \B^*$, $\pr[A(n) = x ] = D_n(x)$.
\end{itemize}
\end{definition}

We will also be interested in a more restricted class of distributions, those for which the {\em cumulative} probability of a given string is efficiently computable. Let $\preceq$ denote the lexicographic ordering between bit strings, then if $D$ is a distribution we define $f_D(x) = D(\{y: y \preceq x\}) = \sum_{y \preceq x} D(y)$.

\begin{definition}[Computable Ensemble] We say that
an ensemble $\D = \{ D_n \}$ is {\em polynomial time computable} if
there is an algorithm that, given an integer $n$ and a string $x$,
runs in time polynomial in $n$  and computes $f_{D_n} (x)$.
\end{definition}

Observe that if $\{ D_n \}$ is a computable ensemble, then in particular the function $D_n(x)$ is computable in time polynomial in $n$.

We let $\psamp$ denote the class of polynomial-time samplable ensembles, and $\pcomp$ denote the class of polynomial time computable ensembles.

The {\em uniform ensemble} $\calU = \{ U_n \}$, where $U_n$ is the uniform
distribution over $\B^n$, is an example of a polynomial time computable ensemble.  Abusing notation, we also denote the class whose only member is the uniform ensemble by $\calU$.

It is not difficult to see that every polynomial-time computable ensemble is also polynomial-time samplable (see \secref{sec:compobs}).  The converse does not hold unless $\p = \p^{\#\p}$.  In fact, $\pcomp = \psamp$ if and only if $\p = \p^{\#\p}$.

\iffalse
More generally, Ben David et al. show that assuming one-way functions exist, there exists a samplable ensemble that is not dominated by any
computable ensemble.
\fi

\paragraph{Distributional Complexity Classes.}
A {\em distributional complexity class} is a collection of distributional decision problems.  For a class of languages $\mathbf{C}$ and a class of ensembles $\frak{D}$, we use $(\mathbf{C}, \frak{D})$ to denote the distributional complexity class consisting of all problems $(L, \calD)$ where $L \in \mathbf{C}$ and $\calD \in \frak{D}$.  In this survey we focus on the distributional complexity classes $(\np, \psamp)$, $(\np, \pcomp)$, and $(\np, \uniform)$.

\iffalse
The following result is easily verified.

\begin{proposition} If $\D$ is a polynomial time computable ensemble then it
is within negligible statistical distance of a polynomial time samplable ensemble.
\end{proposition}

Under complexity-theoretic assumptions, the converse is not true. For example,
the output of a pseudorandom generator is polynomial time samplable but not
polynomial time computable. Distributions that are polynomial time computable
but not polynomial time samplable can be constructed under weaker assumptions
than the existence of pseudorandom generators, see \cite{??}.
\fi

\section{Heuristic and Errorless Algorithms}
\label{sec:definitions}

In this section we define two notions of average-case tractability.

Suppose that we are interested in algorithms that are efficient on average for some samplable ensemble $\calD = \{D_n\}$.  For technical reasons, our algorithms are given, in addition to the input $x$, a parameter $n$ corresponding to the distribution $D_n$ from which $x$ was sampled.  We write $A(x; n)$ to denote the output of algorithm $A$ on input $x$ and parameter $n$.

\subsection{Average Polynomial Time and Errorless Heuristics}
We begin by considering algorithms that never make mistakes and that
are efficient on ``typical instances.'' A simple measure of average-case
complexity of an algorithm $A$ would be its expected running time, and
so we may think of defining an algorithm $A$ as having ``polynomial on average''
running time for a distributional problem $(L,\D)$ if there is a polynomial $p$ such that
\[ \E_{x \sim D_n} [ t_A(x; n) ] = \sum_{x \in \B^*} D_n(x) t_A(x; n) \leq p(n) \]
for every $n$, where $t_A(x; n)$ is the running time of $A$ on input $x$ and parameter $n$.

Such a definition is problematic because there are algorithms that we
would intuitively consider to be ``typically efficient'' but whose expected
running time is superpolynomial. For example, suppose that $A$ is an
algorithm of expected polynomial running time, and let $B$ be an algorithm
that is quadratically slower than $A$. (That is, for every $x$, $t_B(x; n) = (t_A(x; n))^2$.)
Then we should definitely think of $B$ as being typically efficient.
Suppose, however, that $D_n$ is the uniform ensemble and that
$A$ runs in time, say, $O(n^2)$ on all inputs of length $n$, except on a set
of $2^{n/2}$ inputs on which it takes time $O(2^{n/2})$; then the expected
running time of $A$ is $O(n^2)$ (the few ``hard inputs'' only contribute
an additive constant to the average running time). If $B$, however, is quadratically
slower than $A$, then $B$ takes time $O(n^4)$ on all inputs except on $2^{n/2}$
on which it takes time $O(2^n)$. The average expected running time of $B$ is now $O(2^{n/2})$,
dominated by the time taken on the hard inputs.

In order to be less dependent on the running time of exceptional inputs, we
may decide to look at the {\em median} running time instead of the expected
running time. Such a choice would work well with the above example: both $A$
and $B$ have polynomial median running time. More generally, if $A$ is an algorithm
of polynomial median running time and $B$ runs polynomially slower than $A$, then
$B$ must also have polynomial median running time.

Consider, however, an algorithm that runs in time $O(n^2)$ on $\frac 23 \cdot 2^{n}$ inputs
and in time $O(2^n)$ on $\frac 13 \cdot 2^n$ inputs of length $n$. Such an algorithm
has polynomial median running time with respect to the uniform ensemble, but intuitively
we wouldn't consider it to be a ``typically'' efficient algorithm.

We may choose to consider the $99$th percentile instead of the median, but each such
threshold would be arbitrary. What we would really like to capture with a definition
is the notion that a ``typically efficient'' algorithm may take very long, even exponential,
time on some inputs, but that the fraction of inputs requiring larger and larger running time
are smaller and smaller. In formalizing this intuition, it is natural to require
a {\em polynomial trade-off} between running time and fraction of inputs. This
leads us to our first definition.

\begin{definition}[Average Polynomial Running Time -- Trade-off Definition] \label{def:avgp-a}
An algorithm $A$ has average polynomial running time with respect to the
ensemble $\D$ if there is an $\epsilon>0$ and a polynomial $p$  such that for every $n$ and every  $t$:
\[ \pr_{x \sim D_n} [ t_A(x; n) \geq t] \leq \frac {p(n)} {t^\epsilon} \]
\end{definition}

If $A$ satisfies the above definition, then the median
running time of $A$ is polynomial, and, furthermore, $A$ runs in polynomial time
on all but at most a $1/n$ fraction of the inputs, in time at most $O(n^{O(\log n)})$
on all but at most a $1/n^{\log n}$ fraction of the inputs, and so on.
Levin gave the following equivalent definition.

\begin{definition}[Average Polynomial Running Time -- Levin's Definition] \label{def:avgp-b}
An algorithm $A$ has average polynomial running time with respect to the
ensemble $\D$ if there is an $\epsilon>0$  such that
\[ \E_{x \sim D_n}[t_A(x; n)^\epsilon] = O(n) \]
\end{definition}

Naturally, $O(n)$ can be replaced by an arbitrary polynomial in $n$.  The two definitions are easily seen to be equivalent.

\begin{proposition} An algorithm $A$ has average polynomial running time
with respect to the ensemble $\D$ according to Definition \ref{def:avgp-a} if and only
if it does according to Definition \ref{def:avgp-b}.
\end{proposition}
\begin{proof} Suppose that the running time $t_A$ of $A$ satisfies
\[ \pr_{D_n} [ t_A(x; n) \geq t] \leq n^c t^{-\epsilon} \]
for some constants $c,\epsilon$ and for every sufficiently large $n$.
Define $\delta = \epsilon / (c+2)$. Then

\begin{align*} \E_{D_n} [ t_A(x; n)^\delta ] & =  \sum_t \pr_{D_n} [ t_A(x; n)^\delta \geq t] \\
& \leq n + \sum_{t\geq n} \pr_{D_n} [ (t_A(x; n)) \geq t^{1/\delta} ] \\
& \leq n + \sum_{t\geq n}  n^c t^{-\epsilon/\delta} \\
& = n + \sum_{t\geq n} n^c t^{-(c+2)}\\
& \leq n + \sum_t t^{-2}\\
& =  n + O(1)
\end{align*}
This proves if $A$ satisfies Definition \ref{def:avgp-a} then it satisfies Definition \ref{def:avgp-b}.
For the other implication, suppose

\[ \E_{D_n} [ t_A(x; n)^\epsilon ] = O(n) \]
Then, by Markov's inequality
\[ \pr_{D_n} [ t_A(x; n) \geq t ] = \pr_{D_n} [ t_A(x; n)^\epsilon \geq t^\epsilon]
\leq \frac {\E_{D_n} [ t_A(x; n)^\epsilon ] }{t^\epsilon} = O( n t^{-\epsilon} ) \hfill\qed \]
\end{proof}

We now describe a third equivalent way to think of average polynomial time. Suppose that $A$ is an algorithm of average polynomial running time according to the above definitions. If we think about running $A$ ``in practice,'' it is reasonable to assume that we will not be able to run $A$ for more than a polynomial number of steps. We can then think of the inputs on which $A$ takes super-polynomial time as inputs on which $A$ ``fails,'' because we have to stop the computation without being able to recover the result.

The notion of an algorithm that fails on some inputs is captured by the following definition.

\begin{definition}[Errorless Heuristic Scheme]
We say that an algorithm $A$ is a {\em (fully polynomial-time) errorless heuristic scheme} for $(L, \D)$  if there is a polynomial $p$ such that
\begin{itemize}
\item For every $n, \delta > 0$, and every $x$ in the support of $D_n$, $A(x; n, \delta)$ outputs either $L(x)$ or the special failure symbol $\bot$;
\item For every $n, \delta > 0$, and every $x$ in the support of $D_n$, $A(x; n, \delta)$ runs in time at most $p(n/\delta)$;
\item For every $n$ and every $\delta > 0$,
\[ \pr_{x \sim D_n} [ A(x; n, \delta) = \bot ] \leq \delta\]
\end{itemize}
\end{definition}

We now show that errorless heuristic schemes are yet another way to capture
the notion of average-case tractability of Definition \ref{def:avgp-a} and
Definition \ref{def:avgp-b}.

\begin{proposition} A distributional problem $(L,\D)$ admits a fully polynomial
time errorless heuristic scheme if and only if it admits an algorithm whose
running time is {\em average-polynomial} according to Definition \ref{def:avgp-a} and
Definition \ref{def:avgp-b}.
\end{proposition}

\begin{proof} Suppose that $A$ is an algorithm that runs in average-polynomial
time according to Definition \ref{def:avgp-a}, that is, assume that there is a polynomial $p$ and an $\epsilon>0$ such that for every $n$,
\[ \pr_{D_n} [ t_A(x; n) \geq t] \leq \frac {p(n)} {t^\epsilon} \]
Then define the algorithm $A'$ that on input $x$ and parameters $n, \delta$ simulates $A(x; n)$ for $(p(n)/\delta)^{1/\epsilon}$ steps. If the simulation halts within the required number of steps, then $A'(x; n, \delta)$ gives the same output as $A(x; n)$; otherwise $A'(x; n, \delta)$ outputs $\bot$. It is easy to see that $A'$ satisfies the definition of an errorless heuristic scheme.

Suppose now that $A'$ is an errorless heuristic scheme for $(L,\D)$. Define the algorithm $A$ as follows: On input $(x; n)$, simulate $A(x; n, 1/2)$, if $A(x; n, 1/2)\neq \bot$, then return the output of $A(x; n, 1/2)$, otherwise simulate $A(x; n, 1/4)$, and so on, simulating $A(x; n, 1/8),\ldots,A(x; n, 2^{-k}),\ldots$ until we reach a value of $\delta$ such that $A(x; n, \delta) \neq \bot$. Eventually, the algorithm succeeds, because when $\delta < D_n(x)$ then $A(x; n, \delta)$ cannot output $\bot$.  After $k$ iterations, $A$ uses time $\sum_{i=1}^k p(2^i n) = O(k\cdot p(2^kn))$, for a polynomial $p$, and it halts within $k$ iterations on all but a $1/2^k$ fraction of inputs. It is now easy to verify that $A$ runs in average polynomial time according to Definition \ref{def:avgp-a}.
\end{proof}

Having given three equivalent formulations of ``efficient on average'' algorithms, we are ready to define a complexity class of distributional problems.

\begin{definition}[Average Polynomial Time]
We define $\avg \p$ to be the class of distributional problems that admit an errorless heuristic scheme.
\end{definition}

The third approach to the definition leads naturally to a finer quantitative definition.

\begin{definition}[Errorless Heuristic Algorithms]
Let $L$ be a language, $\D$ be an ensemble, and $\delta: \N \to \R^+$. We say that an algorithm $A$ is an {\em errorless heuristic algorithm} for $(L,\D)$ with {\em failure probability} at most $\delta$ if
\begin{itemize}
\item For every $n$ and every $x$ in the support of $D_n$, $A(x; n)$ outputs either $L(x)$ or the special failure symbol $\bot$, and
\item For every $n$, $\pr_{x \sim D_n} [ A(x; n) = \bot  ] \leq \delta(n).$
\end{itemize}

For a function $t: \N \to \N$, we say that $(L,\D) \in \avg_{\delta} \dtime(t(n))$ if there is an errorless heuristic deterministic algorithm $A$ that for every $n$ and every $x \in \supp(D_n)$ runs in time $t(n)$ with failure probability at most $\delta(n)$.

We define $\avg_{\delta} \p$ as the union over all polynomials $p$ of $\avg_\delta \dtime(p(n))$.
\end{definition}

We use $\avg_{\rm neg} \p$ to denote the union of all classes $\avg_\delta \p$, where $\delta$ is a negligible function.  Recall that $\delta$ is negligible if, for every polynomial $p$ and for every sufficiently large $n$, $\delta(n) \leq 1/p(n)$.

Observe that an errorless heuristic scheme for a distributional problem automatically yields errorless heuristic algorithms with error probability $1/p(n)$ for the same problem, for every polynomial $p$.  For certain problems, heuristic algorithms can conversely be turned into heuristic schemes.  We discuss this connection in \secref{sec:compobs}.

\subsection{Heuristic Algorithms}
So far we have considered only algorithms that never make mistakes: they always either produce a correct answer or fail. It is also interesting to consider algorithms that return incorrect answers on a small fraction of inputs, which is what we do next.

\begin{definition}[Heuristic Algorithms]
Let $L$ be a language, $\D$ be an ensemble, and $\delta: \N \to \R^+$. We say that an algorithm $A$ is a {\em heuristic algorithm} for $(L,\D)$ with {\em error probability} at most $\delta$ if for all $n > 0$,
\[ \pr_{x \sim D_n} [ A(x; n) \neq L(x)  ] \leq  \delta(n) \ . \]
\end{definition}

\begin{definition}[Heuristic Polynomial Time]
For functions $t: \N \to \N$ and $\delta: \N \to \R^+$, we say that $(L,\D) \in \heur_{\delta} \dtime (t(n))$ if there is a heuristic deterministic algorithm $A$ that for every $n$ and every $x \in \supp(D_n)$ runs in time $t(n)$ with failure probability at most $\delta(n)$.

We define $\heur_\delta \p$ as the union over all polynomials $p$ of $\heur_\delta \dtime(p(n))$.

We say that an algorithm $A$ is a {\em (fully polynomial-time) heuristic scheme} for $(L,\D)$  if there is a polynomial $p$ such that
\begin{itemize}
\item For every $n$, for every $x$ in the support of $D_n$ and every $\delta>0$,
$A(x; n, \delta)$ runs in time at most $p(n/\delta)$;
\item For $\delta>0$, $A(\cdot; \cdot, \delta)$ is a heuristic algorithm for $(L,\D)$ with error probability at most $\delta$.
\end{itemize}

We define $\heur\p$ to be the class of distributional problems that admit a heuristic scheme.
\end{definition}

We use $\heur_{\rm neg} \p$ to denote the union of all classes $\heur_\delta \p$, where $\delta$ is a negligible function.  

An errorless algorithm can be easily turned into a heuristic algorithm by replacing the failure symbol $\bot$ by an arbitrary output. Thus $\avg\mathbf{C} \subseteq \heur\mathbf{C}$ and $\avg_\delta\mathbf{C} \subseteq \heur_\delta\mathbf{C}$ for all classes of this type described above.

\section{Non-uniform and Randomized Heuristics}
We will also be interested in non-uniform and randomized heuristic algorithms.

Deterministic heuristics turn out to be an inadequate notion in much of average-case complexity, including many of the results stated in this survey.  For instance, the decision-to-search reduction of Ben-David et al. in \secref{sec:decsearch} and the reductions of Impagliazzo and Levin from $(\np, \psamp)$ to $(\np, \uniform)$ in \secref{sec:sampl} are both randomized, so to understand these reductions one must first define the notion of a randomized heuristic.  The results on hardness amplification in \secref{sec:amp} make use of both randomness and non-determinism.

However, the definitions of non-uniform and randomized heuristics contain some subtleties, and the the reader feels overwhelmed by definitions at this point he may skip ahead to \secref{sec:represent}.

\paragraph{Non-Uniform Heuristics.}
For a function $s:\N \to \N$, we define $\heur_{\delta}\size(s(n))$ and $\heur \p/\poly$ in the same way we define $\heur_{\delta} \dtime(t(n))$ and $\heur \p$, respectively, but referring to ``circuits of size $s(n)$'' instead of ``algorithms running in time $t(n)$.''  Similarly, we define the non-uniform errorless heuristic classes $\avg_{\delta} \size(s(n))$ and $\avg \p/\poly$. 

A small technical point is that, when we consider a distributional problem $(L,\{ D_n \})$, the inputs in the support of $D_n$ may have different lengths. In such a case, we need to fix a convention to allow Boolean circuits to accept
inputs of various lengths. Once such a convention is chosen,
then, for example, $(L,\{ D_n \}) \in \avg_{\delta} \size(s(n))$ means
that there is a family of circuits $C_n$ such that, for every $n$: (i) $C_n$
is of size at most $s(n)$; (ii) for every $x$ in the support of $D_n$, $C_n(x)$
outputs either $L(x)$ or $\bot$; (iii) $\pr_{x\sim D_n} [C(x)\neq L(x)] \leq \delta(n)$.

\paragraph{Randomized Heuristics.}
When defining randomized heuristic algorithms, there are two ways in which the algorithm can fail to produce a correct answer:  It can either run on an input on which the heuristic fails, or it can run on an input for which the heuristic is good but make a bad internal coin toss.  It is important to keep this distinction in mind when defining randomized {\em errorless} heuristic algorithms.  Here ``errorless'' refers to the choice of input and not to the internal coin tosses of the algorithm.

In particular, we allow the randomized errorless algorithm to sometimes output incorrect answers, as long as for every instance $x$, the fraction of random strings for which the algorithm outputs the wrong answer is small compared to the fraction of random strings for which it outputs either the right answer or $\bot$.  

\begin{definition}[Randomized Errorless Heuristics]
\label{dfn:randless}
Let $(L,\D)$ be a distributional problem and $\delta: \N \to \R^+$.
We say that a randomized polynomial-time algorithm $A$ is a {\em randomized errorless heuristic algorithm} of failure probability at most $\delta$ if,
for every $n > 0$, and every $x$ in the support of $D_n$,
\[ \pr [ A(x; n) \not\in \{ L(x), \bot \} ] \leq 1/4 \]
where the probability is taken over the coin tosses of $A$, and
\[ \pr_{x\sim D_n} \bigl[ \pr[ A(x; n) = \bot ] \geq 1/4 \bigr] \leq \delta(n) \]
where the inner probability is over the internal coin tosses of $A$.
\end{definition}

To see why this definition makes sense, fix an input $(x; n)$ and imagine running the algorithm $k$ times, for some large $k$.  If substantially more than $k/4$ --- say, $k/3$ --- of these runs return the failure symbol $\bot$, we can interpret this as a sign that the algorithm doesn't know the answer for $x$.  The second condition of \dfnref{dfn:randless}, together with standard Chernoff-type bounds, guarantees that this won't happen for more than a $\delta(n)$-fraction of instances $x \sim D_n$ with high probability over the randomness of the algorithm.  

If, on the other hand, the number of runs that return $\bot$ is smaller than $k/3$, then the first condition of \dfnref{dfn:randless} guarantees that with high probability, a majority of the runs that do not output $\bot$ will output the correct answer, so we obtain the correct answer for $x$ with high probability over the randomness of the algorithm.

This argument shows that the choice of constant $1/4$ is arbitrary, and any constant bounded away from $1/3$ can serve in the definition.  In the other direction, the algorithm $A'$ that simulates $A$ $k = k(n)$ times satisfies:
\begin{equation}
\label{eqn:randampl1}
\pr[ A'(x; n)\not\in \{ L(x), \bot \} ] = 2^{-\Omega(k(n))}
\end{equation}
and
\begin{equation}
\label{eqn:randampl2}
\pr_{x\sim D_n}\biggl[ \pr [ A'(x; n) = \bot ] \geq \frac 1 {2^{k(n)/100}} \biggr] \leq \delta(n).
\end{equation}

If the constant $1/4$ is replaced by $0$ in the first condition of \dfnref{dfn:randless}, we obtain the definition of {\em zero-error} randomized errorless heuristics.

\begin{definition}[Randomized Errorless Classes]
We say that $(L,\D)$ is in $\avg_\delta \bptime(t(n))$ if
there is a randomized errorless algorithm $A$ of failure
probability at most $\delta(n)$ and of running time at most
$t(n)$ on inputs in the support of $D_n$.  If $A$ is zero-error, we say that $(L, \D)$ is in $\avg_\delta \zptime(t(n))$.

We define $\avg_\delta \bpp$, $\avg \bpp$, $\avg_\delta \zpp$, and $\avg \zpp$ in the obvious way.
\end{definition}

If we choose $k(n) = O(n)$ in equations~(\ref{eqn:randampl1}) and (\ref{eqn:randampl2}), the probabilities over the internal coin tosses
of $A'$ can be made smaller than $2^n$, and using Adleman's proof
that $\bpp \subseteq \p/ \poly$~\cite{A78}, we have $\avg_\delta \bpp \subseteq \avg_\delta \p/ \poly$, $\avg \bpp \subseteq \avg \p / \poly$ and so on.

In the case of heuristic algorithms that are allowed to make errors the definition simplifies as we do not have to distinguish between errors owing to bad inputs and errors owing to bad internal coin tosses.

\begin{definition}[Randomized Heuristics]
Let $(L,\D)$ be a distributional problem and $\delta: \N \to \R^+$.
We say that a randomized algorithm $A$ is a randomized
heuristic of failure probability at most $\delta$ if for every $n$,
\[ \pr_{x\sim D_n} \bigl[ \pr [ A(x; n) \neq L(x) ] \geq 1/4 \bigr] \leq \delta(n) \]
where the inner probability is over the internal coin tosses of $A$.
\end{definition}

\begin{definition}[Randomized Heuristic Classes]
We say that $(L,\D)$ is in $\heur_\delta \bptime(t(n))$ if
there is a randomized errorless algorithm $A$ of failure
probability at most $\delta(n)$ and of running time at most
$t(n)$ on inputs in the support of $D_n$.  We define $\heur_\delta \bpp$ and $\heur \bpp$ in the obvious way.
\end{definition}

For all classes of the type $\avg_\delta\mathbf{C}$ and $\heur_\delta\mathbf{C}$ defined above, we define $\avg_{\rm neg}\mathbf{C}$ and $\heur_{\rm neg}\mathbf{C}$ as their union over all negligible functions $\delta$, respectively.

For the non-uniform and randomized heuristic classes, we have the standard containments $\avg\mathbf{C} \subseteq \heur\mathbf{C}$.  For the classes of type $\avg_\delta\mathbf{C}$ and $\heur_\delta\mathbf{C}$ it is possible to improve the containments in the deterministic case, as the algorithm can randomly (or non-uniformly) guess the answer for $\bot$, so that $\avg_\delta\mathbf{C} \subseteq \heur_{\delta / 2}\mathbf{C}$.

\section{Representing Inputs}
\label{sec:represent}

Average-case complexity is more sensitive to how we encode inputs to
algorithms than worst-case complexity.  For instance, operations like changing the alphabet or duplicating an instance do not have much effect in most treatments of worst-case complexity, while in average-case complexity they can considerably modify the distribution on inputs.

It will therefore be convenient to fix an encoding for inputs that is robust for average-case reductions and algorithms.  In the applications described in this survey, it will be necessary to have robust representations of the following types of inputs with respect to the uniform distributions:  tuples of strings, machines, and hash functions.  For instance, one feature of the encodings is that a random string in the uniform distribution will represent a valid tuple or a valid hash function with non-negligible probability.  It is not difficult to imagine why this is crucial for average-case algorithms.  In contrast, many natural encodings of these objects that are perfectly adequate in worst-case complexity do not have these property.

We do not try to optimize our representations in any manner; we simply choose representations that will be adequate for all applications covered in this survey.

\paragraph{Tuples.} We represent inputs to algorithms as strings in $\B^*$.  A good
representation for tuples of strings (in the uniform distribution) should have the property that the probability of generating a tuple $(x_1,\dots,x_t)$
should be roughly $2^{-(\abs{x_1} + \dots + \abs{x_t})}$.  We will adopt the following convention for tuples:  First, write a prefix free encoding of the number $\abs{x_1}$ by repeating every bit twice and ending with $01$.  Then write down $x_1$.  Repeat with $x_2$, $x_3$, up to $x_t$.  Thus the description
length of $(x_1,\dots,x_t)$ is $2\log\abs{x_1} + \dots + 2\log\abs{x_t} +
\abs{x_1} + \dots + \abs{x_t} + O(t)$.  Alternatively, the probability of
generating $(x_1,\dots,x_t)$ in the uniform distribution according to this
representation is $(\abs{x_1} \dots\abs{x_t})^{-2} 2^{-(\abs{x_1} + \dots +
\abs{x_t} + O(t))}$.  \begin{change}Observe that this representation is prefix-free.\end{change}

When all of the strings in the tuple have the same length more compact
representations are of course possible; such representations will be necessary for the results on hardness amplification in \secref{sec:amp}.

\paragraph{Machines.}
Sometimes the input (or a part of it) is the description of a machine.  The
exact way in which machines are represented is irrelevant, so we fix an
arbitrary representation for machines.

\paragraph{Hash functions.}  In \secref{sec:decsearch} and \secref{sec:sampl}, algorithms take as part of their input a description of a hash function $h$.  By "hash function" we mean a random instance from a family of pairwise
independent hash functions mapping $\B^m$ to $\B^n$ for fixed $m$ and $n$.  To
be specific, we can think of the family of affine transformations $h(x) = Ax +
b$, where $A$ is an $m \times n$ matrix, $b$ is an $n$ bit vector, and the
operations are over $\mathbb{Z}_2$.  We represent such transformations by
specifying the tuple $(A, b)$, so that the description length is $2\log m +
4\log n + mn + n + O(1)$.

For a function $h: \B^m \to \B^n$, we use $h|_j$ (where $1 \leq j \leq n$) to denote the function that consists of the first $j$ output bits of $h$. If $h$ is a hash function, then so is $h|_j$.

We will also consider hash functions from $\B^{\leq m}$ (the set of binary
strings of length at most $m$) to $\B^n$.  We will identify such functions
with hash functions from $\B^{m+1}$ to $\B^n$, where $\B^{\leq m}$ is embedded in $\B^{m+1}$ in the natural way:  String $x$ maps to $0^{m - \abs{x}}1x$.

\section{A Distribution for Which Worst-Case and Average-Case Are Equivalent}
\label{sec:nonsamplable}

In this section we show that there exists a (possibly non-samplable) ensemble of distributions with respect to which worst-case and average-case tractability are equivalent notions.  Thus the study of average-case complexity with respect to {\em all} ensembles reduces to the study of worst-case complexity, and in this sense it is natural to consider restricted classes such as computable and samplable ensembles, as we do in the remainder of this survey.

\begin{theorem}[Levin, Li and Vit\'anyi]
\label{th:LV} 
There is an ensemble $\D$ such that if $L$ is a decidable language and  the distributional
problem $(L,\D)$ is in $\heur_{1 / n^3} \p$,
then $L \in \p$.
\end{theorem}

We present a proof due to Li and Vit\'anyi \cite{LV92} that relies on Kolmogorov complexity.

\iffalse % now we have a section on encoding inputs
Fix a prefix-free binary encoding $\langle M,w\rangle$ of pairs $(M,w)$ where $M$ is a Turing machine and $w$
is a binary string. That is, if $(M,w)$ and $(M',w')$ are two different pairs then the binary string $\langle M,w\rangle$ is not a prefix of the binary string $\langle M',w' \rangle$.

For example, suppose that a binary description of $M$ is $\ell$ bits long and that
$w$ is $n$ bits long. Then $\langle M,w\rangle$ can be defined as follows: write the number $\ell$
in binary, but using $00$ to represent $0$ and $11$ to represent $1$; then write the string $01$ to
denote the end of the description of $\ell$, then write the binary representation of $n$ as above,
then write $01$, and finally write the binary description of $M$ followed by $w$. Overall, $\langle M,w\rangle$
has length $2\log \ell + 2\log n + \ell + n +O(1)$.
\fi

We consider pairs $(M, w)$, where $M$ is a machine and $w$ is a string.  Recall that if $M$ is $\ell$ bits long and $w$ is $n$ bits long, then $(M, w)$ has length $\ell + n + 2\log \ell + 2\log n + O(1)$.

For a binary string $x$, denote $K(x)$ as the length of the shortest pair $(M, w)$ such that $M$ on input $w$ outputs $x$. The value $K(x)$ is called the ({\em prefix-free}) {\em Kolmogorov complexity} of $x$.

The {\em universal probability distribution} $\cal K$ is defined so that the probability of a string $x$ is $2^{-K(x)}$.  \begin{change} Observe that $\sum_x 2^{-K(x)} \leq 1$ since the representation of $(M, w)$ is prefix-free.  (In fact, $\sum_x 2^{-K(x)} < 1$ so $K$ is technically not a probability distribution, but we can correct this by assigning, say, to the empty string $\epsilon$ the probability $1 - \sum_{x\neq 0} 2^{-K(x)}$.)\end{change}
Finally, let $\{ K_n \}$ be the ensemble of distributions where $K_n$ is the distribution $\cal K$
conditioned on strings of length $n$.

It turns out that for every language $L$, solving $L$ well on average with a heuristic algorithm
is as hard as solving $L$ well on the worst case.

\begin{proof*}[Proof of Theorem \ref{th:LV}]
We use the ensemble $\{K_n\}$ defined above.

Let $A$ be the polynomial time heuristic algorithm that witnesses $(L,\{ K_n \})\in \heur_{1/n^3} \p$.
We will argue that there is only a finite number of inputs $x$ such that $A(x; \abs{x}) \neq L(x)$, which implies that $L\in \p$.

We first need to understand the distributions $K_n$ in the ensemble. By
definition, \[ K_n(x) = \frac{2^{-K(x)}}{\sum_{y\in \B^n} 2^{-K(y)}} \]
and we can see that $\sum_{y\in \B^n} 2^{-K(y)} = \Omega(1 / n (\log n)^2)$
because the string $0^n$ has Kolmogorov complexity at most $\log n + 2\log\log n + O(1)$ and so contributes at least $\Omega(1 / n (\log n)^2)$ to the sum.

This implies
\[ K_n(x) = O(n(\log n)^2 \cdot 2^{-K(x)} ) = 2^{-K(x) + \log n + 2\log\log n + O(1)} \]
Let now $x$ be a string of length $n$ such that $A(x; n)\neq L(x)$; since the overall
probability of all such strings is at most $1/n^3$, in particular we must have $K_n(x) \leq 1/n^3$, and
\begin{align} \label{eq:LV-one} K(x) = \log \frac 1 {K_n(x)} - \log n -2\log\log n - O(1)
\geq 2\log n - 2\log\log n - O(1) \end{align}
Consider now the lexicographically  first string $x$ in $\B^n$ (if any) such that
$A(x; n)\neq L(x)$. Such a string can be computed by an algorithm that, given $n$,
computes $A(x; n)$ and $L(x)$ for all strings $x\in \B^n$ and outputs the lexicographically
first $x$ for which $A(x; n) \neq L(x)$. (Here we are using the assumption that $L$
is decidable.) Such an algorithm proves that $K(x) \leq \log n + 2\log\log n +O(1)$,
and, for sufficiently large $n$, this is in contradiction with (\ref{eq:LV-one}).

We conclude that there can only be a finite number of input lengths on which $A$
and $L$ differ, and so a finite number of inputs on which $A$ and $L$ differ.
\end{proof*}

\iffalse
\subsection{Proof Using Hard-Core Distributions}
The following proof is due to  Lipton and Young \cite{LY94}.
%Omitted because it requires non-uniform definitions
\fi

\chapter{A Complete Problem for Computable Ensembles}
\label{sec:complete}

In this section we give a definition of reduction
that preserves average-case tractability and we
prove the existence of a problem complete
for $(\np,\pcomp)$. We follow the treatment provided by Goldreich in \cite{G97}.

\section{Reductions Between Distributional Problems}

We begin by defining an appropriate notion of reduction.  Besides the usual correctness requirement for reductions in worst-case complexity, a reduction in average-case complexity must in some sense match the distributions on instances of the two problems.  Namely, in a reduction from $(L, \calD)$ to $(L', \calD')$, we want that the process of sampling an instance from $\calD$, then applying the reduction to it, roughly yields the distribution $\calD'$.

\begin{definition}[Reduction Between Distributional Problems] \label{def:reduction}
Let $(L,\D)$ and $(L',\D')$ be two distributional problems. We say that
$(L,\D)$ reduces to $(L',\D')$, and write $(L,\D) \ared (L',\D')$ if there is
a function $f$ that for every $n$, on input $x$ in the support of $D_n$ and
parameter $n$ can be computed in time polynomial in $n$ and

\begin{enumerate}
\item (Correctness) $x \in L$ if and only if $f(x; n)\in L'$
%\item There is an $\epsilon>0$ such that if $x$ is in the support of $D_n$ and $f(x)$
%is in the support of $D'_m$,
%then $m\geq n^\eps$
\item (Domination) There are polynomials $p$ and $m$ such that, for every $n$ and every $y$ in the support of $D'_{m(n)}$,
\begin{align*} \sum_{x: f(x; n) = y} D_n(x) \leq p(n) D'_{m(n)}(y) \end{align*}
\end{enumerate}
\end{definition}

Part (1) of the definition is the standard requirement of mapping reductions.
The intuition for part (2) is that if we sample a string $x$ from $D_n$ and
then compute $y=f(x; n)$, we generate $y$ with probability not much larger
than if $y$ had been sampled according to $D'_{m(n)}$.

The reduction preserves the notions of average-case tractability as defined in Section \ref{sec:defavg}.

\begin{lemma}
If $(L,\D) \ared (L',\D')$ and $(L',\D') \in \mathbf{C}$, where $\mathbf{C}$
is one of the distributional classes $\avg\p, \avg_{\rm neg} \p, \heur\p,
\heur_{\rm neg}\p, \avg\bpp,\heur\bpp,\avg\p/\poly,\heur\p/\poly$, then
$(L,\D) \in \mathbf{C}$.
\end{lemma}

\begin{proof}
For concreteness, we show the case $\mathbf{C} = \avg\p$, but the same proof
works for all the other cases.  Suppose that $(L',\D')$ is in $\avg\p$ and let
$A'$ be the fully polynomial time errorless heuristic scheme for $(L',\D')$,
let $f$ be the reduction from $(L,\D)$ to $(L',\D')$, let $p$ and $m$ be the
polynomials as in the definition of reduction.

We claim that $A(x; n, \delta) := A'(f(x; n); m(n), \delta / p(n))$ is a fully
polynomial time errorless heuristic scheme for $(L, \D)$.

To prove the claim, we bound the failure probability of $A$. Let us fix
parameters $n$ and $\delta$, and let us define $B$ to be the set of ``bad''
strings $y$ such that $A'(y; m(n), \delta /p(n))=\bot$, and let $B_m$ be $B$
restricted to the support of $D'_m$. Observe that $D'_{m(n)}(B_{m(n)}) \leq \delta/p(n)$. Then

\begin{align*}
\pr_{x\sim D_n} [ A(x; n, \delta) = \bot ] & = \sum_{x:  f(x; n) \in B_{m(n)}} D_n (x) \\
&\leq \sum_{y \in B_{m(n)}} p(n) D'_m(y)\\
& = p(n) \cdot D'_{m(n)}(B_{m(n)})\\
& \leq \delta
\end{align*}
This establishes the claim and proves that $(L,\D)\in \avg\p$.
\end{proof}

\section{The Completeness Result}

In this section we prove the existence of a complete
problem for $(\np,\pcomp)$, the class of all distributional
problems $(L,\D)$ such that $L$ is in $\np$ and $\D$
is polynomial time computable.  Our problem is the following ``bounded halting'' problem for non-deterministic Turing machines:
\begin{align} \bh = \{ (M,x,1^t): \text{$M$ is a non-deterministic
  Turing machine that accepts $x$ in $\leq t$ steps.} \}
\end{align}
Note that $\bh$ is \np-complete: Let $L$ be a language in \np\ and $M$ be a non-deterministic Turing machine that decides $L$ in time at most $p(n)$ on inputs of length $n$. Then a reduction from $L$ to $\bh$ is
simply the mapping that takes a string $x$ of length $n$ to the triple $(M,x,1^{p(n)})$.

We would like to show that the distributional problem $(\bh, \calubh)$, where $\calubh = \{ \ubh_n \}$ is the ``uniform'' ensemble of inputs for $\bh$ (we will get to the exact definition of this ensemble shortly) is complete for $(\np,\pcomp)$. The standard reduction is clearly inadequate, because, if $(L,\D)$ is a distributional problem in $(\np,\pcomp)$ and $\D$ is
a distribution that is very far from uniform, then the triples $(M,x,1^{p(n)})$ produced by the reduction will not be uniformly distributed.

The key idea in the reduction is to find an injective mapping $C$ such that
if $x$ is distributed according to $\D$ then $C(x)$ is distributed ``almost'' uniformly.
The reduction then maps $(x; n)$ into $(M',C(x),1^{p'(n)})$, where $M'$ is a machine
that on input $C(x)$ computes $x$ and then runs $M$ on $x$, and where $p'(n)$ is
a polynomial upper bound to the running time of $M'$. We will show that such a mapping
exists whenever $\D$ is a polynomial time computable ensemble.

Before moving on, let us define the ``uniform distribution'' of inputs for $\bh$. The instances of the problem are triples $(M, x, 1^t)$, so if the representation of $M$ has length $\ell$ and $x$ has length $n$, then the length of the representation of $(M, x, 1^t)$ is $2\log\ell + 2\log n + 2\log t + \ell + n + t + \Theta(1)$.

We think of the ``uniform distribution'' over inputs of length $N$
as follows: we flip random bits $b_1,\ldots,b_i$ until either $i=N$
or we have generated a valid prefix-free representation (according
to the above rules) of $M,x$. In the former case, we output
$b_1,\ldots,b_N$, in the latter case we output $(M,x,1^{N-i})$. We
denote this distribution by $\ubh_N$. In $\ubh_N$, an instance
$(M,x,1^t)$  has probability $2^{-(2\log\ell + 2\log n + \ell + n  +
\Theta(1))}$, where $\ell$ is the length of the representation of
$M$ and $n$ is the length of $x$.  (By convention, we declare that 
outputs not of the proper form $(M, x, 1^t)$ are not in the language $\bh$.)

We now prove the following completeness result, which is due to Gurevich \cite{G91}.

\begin{theorem}
\label{thm:comppcomp}
The distributional problem $(\bh, \calubh)$ is complete in $(\np,\pcomp)$ under
the reductions of Definition \ref{def:reduction}.
\end{theorem}

\begin{proof}
Let $(L,\D)$ be a distributional problem in $(\np,\pcomp)$.

\begin{claim}
Suppose $\calD = \{D_n\}$ is a polynomial-time computable distribution over
$x$.  Then there exists an algorithm $C(x)$ such that for all $n$, 
$C(x)$ runs in time polynomial in $n$ and
\begin{enumerate}
\item For every fixed $n$, for all $x$ in the support of $D_n$, $C(x)$ is injective as a function of $x$, and
\item $\abs{C(x)} \leq 1 + \min\left\{\abs{x}, \log\frac{1}{D_n(x)}  \right\}$.
\end{enumerate}
\end{claim}

Observe that since $D_n$ is polynomial-time computable, there exists
a polynomial $m(n)$ such that no string in the support of $D_n$ can 
be more than $m(n)$ bits long.  

\begin{proof}
Fix an $x \in \supp D_n$.  If $D_n(x) \leq2^{-\abs{x}}$\ then simply let $C(x) = 0x$, that is, $0$ concatenated with $x$.

If, on the other hand, $D_n(x) >2^{-\abs{x}}$,  let $y$ be the string that
precedes $x$ in lexicographic order among the strings in $\B^n$ and
let $p = f_{D_n}(y)$ (if $x$ is the empty string, then we let $p=0$.)  
Then we define
$C(x; n) = 1z$.  Here $z$ is the longest common prefix of
$f_{D_n}(x) $ and $p$ when both are written out in
binary. Since $f_{D_n}$ is computable in polynomial time, so is $z$.
$C$ is injective because only two binary strings $s_{1}$ and $s_{2}$
can have the same longest common prefix $z$; a third string $s_{3}$\
sharing $z$ as a prefix must have a longer prefix with either
$s_{1}$ or $s_{2}$.  Finally, since $D_n(x) \leq2^{-\abs{z}}$, $\abs{C(x)} \leq
1 + \log\frac{1}{D_n(x)}$.
\end{proof}

Let $M$ be the nondeterministic Turing machine that, on input $y$,
accepts if and only if there exists a string $x$ such that $y=C(x)$
and $x\in L$. Since $L$ is in \np, machine $M$ can be implemented so
that, on input $C(x)$, where $x$ is of length $n$, $M$ runs in time
at most $q(n)$, where $q$ is a polynomial.

We can now describe the reduction. On input $x$ and parameter $n$, the
reduction outputs the instance $(M,C(x),1^{t(x)})$ of length $N(n)$;  here, $N(n)$ is chosen large enough so that when $\abs{C(x)} \leq m(n)$, we have $t(x) \geq q(n)$ (for instance, $N(n) = m(n) + q(n) + 2\log m(n) + 2\log q(n) + O(1)$ suffices.)

It is immediate to see that $x\in L$ if and only if $(M,C(x),1^{t(x)}) \in \bh$. Regarding the domination condition, we observe that the reduction is injective, and so we simply need to check that for every $n$ and $x \in \supp D_n$ we have
\[ D_n(x) \leq \poly(n) \cdot \ubh_{N(n)}(M,C(x),1^{t(x)}). \]

To verify the inequality, let $\ell$ be the length of the binary representation of $M$.  We have
\[ \ubh_{N(n)}(M,C(x),1^{q(n)}) = 2^{-(2\log\ell + 2 \log |C(x)| + \ell + |C(x)| + \Theta(1))} \]
We observe that $\log\abs{C(x)} \leq \log(m(n)+1)$ and that $\abs{C(x)} \leq
\log (1/D_n(x)) + 1$, and so
\[ \ubh_{N(n)}(M,C(x),1^{q(n)}) \geq 2^{-(2\log\ell + \ell)} \cdot (m(n) + 1)^{-2} \cdot D_n(x)
\cdot \Omega(1)
\]
as desired.
\end{proof}

\section{Some Observations}
\label{sec:compobs}

\subsection{Completeness of Bounded Halting: A Perspective}
The main idea in the proof of \thmref{thm:comppcomp} is that it is possible to 
extract the randomness from samples in a computable ensemble.  In the proof of \thmref{thm:comppcomp}, the randomness is extracted through compression:  Indeed, the algorithm $C$ compresses samples $x$ from $D_n$ in such a way that the output $C(x)$ is dominated by the uniform distribution.

Another possible way to extract the randomness from samples of a computable ensemble is by inversion.  Namely, if one views an instance $x \sim D_n$ as the output of some sampler $S$, then the problem of extracting the randomness from $x$ can be solved by {\em inverting} $S$.  More precisely, one arrives at the following question:  Given $x$, is there an efficient procedure that produces a random $r$ such that $S(n; r) = x$?  Such a procedure would map samples of $D_n$ to samples of the uniform distribution and can be used to reduce the distributional problem $(L, \calD)$ to some distributional problem $(L', \calU)$.  This perspective leads to an alternate proof of \thmref{thm:comppcomp}.\footnote{The statement is actually weaker as the alternate reduction is randomized.}

\begin{proof*}[Alternate proof of \thmref{thm:comppcomp}]
First, it is not difficult to see that every polynomial-time computable ensemble
$\calD = \{D_n\}$ is also polynomial-time samplable.  To sample from a distribution $D_n$, the sampling algorithm $S(n)$ generates random bits $r_1,r_2,\dots,r_{m(n)}$ and, using binary search, returns the lexicographically smallest $x$ such that $f_{D_n}(x) > 0.r_1r_2\dots\/r_{m(n)}$.  Here, $m(n)$ is the running time of the algorithm that computes $f_{D_n}$, and we assume without loss of generality (for technical reasons) that $m$ is injective.  It is easy to check that each sample is produced with the correct probability.

Observe that the sampler $S$ is efficiently invertible in the following sense:  There exists an algorithm $I$ that on input $x \in \supp(D_n)$ runs in time polynomial in $n$ and outputs a uniformly random $r \in \B^{m(n)}$ conditioned on $S(n; r) = x$ (meaning that $S(n)$ outputs $x$ when using $r$ for its internal coin tosses.)  The algorithm $I$ first determines $f_{D_n}(x)$ and $D_n(x)$ using binary search and oracle calls to $f_{D_n}$, then samples a $m(n)$-bit number uniformly from the interval $(f_{D_n}(x) - D_n(x), f_{D_n}(x)]$.

Now consider the language $L'$ that contains all $r$ such that $S(n; r) \in L$, where $\abs{r} = m(n)$ (recall that $m$ is injective.)  Then $L'$ is an $\np$ language, and moreover $(L, \calD)$ reduces to the distributional problem $(L', \calU)$:  The reduction is implemented by the inversion algorithm $I$, and both the correctness and domination properties are straightforward from the definition.

Finally, consider the canonical reduction from $(L', \calU)$ to $(\bh, \calubh)$ which maps instance $r$ of $L'$ to instance $(M', r, 1^{q(\abs{x})})$ of $\bh$, where $M'$ is a non-deterministic Turing machine for L', and $q(n)$ is the running time of $M'$ on inputs of length $n$.  Let $\ell$ denote the size of $M'$, and $\abs{r} = m$.  Then for an appropriate choice of $N$, we have
$$\ubh_N(M', r, 1^{q(m)}) = 2^{-(2\log\ell + 2\log m + \ell + m + \Theta(1))}
  = 2^{-(2 \log\ell + \ell)} \cdot m^{-2} \cdot U_m(r) \cdot \Omega(1),$$
and this reduction also satisfies the domination condition (as $\ell$ does not grow with input size).  
\end{proof*}

The two proofs of \thmref{thm:comppcomp} are not that different, as the encoding function $C$ in the original proof plays much the same role as the inverter $I$ in the alternate proof.  However despite the somewhat artificial technical distinction, the perspectives are quite different:  To ``recover'' the uniform ensemble from a computable ensemble $\calD$, one may either attempt to compress $\calD$ or to invert its sampler. Indeed, the two approaches lead to different insights and different proofs (and even somewhat different theorems) when we extend these arguments to the case of polynomial-time samplable ensembles in \secref{sec:sampl}.

\subsection{Heuristic Algorithms versus Heuristic Schemes}
When defining average-case complexity classes we distinguished between heuristic algorithms and heuristic schemes:  For heuristic algorithms, we fix a failure probability $\delta$ and require that the algorithm succeeds on all but a $\delta$-fraction of the instances.  For heuristic schemes, we require a single algorithm that works for all $\delta$, but we allow the running time to grow as a function of $1/\delta$.

It is clear that if a distributional problem has a heuristic scheme, then it has heuristic algorithms with failure probability $\delta(n) = n^{-c}$ for every $c > 0$.  In other words, for every $c > 0$, $\heur\p \subseteq \heur_{n^{-c}}\p$, $\heur\bpp \subseteq \heur_{n^{-c}}\bpp$, $\avg\p \subseteq \avg_{n^{-c}}\p$, and so on.

In general the containments do not hold in the other direction:  For instance, $\heur_{n^{-c}}\p$ contains undecidable problems but $\heur\p$ doesn't.  However, the class $(\np, \pcomp)$ as a whole admits heuristic schemes if and only if it admits heuristic algorithms, as formalized in the following proposition.

\begin{proposition}
\label{prop:fraction}
If $(\bh, \calubh) \in \avg_{1/n}\mathbf{C}$ (respectively, $\heur_{1/n}\mathbf{C}$), then $(\np, \pcomp) \subseteq \avg\mathbf{C}$ (respectively, $\heur\mathbf{C}$).  Here, $\mathbf{C}$ is one of $\p$, $\bpp$, or $\zpp$.
\end{proposition}
\begin{proof}
For concreteness, let us show that if $(\bh, \calubh)$ is in $\avg_{1/n}\p$, then $(\np, \pcomp) \in \avg\p$.  By completeness of $(\bh, \calubh)$ with respect to distributional reductions, it is sufficient to show that $(\bh, \calubh) \in \avg\p$.

Let $A$ be an errorless heuristic algorithm for $(\bh, \calubh)$ with failure probability $1/n$.  Using $A$, we construct an errorless heuristic scheme $A'(\cdot; \cdot)$.  The idea is to use self-reducibility and padding in order to map short instances of $\bh$ into longer ones.  Since the error probability of $A$ decreases with instance length, the scheme $A'$ can solve any desired fraction of instances by choosing a padding of appropriate length.

We claim that the following $A'$ is an errorless heuristic scheme for $(\bh, \calubh)$:  $A'((M, x, 1^t); N, \delta) = A((M, x, 1^{t + \ceil{1/\delta}}); N + \ceil{1/\delta})$, where $N$ is the length of the instance $(M, x, 1^t)$.  (When the input is not of the proper form $(M, x, 1^t)$, $A'$ rejects it.)  From the definition of the ensemble $\calubh$, we have that for all $N$,
$$\ubh_{N + \ceil{1/\delta}}(M, x, 1^{t + \ceil{1/\delta}})
  = \ubh_N(M, x, 1^t).$$
On inputs from distribution $\ubh_{N + \ceil{1/\delta}}$, $A$ outputs $\bot$ on at most a $1 / (N + \ceil{1/\delta}) < \delta$ fraction of instances, so it follows that $A'$ outputs $\bot$ on at most a $\delta$ fraction of instances from $\ubh_N$.
\end{proof}

In fact, the error parameter $1/n$ in \propref{prop:fraction} can be replaced with $1/n^\epsilon$ for any fixed $\epsilon > 0$.

\subsection{Other Completeness Results}

Levin \cite{L86} established the first completeness result for $(\np,\pcomp)$, which applied to  a bounded tiling problem with respect to the uniform distribution, introducing the compression argument used in the proof of  \thmref{thm:comppcomp}. Gurevich  \cite{G91} applied the proof to the Bounded Halting problem. A number of other distributional problems have been shown to be complete for $(\np,\pcomp)$, including a graph coloring problem \cite{VL88,LV18}, a matrix decomposition problem \cite{G90,BG95},
a bounded version of the Post correspondence problem \cite{G91},  and diophantine matrix  problems \cite{VR92}.

\chapter{Decision versus Search and One-Way Functions}
\label{sec:decsearch}

In worst-case complexity, a search algorithm $A$ for an $\np$-relation
$V$ is required to produce, on input $x$, a witness $w$ of length
$\poly(\abs{x})$ such that $V$ accepts $(x; w)$, whenever such a $w$
exists.  Abusing terminology, we sometimes call $A$ a search algorithm for
the $\np$-language $L_V$ consisting of all $x$ for which such a witness
$w$ exists.  Thus, when we say ``a search algorithm for $L$'' we mean an
algorithm that on input $x \in L$ outputs an $\np$-witness $w$ that $x$
is a member of $L$, with respect to an implicit $\np$-relation $V$ such
that $L = L_V$.

Designing search algorithms for languages in $\np$ appears to be in general a
harder task than designing decision algorithms.  An efficient search algorithm
for a language in $\np$ immediately yields an efficient decision algorithm for
the same language.  The opposite, however, is not believed to be true in
general (for instance, if one-way permutations exist, even ones that are hard
to invert in the worst case).  However, even though search algorithms may be
more difficult to design than decision algorithms for specific problems, it is
well known that search is no harder than decision for the class $\np$ as a
whole:  If $\p = \np$, then every language in $\np$ has an efficient
(worst-case) search algorithm.

In this section we revisit the question of decision versus search in the
average-case setting: If all languages in distributional $\np$ have good
on average decision algorithms, do they also have good on average search
algorithms?  The answer was answered in the affirmative by Ben-David et
al., though for reasons more subtle than in the worst-case setting.
Their argument yields search to decision connections even for
interesting subclasses of distributional $\np$.  For instance, if every
language in $\np$ is easy on average for decision algorithms with
respect to the uniform distribution, then it is also easy on average for
search algorithms with respect to the uniform distribution.  We present
their argument in \secref{sec:std}.

From a cryptographic perspective, the most important distributional
search problem in $\np$ is the problem of inverting a candidate one-way
function.  By the argument of Ben-David et al., if all problems in
distributional $\np$ are easy on average, then every candidate one-way
function can be inverted on a random {\em output}.  In
\secref{sec:searchowf} we will see that this conclusion holds even
under the weaker assumption that every problem in $\np$ is easy on
average with respect to the uniform distribution.  Thus cryptographic
one-way functions can exist only if there are problems in $(\np, \uniform)$
that are hard on average for decision algorithms.

The search-to-decision reduction presented in this Section yields {\em
randomized} search algorithms for distributional $\np$.  We begin by
defining the types of search algorithms under consideration.

\section{Search Algorithms}
\label{sec:searchdef}

By analogy with worst-case complexity, it is easiest to define search
algorithms for $\np$ whose running time is polynomial on average.  For
illustration, we present the definition for deterministic algorithms.

\begin{definition}[Average polynomial-time search]
For an $\np$ language $L$ and ensemble of distributions $\calD$, we say $A$ is
a {\em deterministic average polynomial-time search algorithm} for $(L,
\calD)$ if for every $n$ and every $x$ in $L$ and in the support of $D_n$,
$A(x; n)$ outputs an $L$-witness for $x$ and there exists a constant
$\epsilon$ such that for every $n$, $\E_{x \sim D_n}[t_A(x; n)^\epsilon] =
O(n)$.
\end{definition}

As in the case of decision algorithms, the existence of average polynomial-time
search algorithms is equivalent to the existence of errorless heuristic
search algorithms, which we define next.  In the case of randomized
algorithms, the adjective ``errorless'' refers to the random choice of
an input from the language, and not to the choice of random coins by the
algorithm.  To make this distinction clear, we first define ``errorless
search'' in the deterministic case, then extend the definition
to the randomized case.

\begin{definition}[Deterministic errorless search]
We say $A$ is a {\em deterministic errorless search scheme} for $(L,
\calD)$, where $L \in \np$, if there is a polynomial $p$ such that
\begin{itemize}
\item For every $n, \delta > 0$, and every $x$ in the support of $D_n$, $A(x; n, \delta)$ runs in time at most $p(n/\delta)$;
\item For every $n, \delta > 0$, and every $x$ in $L$ and in the support of $D_n$, $A(x; n, \delta)$ outputs either an $L$-witness $w$ for $x$ or $\bot$;
\item For every $n$ and every $\delta > 0$, $\pr_{x \sim D_n}[A(x; n, \delta) = \bot] \leq \delta$.
\end{itemize}
\end{definition}

Observe that when $x \not\in L$, the output of the algorithm can be arbitrary.  If the algorithm outputs anything other than the special symbol $\bot$, this provides a certificate that $x$ is not in $L$, as it can be efficiently checked that the output of the algorithm is not a witness for $x$.

In the case of randomized algorithms, we can distinguish different types
of error that the algorithm makes over its randomness.  A ``zero-error''
randomized search algorithm is required to output, for all $x \in L$,
either a witness for $x$ or $\bot$ with probability one over its
randomness.  The type of search algorithm we consider here is allowed to
make errors for certain choices of random coins; namely, even if $x
\in L$, the search algorithm is allowed to output an incorrect witness
with probability bounded away from one.

\begin{definition}[Randomized errorless search]
We say $A$ is a {\em randomized errorless search algorithm} for $(L, \calD)$,
where $L \in \np$, if there is a polynomial $p$ such that

\begin{itemize}
\item For every $n, \delta > 0$, $A$ runs in time $p(n / \delta)$ and outputs either a string $w$ or the special symbol $\bot$;
\item For every $n, \delta > 0$, and $x \in L$, 
\[ \pr_A[\text{$A(x; n, \delta)$ outputs a witness for $x$ or $A(x; n, \delta) = \bot$}] > 1/2; \]
\item For every $n$ and $\delta > 0$, 
\[ \pr_{x \sim D_n}\bigl[\pr_A[A(x; n, \delta) = \bot] > 1/4\bigr] \leq \delta. \]
\end{itemize}
\end{definition}
This definition is robust with respect to the choice of constants $1/2$ and $1/4$; it would remain equivalent if $1/2$ and $1/4$ were replaced by any two constants $c$ and $c'$, respectively, where $0 < c' < c < 1$.  Using standard error reduction be repetition, the constants $1/2$ and $1/4$ can be amplified to $1 - \exp(-(n/\delta)^{O(1)})$ and $\exp(-(n/\delta)^{O(1)})$, respectively.

Finally, we define heuristic search algorithms:  Such algorithms are
allowed to output incorrect witnesses on a small fraction of inputs.

\begin{definition}[Randomized heuristic search]
We say $A$ is a {\em randomized heuristic search algorithm} for $(L, \calD)$,
where $L \in \np$, if for every $n$, on input $x$ in the support of $D_n$ and
parameter $\delta > 0$, $A$ runs in time polynomial in $n$ and $1/\delta$, and
$$\pr_{x \sim D_n}\bigl[\text{$x \in L$ and $\pr_A[\text{$A(x; n, \delta)$
    is not a witness for $x$}] > 1/4$}\bigr] \leq \delta.$$
\end{definition}

\section{Reducing Search to Decision}
\label{sec:std}

It is well known in worst-case complexity that the hardness of search and
decision versions of $\np$-complete problems are equivalent. Namely, if any
$\np$-complete problem has an efficient decision algorithm (on all instances),
then not only does all of $\np$ have efficient decision algorithms, but all of
$\np$ has efficient search algorithms as well.  The same question can be asked
for distributional $\np$:  If every decision problem in $\np$ has good on
average algorithms with respect to, say, the uniform distribution, does every
search problem in $\np$ also have efficient algorithms with respect to the
uniform distribution?

We show a result of Ben-David et al. that establishes the equivalence of
search and decision algorithms for $\np$ with the uniform distribution.
We focus on the uniform distribution not only because it is the most
natural distribution of instances, but also because the equivalence of
search and decision complexities for the uniform distribution will be
used to establish a much more general result in \secref{sec:il}.

Let us recall the common argument used to establish the
equivalence of $\np$-hardness for search and decision problems in the
worst-case setting, and see why this argument fails to carry over
directly to the average-case setting.  Given a decision oracle for
$\np$, and an instance $x$ of an $\np$-language $L$, a search algorithm
for $x$ finds a witness by doing binary search for the lexicographically
smallest $w$ such that the oracle answers ``yes'' on the $\np$-query:
\begin{quote}
$(x, w)$: Is there an $L$-witness for $x$ that is lexicographically at
most $w$?
\end{quote}

To see why this reduction is useless in the average-case setting with
respect to the uniform distribution, fix the lexicographically smallest
witness $w_x$ for every $x \in L$, and suppose that the average-case
decision oracle answers all queries correctly, except those $(x, w)$
where the distance between $w$ and $w_x$ in the lexicographic order is
small.  Then the algorithm obtains only enough information from the
oracle to recover the first few significant bits of $w_x$ and cannot
efficiently produce a witness for $x$.

To understand the idea of Ben-David et al., let us first consider the
special case when $L$ is an $\np$ language with {\em unique} witnesses.
Given an input $x$, the reduction attempts to recover a witness for $x$
by making oracle queries of the type
\begin{quote}
$(x, i)$: Does there exists a witness $w$ for $x$ such that the $i$th
bit $w_i$ of $w$ is $1$?
\end{quote}
for every $i = 1,\dots,m(\abs{x})$, where $m(n)$ is the length of a
witness on inputs of length $n$.  (Since $L \in \np$, we have that $m(n) = \poly(n)$.)  Given a worst-case decision oracle
for this $\np$ language, the sequence of oracle answers on input $x \in
L$ allows the search algorithm to recover all the bits of the unique
witness $w$.  In this setting, the reduction also works well on average:
Given an average-case decision oracle that works on a $1 - \delta/m(n)$
fraction of inputs $(x, i)$ where $\abs{x} = n$ and $i \leq m(n)$, the
search algorithm is able to recover witnesses (if they exist) on a $1 -
\delta$ fraction of inputs $x \sim U_n$.

In general, witnesses need not be unique.  However, using the isolating
technique of Valiant and Vazirani~\cite{VV86} it is possible to (randomly) map
instances of $L$ to instances of another $\np$-language $L'$ in such a
way that (1) The distribution of each query is dominated by uniform; (2)
If $x$ maps to $x'$, then any witness that $x' \in L'$ is also a witness
that $x \in L$, and (3) If $x \in L$, then $x$ maps to an instance $x'
\in L'$ with a unique witness with non-negligible probability.

The language $L'$ is defined as follows:
$$L' = \{(x, h, i, j): \text{there exists an $L$-witness $w$ for $x$
  such that $w_i = 1$ and $h|_j(w) = 0^j$}\},$$
where $i$ and $j$ are numbers between $1$ and $m(n)$, and $h$ is a hash
function mapping $\B^{m(n)}$ to $\B^{m(n)}$.  The argument of Valiant and
Vazirani guarantees that if $j$ is the logarithm of the number of
$L$-witnesses for $x$, there is a unique $w$ satisfying $h|_j(w) = 0$ with
constant probability over the choice of $h$.  The reduction $R$, on input $x
\sim U_n$, chooses a random hash
function $h: \B^{m(n)} \to \B^{m(n)}$ and queries the average-case oracle for
$L'$ on instances $(x, h, i, j)$, for all $i, j$ between $1$ and $m(n)$.

If, for any $j$, the sequence of answers to the queries $(x, h, i, j)$
received from the oracle is an $L$-witness for $x$, the search algorithm for
$L$ outputs this witness.  If no witness is found, a heuristic search
algorithm outputs an arbitrary string.  An errorless algorithm outputs the
special symbol $\bot$ if this symbol was ever encountered as an answer to a
query and an arbitrary string otherwise.

\begin{theorem}[Ben-David et al.]
\label{thm:search}
If $(\np, \uniform) \subseteq \avg\bpp$ (respectively, $\heur\bpp$),
then every problem in $(\np, \uniform)$ has an errorless (respectively,
heuristic) randomized search algorithm.
\end{theorem}

Observe that the search-to-decision reduction only applies to decision algorithms that succeed on most instances.  For the argument to achieve non-trivial parameters, the fraction of instances on which the decision algorithm fails must be smaller than $1 / m(n)^2$.

\section{Average-Case Complexity and One-Way Functions}
\label{sec:searchowf}

If every problem is easy on average for the uniform ensemble, can
one-way functions exist?  The above arguments show that in the case for
one-way permutations, the answer is no.  Given any efficiently
constructible family of permutations $f_n: \B^n \to \B^n$ solving the
search problem ``Given $y$, find $f^{-1}_n(y)$'' on most $y$
chosen from the uniform ensemble gives the ability to invert $f_n(x)$ on
a randomly chosen $x \sim U_n$.

In the general case, the answer is not immediately clear; to illustrate,
consider the case of a function $f_n: \B^n \to \B^n$ whose image has
density $2^{-n/2}$ in $\B^n$ under the uniform distribution.  An
average-case inversion algorithm for $f_n$ may fail to answer any
queries that fall into the image of $f_n$, yet be efficient with respect
to the uniform distribution by not failing on the other queries.

To rule out the existence of general one-way functions in this setting, it is
sufficient by H\aa stad et al. to show that no pseudo-random generators exist.
We argue that this is the case in the errorless setting, that is under the
assumption $(\np, \uniform) \subseteq \avg\bpp$.  Given a candidate
pseudo-random generator $G_n: \B^{n-1} \to \B^n$, consider the $\np$ decision
problem ``Is $y$ in the image set of $G_{\abs{y}}$?''  An errorless algorithm
$A$ for this problem must always answer ``yes'' or $\bot$ when the input is
chosen according to $G_n(U_{n-1})$.  On the other hand, $A(y; n, 1/4)$
must answer ``no'' on at least a $1/4$ fraction of inputs $y \sim U_n$,
since at least a $1/2$ fraction of such inputs are outside the image of $G_n$, and the algorithm is allowed to fail on no more than a $1/4$ fraction of inputs.  Hence $A$ distinguishes $G_n(U_{n-1})$ from the uniform distribution,
so $G_n$ is not a pseudo-random generator.

In the case of heuristic algorithms, this argument fails because there
is no guarantee on the behavior of $A$ on inputs that come from
$G_n(U_{n-1})$.  However, a different argument can be used to rule out
one-way functions under this more restrictive assumption.  H\aa stad et
al. show that if one-way functions exist, then a form of ``almost
one-way permutations'' exists: There is a family of strongly one-way
efficiently constructible functions $f_n: \B^n \to \B^n$ such that the
image of $f_n$ has non-negligible density in $\B^n$, that is
$U_n(f_n(\B^n)) = \sum_{x \in \mathrm{Image}(f_n)} U_n(x) \geq n^{-O(1)}$.  By choosing parameters appropriately, every such family of functions can be inverted on a large fraction of the image set $f_n(\B^n)$.  This gives an algorithm that
inverts $f_n(x)$ on a non-neglibible fraction of {\em inputs} $x$ and
contradicts the assumption that $f_n$ is strongly one-way.

In \secref{sec:sampl}, we give a different proof of this result
that bypasses the analysis of H\aa stad et al.  Summarizing, and using the
equivalence of weakly and strongly one-way functions, we have the
following:

\begin{theorem}
\label{thm:noowf}
If $(\np, \uniform) \subseteq \heur\bpp$, then for every polynomial-time
computable family of functions $f_n: \B^n \to \B^*$ there is a
randomized algorithm $I(y; n, \delta)$ running in time polynomial in
$n$ and $1/\delta$ such that for every $n$ and $\delta > 0$,
$$\pr_{x \sim U_n}[I(f_n(x); n, \delta) \in f_n^{-1}(f_n(x))] \geq 1 - \delta.$$
\end{theorem}

\chapter{Samplable Ensembles}
\label{sec:sampl}

The worst-case $\np$ hardness of computational problems does not always
reflect their perceived difficulty in practice.  A
possible explanation for this apparent disconnect is that even if a
problem may be hard to solve in the worst-case, hard instances of the
problem are so difficult to generate that they are never encountered.
This raises the intriguing possibility that an $\np$ hard problem, for
instance SAT, does not have an efficient algorithm in the worst
case, but generating a hard instance of SAT is in itself an infeasible
problem.  More precisely, for every sampler of presumably hard
instances from SAT, there is an efficient algorithm that solves SAT
on most of the instances generated by the sampler.

When the distribution of instances is known in advance, it makes sense
to restrict attention to a fixed sampler and design algorithms that work
well with respect to the output distribution of this sampler.  This is a
viewpoint commonly adopted in average-case algorithm design, where newer
algorithms for problems such as $k$SAT are designed that work well on
average for larger and larger classes of distributions on inputs.  From
a complexity theoretic perspective, on the other hand, one is more
interested in the inherent limitations of average case algorithms, and
it is natural to think of the sampler as chosen by an adversary that
tries to generate the hardest possible instances of the problem.

How much computational power should such a sampler of ``hard'' instances
be allowed?  It does not make sense to give the sampler more
computational power than the solver, since the solver must have at least
sufficient time to parse the instance generated by the sampler.  On the
other hand, in practice the sampler will have access to the same
computational resources as the solver, so if our notion of ``efficient
on average'' solver is that of a polynomial-time algorithm, the sampler
should also be allowed to perform arbitrary polynomial-time
computations.  This motivates the study of the distributional class $(\np, \psamp)$.

Even though instances drawn from a samplable ensemble
may be harder than instances drawn from a computable (or from the
uniform) ensemble for a specific problem in $\np$, it turns out
this is not the case for the class $\np$ as a whole: If uniformly
distributed inputs are easy for every problem in $\np$, then so are
inputs drawn from an arbitrary samplable ensemble.

\subsection{Samplable Ensembles versus Samplable Distributions.}
In the work of Ben-David et al.~\cite{BCGL92} that explains and extends Levin's original definitions from~\cite{L86}, a distribution over $\B^*$ is considered samplable if it is generated by a randomized algorithm $S$ that runs in time
polynomial in the length of its {\em output}.

Working with ensembles of samplable distributions instead of a single samplable distribution does not incur any loss of generality: In fact, for every samplable
distribution $\calD$ there exists a samplable ensemble $\{D_n\}$ such that $A$ is a heuristic scheme with respect to $\calD$ if and only if some
algorithm $A'$ (a slight modification of $A$) is a heuristic scheme with respect to $\{D_n\}$.  (The equivalence preserves the errorless property of heuristic schemes.)

To sketch the proof, let $X_n$ be the set of all $x \in \B^*$ such
that the sampler $S$ for $\calD$ outputs $x$ in $n$ or fewer steps.  Let
$D_n$ be the distribution $\calD$ conditioned on the event $x \in X_n$, so
that for every $x \in X_n$, $D_n(x) = \calD(x) / \calD(X_n)$.  Let $n_0$ be the smallest $n$ for which $\calD(X_n) \geq 1/2$.
The ensemble $\{D_n\}$ is samplable,\footnote{When $n \geq n_0$, run $S$ for $n$ steps repeatedly until a sample is produced; for smaller $n$, the distribution $D_n$ can be hard-coded in the sampler.  This sampler runs in {\em expected} polynomial-time, so $D_n$ does not in fact satisfy the definition on perfect samplability; however, it is within statistical distance $2^{-\poly(n)}$ of a samplable distribution, and we will ignore the distinction.} the support of $D_n$ is contained in
$\B^{\leq n}$, and $\calD(X_n) = 1 - o_n(1)$.  

Given an algorithm $A$ that is good on average for $\calD$, we define
\[ A'(x; n, \delta) = \begin{cases}
    A(x; \delta / 2),& \text{if $n \geq n_0$}, \\
    L(x),& \text{otherwise.} \end{cases} \]
For $n < n_0$, the distribution $D_n$ contains strings of length at most $n_0$, and the answers for these inputs are hardcoded into $A'$.  For $n \geq n_0$, we have
$$\pr_{x \sim D_n}[A'(x; n, \delta) \neq L(x)] \leq
  \pr_{x \sim \calD}[A'(x; n, \delta) \neq L(x)] / \calD(X_n) \leq
  \pr_{x \sim \calD}[A(x; \delta / 2) = \bot] / \tfrac12 \leq \delta.$$

Conversely, given an algorithm $A'$ that is good on average for $\{D_n\}$, we define
$$A(x; \delta) = A'(x; p(\abs{x}), \delta / 2\abs{x}^2),$$
where $p(n)$ is an upper bound on the time it takes $S$ takes to output a string of length $n$.  We have
\begin{align*}
\pr_{x \sim \calD}[A(x; \delta) \neq L(x)] &=
  \pr_{x \sim \calD}[A'(x; p(\abs{x}), \delta / 2\abs{x}^2) \neq L(x)] \\ &=
  \sum_{n=0}^\infty \pr_{x \sim \calD}[\text{$A'(x; p(n), \delta / 2n^2) \neq L(x)$ and $\abs{x} = n$}] \\ &\leq
  \sum_{n=0}^\infty \pr_{x \sim \calD}[\text{$A'(x; p(n), \delta / 2n^2) \neq
    L(x)$ and $S \to x$ in $p(n)$ steps}] \\ &\leq
  \sum_{n=0}^\infty \pr_{x \sim D_{p(n)}}[A'(x; p(n), \delta / 2n^2) \neq
    L(x)] \\ &\leq
  \sum_{n=0}^\infty \delta / 2n^2 < \delta.
\end{align*}

\section{The Compressibility Perspective}
\label{sec:il}

In \secref{sec:complete} we showed that the distributional problem $(\bh, \calubh)$ is complete for the class $(\np, \pcomp)$.  We did so by giving a reduction that maps instances of an arbitrary distributional problem $(L, \D)$ in $(\np, \pcomp)$ to instances of $(\bh, \calubh)$.

Recall that the key idea of the proof was to find an efficiently computable mapping $C$ with the following properties:
\begin{enumerate}
\item The map $C$ is injective, or equivalently, the encoding computed by
  $C$ is uniquely decodable.
\item When $x$ is distributed according to $\calD$, the output $C(x)$ is
  distributed ``almost'' uniformly.  If we think of $C$ as a compression
  procedure, it means that the rate of $C$ is close to optimal.
\end{enumerate}

In general it is not clear if an encoding $C$ with such properties exists for arbitrary samplable ensembles.  Our approach will be to gradually relax these properties until they can be satisfied for all samplable ensembles $\calD$.

To relax these properties, we look at randomized encodings.  First, observe that randomness can be added to the encoding without affecting the correctness of the reduction:  Suppose that $C$ is a mapping such that when $x$ is chosen according to the ensemble $\calD$, the image $C(x)$ is distributed almost uniformly.  Define a random mapping $C'$ that, on input $x$, chooses a uniformly random string $r$ of some fixed length and outputs the pair $(C(x), r)$.  It is evident that if the mapping $C$ satisfies conditions (1)-(3), then so does the mapping $C'$.  We use $C'(x; r)$ to denote the output of $C'$ on input $x$ and randomness $r$; thus $C'(x; r) = (C(x), r)$.

The advantage of a randomized encoding is that it allows for a natural relaxation of condition (1): Instead of requiring that the mapping be injective, we can now consider encodings that are ``almost injective'' in the sense that given $C'(x; r)$, the encoding needs to be uniquely decodable only with high probability over $r$.

In fact, we will further weaken this requirement substantially, and only require that $C'(x; r)$ be uniquely decodable with non-negligible probability.  Then a query made by the reduction is unlikely to be uniquely decodable, but by running the reduction several times we can expect that with high probability, at least one run of the reduction will yield a uniquely decodable query.

To summarize, we have the following situation:  We are given a reduction that queries $(\bh, \calubh)$ on several instances, and which expects to obtain the correct answer for at least one of these instances.  We do not know which of the instances produced by the reduction is the good one, but since $\bh$ is an $\np$ problem, instead of asking for a yes/no answer to the queries we can in fact ask for a witness that at least one of the queries produced by the reduction is a ``yes'' instance of $\bh$.  In fact, the search to decision reduction from \secref{sec:decsearch} shows that obtaining a witness is no harder than obtaining a membership answer (for randomized reductions.)

There is one important complication that we ignored in the last paragraph.  Many of the queries produced by the reduction may not be uniquely decodable.  Such queries may turn out to be ``yes'' instances of $\bh$ even if $x$ was a ``no'' instance of $L$, so certifying that a query $y$ is a ``yes'' instance $\bh$ is not sufficient to conclude that $x \in L$.  Indeed, we will need to certify
not only that $y \in \bh$, but also that $y$ is uniquely decodable.

\subsection{Reductions Between Search Problems}

We now formalize the properties of the reduction from the above discussion.  Since the reduction needs to access witnesses for membership of its queries, we formalize it as a reduction between search problems.  We only consider the case when one is reducing to a problem with respect to the uniform distribution, as this is our case of interest.

For two distributional problems $(L, \calD)$ and $(L', \calU)$ in $(\np,
\psamp)$, a {\em randomized heuristic search reduction} from $(L, \calD)$ to
$(L', \calU)$ is an algorithm $R$ that takes an input $x$ and a parameter $n$ and runs in time polynomial in $n$, such that for every $n$ and every
$x$, there exists a set $V_x \subseteq \supp{R(x; n)}$ (corresponding to the ``uniquely decodable'' queries) with the following properties:
\begin{enumerate}
\item Disjointness: There is a polynomial $p$ such that for every $n$, $V_x \subseteq \B^{p(n)}$ and the sets $V_x$ are pairwise disjoint.

\item Density: There is a polynomial $q_1$ such that for every $n$ and every
$x$ in the support of $D_n$, \[\pr_R[R(x; n) \in V_x] \geq 1/q_1(n).\]

\item Uniformity: For every $n$ and every $x$ in the support of $D_n$, the distribution of queries $y \sim R(x; n)$ conditioned on $y \in V_x$ is uniform.

\item Domination: There is a polynomial $q_2$ such that for every $n$ and every $x$, \[D_n(x) \leq q_2(n) \cdot U_{p(n)}(V_x).\]

\item Certifiability: There exists a polynomial-time algorithm $Q$ such that
for every $n$, if $x \in L$ and $y \in V_x$, then for every $L'$-witness 
$w$ for $y$, $Q(w)$ is an $L$-witness for $x$.
\end{enumerate}

A randomized search reduction is weaker than a reduction between
decision problems in that it is only guaranteed to work with small
probability, and only on ``yes'' instances.  However, if we are given a
randomized search algorithm for $L'$, it gives a randomized search
algorithm for $L$ as well, since it allows us to recover witnesses for
$L$ from witnesses for $L'$.  If we run the reduction several times, the
probability we hit a witness for $L'$ becomes exponentially close to
one, so the search algorithm for $L$ can be made to work with very high
probability on all instances.

\begin{claim}
\label{claim:reduction}
If there is a randomized search reduction from $(L, \calD)$ to $(L', \calU)$
and $(L', \calU)$ has a randomized heuristic search scheme, then $(L,
\calD)$ has a randomized heuristic search scheme.
\end{claim}
\begin{proof}
Let $A'$ be a randomized heuristic search scheme for $(L', \calU)$.
The search scheme $A$ for $(L, \calD)$ will run the reduction $N$ times, producing $N$ search queries for $A'$.  For each witness $w_i$ returned by $A'$,  $A$ will check whether $w_i$ yields a witness for $L$.

Specifically, on input $x$ and parameters $n$ and $\delta$, $A$ does the following:
\begin{enumerate}
\item Run $R(x; n)$ independently $N = 16 q_1(n)$ times, producing queries $y_1,\dots,y_N$.  
\item Compute $w_i = A'(y_i; p(n), \delta / 2q_2(n))$ for every $i$.  
\item If, for some $i$, $Q(w_i)$ is an $L$-witness for $x$, output $Q(w_i)$ (and otherwise output an arbitrary string.)
\end{enumerate}

Assume $x \in L$, and denote by $F$ the set of all $y$ on which $A'(y; \cdot)$ behaves incorrectly.  Specifically, let $F$ be the set of all $y$ such that $y \in L'$ but $A'(y; p(n), \delta / 2q_2(n))$ fails to return a witness of $y$ with probability $1/4$ or more.  Since $A'$ is a heuristic scheme for $L'$, we have that $U_{p(n)}(F) \leq \delta/2q_2(n)$.

Let $B$ be the set of all $x \in L \cap \supp{D_n}$ for which a large portion of the uniquely decodable queries $V_x$ are ``bad'' for $A'$ in the sense that they fall inside $F$.  Specifically, define $B$ as the set of all $x$ such that
$$U_{p(n)}(V_x \cap F) \geq U_{p(n)}(V_x)/2.$$

The set $B$ cannot have much weight according to $D_n$, since every $x \in B$ is ``responsible'' for many bad queries in $V_x \cap F$, and if there were many such queries then $F$ would be large.  In particular,
\begin{align*}
D_n(B) &= \sum_{x \in B} D_n(x) \\
 &\leq \sum_{x \in B} q_2(n) U_{p(n)}(V_x) \qquad \text{(by domination)} \\
 &\leq \sum_{x \in B} 2q_2(n) U_{p(n)}(V_x \cap F) \\
 &\leq 2q_2(n) U_{p(n)}(F) \leq \delta \qquad \text{(by disjointness.)} 
\end{align*}

Now fix $x \not\in B$, and consider one of the queries $y_i$ generated by $A$ in step (1).  We have that
\begin{multline*}
\pr[\text{$Q(w_i)$ is an $L$-witness for $x$}] \\
\begin{aligned}
  &\geq \pr[\text{$y_i \in V_x$ and $w_i$ is an $L'$-witness for $y_i$}]
    \qquad \text{(by certifiability)} \\ &\geq
  \pr[\text{$y_i \in V_x - F$ and $w_i$ is an $L'$-witness for $y_i$}] \\ &=
  \pr[y_i \in V_x] \cdot \pr[y_i \in V_x - F \cond y_i \in V_x] \\
    & \qquad \cdot
    \pr[\text{$w_i$ is an $L'$-witness for $y_i$} \cond y_i \in V_x - F] \\
     &\geq \frac{1}{q_1(n)} \cdot \frac12 \cdot \frac14 = \frac1{8q_1(n)},
\end{aligned}
\end{multline*}
by density, uniformity, and the definition of $F$.  By the choice of $N$, it follows that at least one of $Q(w_1),\dots,Q(w_N)$ is an $L$-witness for $x$ with probability $1/2$. 
\end{proof}

This claim shows randomized search reductions can be used to prove completeness results for $\heur\bpp$.  However, the proof of the claim does not extend to the class $\avg\bpp$, the reason being that the domination condition is too weak.  For heuristic algorithms, this condition guarantees that the algorithm $A'$ for $(L', \calU)$ will provide witnesses to most of the ``yes'' instances of $(L, \calD)$.  The ``evidence'' that an instance of $(L, \calD)$ is a ``no'' instance is that no such witness is found.

In the case of errorless algorithms, however, we need to certify ``no''
instances of $(L, \calD)$.  It is reasonable to attempt the following:
First, run the reduction several times to estimate the fraction of
queries that $A'$ answers by $\bot$.  If this fraction turns out too
large, this is evidence that $A'$ is unable to provide witnesses
reliably for this instance, so we answer $\bot$.  Otherwise, we look for
a witness and answer accordingly.  Unfortunately, the definition is 
insufficient to guarantee that $\bot$ won't be
answered too often, since it may be that the distribution of queries is
skewed in such a way that, whenever a query for $x$ falls outside $V_x$, the answer to this query is very likely to be $\bot$.

\subsection{Compressing Arbitrary Samplable Distributions}

Let $S$ be a polynomial time sampler that on input $n$ runs in time $m(n)$, where $m$ is some polynomial, and $D_n$ denote the distribution of the random variable $S(n)$.  As for computable distributions, our goal is to extract a sufficient amount of randomness from $S(n)$ so that the output of the extraction procedure is dominated by the uniform distribution.

To describe the approach, it is convenient to begin by considering the problem for certain restricted classes of distributions $D_n$, then gradually remove the restrictions until all samplable distributions are encompassed.  

We begin by considering the case of flat distributions:  We say that $D_n$ is {\em $k_n$-flat} if for each $x$ in the support of $D_n$, $D_n(x) = 2^{-k_n}$.  Flat distributions are convenient to consider because their randomness can be extracted via the Leftover Hash Lemma:  In particular, when $x$ is chosen from a $k_n$-flat distribution and $h$ is a random hash function from $\B^{< m(n)}$ into $\B^{k_n + 7}$, the output of the mapping $C_n(x; h) = (h, h(x))$ is dominated by the uniform distribution.  It is not difficult to check that $C_n$ satisfies the properties of randomized heuristic search reductions: The ``uniquely decodable'' strings $V_x$ are those pairs $(h, y)$ for which $h^{-1}(y) = \{x\}$.  By the choice of parameters, for every $x$ in the support of
$D_n$, $(h, h(x)) \in V_x$ for all but a small fraction of possible $h$, giving both density and domination. (Uniformity and certifiability are trivial.)

Now we consider a small but important generalization of flat distributions:  Instead of requiring that all samples in the support of $D_n$ have the same probability, we allow their probabilities to vary, but require that these probabilities be efficiently computable in the following sense:  There is an algorithm that on input $x$ and parameter $n$, runs in time polynomial in $n$ and computes the approximate entropy of $x$, which is the value
$$k_n(x) = \floor{-\log_2 D_n(x)} =
    m(n) - \ceil{\log_2 \#\{r: S(n; r) = x\}}.$$
Notice that $k_n(x)$ is an integer between $0$ and $m(n)$.  This scenario subsumes the previous one, where $k_n(x)$ was the same for all $x$ in the support of $D_n$.  The reasoning for flat distributions extends to this scenario, as long as we tailor the length of the output of the hash function to depend on the entropy $k(x)$.  Namely, the mapping $C_n(x; h) = (h, h|_{k_n(x) + 7}(x))$, where $h$ is a function mapping $\B^{< m(n)}$ to $\B^{m(n) + 7}$ satisfies the properties of randomized heuristic search reductions.

For arbitrary $S$, $k_n(x)$ could be difficult to compute and it is not clear if the approach of compressing samples via hashing can be extended.  One idea is for the reduction to attempt all possible values for $k_n(x)$, and declare $V_x$ to be the subset of encodings for which the guess was correct.  However, it
is now possible that strings of higher entropy (lower probability) than $x$ become possible decodings of $(h, h(x))$: There may be many such strings, and it is likely that some of them collide with $x$ under $h$.

The solution is to append the encoding $C_n(x)$ of $x$ with a ``certificate''
that the entropy of $x$ is not too high, namely that $k_n(x) \leq k$.  This roughly
amounts to certifying that the size of the set $\{r: S(n; r) = x\}$ is at least $2^{m(n) - k_n}$.
The certificate of this statement will be randomized: We ask to see a
string $r$ such that $S(r) = x$ and $g(r) = 0$ for a random hash
function $g$ that is approximately $2^{k_n}$-to-one.  Such a certificate is only 
approximately correct, but this is sufficient to guarantee that with constant 
probability, for a random $h$, $h(x)$ has a unique preimage for $h$ 
mapping $\B^{< m(n)}$ to $\B^{k_n + 7}$.

\subsection{The Construction}

Putting everything together, the encoding for $x$ chosen from distribution $D_n$ is
$$C_n(x; h, g, k) = (h(x), h, g, k),$$
where $k$ is a number between $0$ and $m(n)$, $h$ is a hash function
mapping $\B^{< m(n)}$ to $\B^{k + 7}$, and $g$ is a hash function
mapping $\B^{m(n)}$ to $\B^{m(n) - k - 4}$.  (In reality, $h$ maps to $\B^{m(n) + 7}$ 
and $g$ maps to $\B^{m(n) - 4}$ and we use the truncated versions $h|_{k+7}$ and 
$g|_{m(n) - k - 4}$ but for simplicity of notation we will not make this distinction.)  Let $p(n)$ 
denote the output length of $C_n$.

The ``uniquely decodable'' encodings are defined as follows:
\begin{quote}
$V_x$ is the set of all $(y, h, g, k)$ such that $k = k_n(x)$, $h(x) = y$, and
\begin{enumerate}
\item There is an $r$ such that $S(n; r) = x$ and $g(r) = 0$.
\item If $h(S(n; r)) = y$ and $g(r) = 0$, then $S(n; r) = x$.
\end{enumerate}
\end{quote}

The reduction $R$ maps instance $(x; n)$ to instance $(h(x), h, g, k)$ of the following $\np$-language $L'$:
\begin{quote}
$(y, h, g, k) \in L'$ if there exists an $r$ of length $< m(n)$ such that
$S(n; r) \in L$ and $h(S(n; r)) = y$ and $g(r) = 0^{m(n) - k - 4}$.
\end{quote}
Observe that a certificate that $(y, h, g, k) \in L'$ in particular contains a certificate that $S(n; r) \in L$, so under appropriate conditions witnesses for membership in $L$ can be extracted from the corresponding witnesses for $L'$.

\begin{theorem}[Impagliazzo and Levin]
\label{thm:il}
$(L, \calD)$ reduces to $(L', \calU)$ via a randomized search reduction.
\end{theorem}

Combining this result with the completeness of $(\bh, \calubh)$ for problems in $(\np, \calU)$, which follows from Cook's reduction (or as a special case of \thmref{thm:comppcomp}), and also using the search-to-decision equivalence of \thmref{thm:search}, we obtain the following corollary.

\begin{corollary}
\label{cor:il}
If $(\bh, \calubh) \in \heur\bpp$, then $(\np, \psamp) \subseteq \heur\bpp$.
\end{corollary}

\begin{proof*}[Proof of \thmref{thm:il}]
We show that the reduction $R$ satisfies the five conditions for
randomized heuristic search reductions.  Let us fix $n$.  Disjointness, uniformity, and certifiability follow from the definitions, so we focus on density and closeness.

Let $k_n(x) = \floor{-\log_2 D_n(x)} = m(n) -  \ceil{\log_2 \abs{\{r: S(n;
r) = x\}}}$.  Let $p(n)$ denote the length of the output of the reduction
when $x$ is chosen from $D_n$.

\noindent{\bf Density:} We show that $\pr_{h,g}[(h(x), h, g, k) \in V_x]$ is lower bounded by a constant conditioned on $k = k_n(x)$. 
Since $k = k_n(x)$ with probability at least $1/m(n)$, it will follow that
\[\pr_{R}[(h(x), h, g, k) \in V_x] = \Omega(1/m(n)).\]

We first show that with probability $7/8$, there exists an $r$ such that
$S(n; r) = x$ and $g(r) = 0$.  Observe that the number of $r$s
satisfying $S(n; r) = x$ is at least $2^{m(n) - k - 1}$.  Since the
range of $g$ is $\B^{m(n) - k - 4}$, in expectation there are at least
eight $r$s such that $S(n; r) = x$ and $g(r) = 0$.  By the pairwise
independence of $g$, at least one $r$ satisfies these conditions with
probability $7/8$.

We now show that there are at most $1/8$ fraction of pairs $h, g$ such
that $h(S(n; r)) = y$ and $g(r) = 0$ for some $r$ with $S(n; r) \neq x$.
Indeed,
\begin{multline*}
\pr_{h,g}[\text{$\exists r: S(n; r) \neq x$ and
  $h(S(n; r)) = h(x)$ and $g(r) = 0$}] \\
\begin{aligned}
  & \leq \sum_{r: S(n; r) \neq x} \pr_h[h(S(n; r)) = h(x)]
      \pr_g[g(r) = 0] \\
  & \leq \sum_{r \in \B^{< m(n)}} 2^{-k-7}2^{-m(n)+k+4} = 1/8.
\end{aligned}
\end{multline*}
It follows that each of conditions (1) and (2) in the definition of
$V_x$ is satisfied with probability $7/8$ separately, so that
$$\pr_{h,g}[(h(x), h, g, k) \in V_x \cond k = k_n(x)] \geq 3/4.$$

\noindent{\bf Domination:}  Observe that for given $n$, a random instance of $\ubh_{p(n)}$ is a $4$-tuple of the correct form $(y, h, g, k)$ with probability at least $1/\poly(p(n))$.  Therefore
\begin{align*}
\ubh_{p(n)}(V_x) &= 
  \pr_{y, g, h, k}[(y, h, g, k) \in V_x] \cdot 1 / \poly(p(n)) \\
  &\geq \pr_{h,g}[(h(x), h, g, k) \in V_x \cond k = k_n(x)] \\
  & \qquad \pr_y[y = h(x) \cond k = k_n(x)]\pr_{k}[k = k_n(x)] \cdot 1 / \poly(p(n)) \\
  &\geq 3/4 \cdot 2^{-k_n(x) - 7} \cdot 1 / (m(n) \poly(p(n))) \\
  &= \Omega(D_n(x) / m(n)\poly(p(n))). \hfill \qed
\end{align*}
\end{proof*}

An important example of a problem in $(\np, \psamp)$ is the problem of
inverting a supposed one-way function $f_n: \B^n \to \B^*$:  The
question of finding an inverse $f_n^{-1}(y)$ is an $\np$ question, and
the distribution ensemble on which the function ought to be inverted is
$\{f_n(U_n)\}$.  Therefore, if $(\bh, \calubh)$ has a heuristic scheme, then no one-way functions exist.

\section{The Invertibility Perspective}
\label{sec:invert}

In this section we present a different proof that $(\np, \psamp)$ is
no harder on average than $(\np, \uniform)$ for randomized algorithms.
This proof works for heuristic as well as errorless algorithms.

Ignoring efficiency considerations for the moment, given an $\np$
language $L$ and a polynomial-time sampler $S$, the distributional
problem ``Compute $f$ on input $x$'', where $x \sim S(n; U_{m(n)})$,
can be solved by first sampling a random $r \sim U_{m(n)}$ conditioned
on $S(n; r) = x$, and then solving the distributional problem ``Compute
$f(S(r))$ on input $r$.''  Observe that given an algorithm that solves
the latter problem well on average with respect to the uniform ensemble
yields an algorithm for the original problem with respect to the
ensemble $S(n; U_{m(n)})$.

The difficulty, of course, is in efficiently carrying out the step of
sampling a random $r$ conditioned on $S(n; r) = x$.  In a general
setting this does not seem possible, as $S(n; r)$ may be a one-way
function of $r$, in which case finding any, let alone a random preimage
of $x$, is an impossible task.

However, if all of $(\np, \uniform)$ has efficient on average algorithms, by
\thmref{thm:noowf} there are no one-way functions.  Impagliazzo and Luby~\cite{IL89}
show that if there are no one-way functions then there are no {\em
distributionally} one-way functions: Given any efficiently computable
family of functions $f_n: \B^n \to \B^*$, for most $y$ it is possible to
efficiently sample an $x$ such that $f_n(x) = y$ and the distribution of
$x$ conditioned on $f_n(x) = y$ is close to uniform.  More precisely,
there exists a (randomized) algorithm $I$ running in time polynomial in
$n$ and $1/\delta$ such that the statistical distance between the
distributions $(x, f_n(x))$ and $(I(f_n(x); n, \delta), f_n(x))$ is at
most $\delta$.  In particular, given an input $x \sim S(n; U_{m(n)})$,
it is possible to sample an almost uniform $r$ such that $S(n; r) = x$.

\begin{theorem}[Impagliazzo and Levin]
\label{thm:ilalt}
If $(\np, \uniform) \subseteq \avg\zpp$ (respectively, $\heur\bpp$),
then $(\np, \psamp) \subseteq \avg\zpp$ (respectively, $\heur\bpp$).
\end{theorem}
\begin{proof}
Consider an arbitrary problem $(L, \calD) \in (\np, \psamp)$.  Let $S$ be the 
polynomial-time sampler for $\calD$.  Assume without loss of generality that
on input $n$, $S$ uses exactly $m(n)$ random bits and that $m$ is an injective function.
Under the assumption of the theorem, by \thmref{thm:noowf} and the
result of Impagliazzo and Luby, there is an algorithm $I$ running in
time polynomial in $n$ and $1/\delta$ and such that for every $n$, the
statistical distance between the distributions
\begin{equation}
\label{eqn:ind}
\{(r, S(r)): r \in \B^{m(n)}\} \qquad \text{and} \qquad
  \{(I(S(r)), S(r)): r \in \B^{m(n)}\}
\end{equation}
is at most $\delta / 3$.  (For simplicity of notation, we omit the parameters
$n$ and $\delta$ in parts of the proof.)  Let $A$ be a heuristic
scheme for the distributional problem $(L \circ S, \uniform)$, where $L
\circ S$ is the $\np$ language $\{r: \text{$S(r)$ is a yes instance of
$L$}\}$.

We show that the algorithm
$$B(x; n, \delta) = A(I(x); m(n), \delta / 3)$$
is a heuristic scheme for $(L, \calD)$.  Observe that if $A$ is
errorless then $B$ is also errorless (since $I$ can be made errorless by
checking that $S$ maps its input to its output, and outputing $\bot$ if
this is not the case.)  Now, it is sufficient to show that
$$\pr_{x \sim S(n; U_{m(n)})}[B(x) = L(x)] = \pr_{r \sim
    U_{m(n)}}[B(S(r)) = L(S(r))] \geq 1 - \delta.$$
We relate the probability of the event $B(S(r)) = L(S(r))$ to the
probability of the event $A(r) = L(S(r))$.  By indistinguishability
(\ref{eqn:ind}), for any event $E$, the probabilities of $E(r)$ and
$E(I(S(r)))$ when $r \sim U_{m(n)}$ can differ by at most $\delta /
3$, so in particular
\begin{align*}
\pr_{r \sim U_{m(n)}}[A(r) = L(S(r))]
    &\leq \pr_{r \sim U_{m(n)}}[A(I(S(r))) = L(S(I(S(r))))] + \delta/3 \\
    &= \pr_{r \sim U_{m(n)}}[B(S(r)) = L(S(I(S(r))))] + \delta / 3.
\end{align*}
Applying indistinguishability (\ref{eqn:ind}) again, the
distributions $(S(r), S(r))$ and $(S(I(S(r))), S(r))$ are $\delta / 3$
statistically close, so in particular $\pr_r[S(r) \neq S(I(S(r)))] <
\delta / 3$ and
\begin{multline*}
\pr_{r \sim U_{m(n)}}[B(S(r)) = L(S(I(S(r))))] \\
\begin{aligned}
  &\leq \pr_{r \sim U_{m(n)}}[\text{$B(S(r)) = L(S(I(S(r))))$
       and $S(r) = S(I(S(r)))$}] \\
  &\qquad + \pr_{r \sim U_{m(n)}}[S(r) \neq S(I(S(r)))] \\
  &\leq \pr_{r \sim U_{m(n)}}[B(S(r)) = L(S(r))] + \delta / 3.
\end{aligned}
\end{multline*}
Putting the last two equations together, we obtain
$$\pr_{r \sim U_{m(n)}}[B(S(r)) = L(S(r))]
  \geq \pr_{r \sim U_{m(n)}}[A(r) = L(S(r))] - 2\delta / 3
  \geq 1 - \delta. \hfill \qed$$
\end{proof}

Notice that the assumption that $(\np, \uniform)$ has good on
average algorithms was used twice in the proof:  Once to invert the
sampler $S$ and once to solve $L \circ S$ on the uniform distribution.
In other words, given an average-case oracle for $(\bh, \calubh)$, to obtain
an algorithm for a problem in $(\np, \psamp)$ one needs to place two
rounds of queries to the oracle.  The first round of queries is used to
obtain a preimage $r$ of $x$ under $S$, and the second round (in fact, a
single query) is used to solve $L \circ S$ on input $r$.  In contrast,
\thmref{thm:il} solves problems in $(\np, \psamp)$ using a single round
of oracle queries.

\chapter{Hardness Amplification}
\label{sec:amp}

Generally speaking, the goal of {\em hardness amplification} is to start
from a problem that is known (or assumed) to be hard on average in a weak sense
(that is, every efficient algorithm has a noticeable probability of making a mistake
on a random input) and to define a related  new problem that is hard on average in the
strongest possible sense (that is, no efficient algorithm can solve the problem noticeably
better than by guessing a solution at random).

\section{Yao's XOR Lemma}

For decision problems, Yao's XOR Lemma \cite{Y82} is a very powerful
result on amplification of hardnes. In the XOR Lemma, we start
from a Boolean function $f:\B^n \to \B$
 and define a new function $f^{\xor k} (x_1,\ldots,x_k) := f(x_1)\xor \cdots \xor f(x_k)$,
and the Lemma says that if every circuit of size $\leq S$ makes at least a $\delta$ fraction
of errors in computing $f(x)$ for a random $x$, then every circuit of size $\leq S \cdot \poly(\delta\epsilon/k)$
makes at least a $1/2 - \epsilon$ fraction of errors in computing $f^{\xor k}$,
where $\epsilon$ is roughly $\Omega((1-\delta)^k)$.

Various proofs of the XOR Lemma are known~\cite{L87,BL93,I95,GNW95,IW97}. In this
section we describe Impagliazzo's proof \cite{I95}, because it is based on
a tool, Impagliazzo's ``hard core distribution'' theorem, that will be
very useful later.

For simplicity, we will restrict ourselves to results in the non-uniform (circuit complexity)
setting. The following definition will be useful.

\begin{definition} We say that a Boolean function $f:\B^n \to \B$ is $(S,\delta)$-hard
with respect to a distribution $D$ if, for every circuit $C$ of size $\leq S$, we have
\[ \pr_{x\sim D} [ f(x)\neq C(x) ] > \delta \]
\end{definition}

To relate this definition to our previous definitions, observe that
$(L,\{ D_n \} )\in \heur_{\delta(n)}\size(S(n))$
if and only if, for every $n$,  $L_n$ is not $(S(n),\delta(n))$-hard with respect to $D_n$, where 
$L_n: \B^n \to \B$ is the characteristic function of the set $L \cap \B^n$.

Impagliazzo  \cite{I95} proves that, if a Boolean function is ``mildly'' hard
on average with respect to the uniform distribution, then there is a large
set of inputs such that the function  is ``very'' hard on average on inputs
coming from that set.

\begin{lemma}[Impagliazzo]\label{lm:hardcore}
Let $f:\B^n \to \B$ be a $(S,\delta)$-hard function with respect
to the uniform distribution. Then, for every
$\epsilon$, there is a set $H\subseteq \B^n$ of size $\delta 2^n$ such that $f$
is $(S\cdot \poly(\epsilon,\delta),\frac 12 - \epsilon)$-hard
with respect to the uniform distribution over $H$.
\end{lemma}

We can now present Impagliazzo's proof of the XOR Lemma.

\begin{theorem}[XOR Lemma, Impagliazzo's version] \label{th:xorimp}
Let $f:\B^n \to \B$ be $(S,\delta)$-hard with respect to the uniform distribution,
let $k$ be an integer, and define $g:\B^{nk} \to \B$ as 
\[ g(x_1,\ldots,x_k) := f(x_1) \xor \cdots \xor f(x_k) \ . \] 
Then, for every $\epsilon>0$, $g$ is $(S\cdot \poly(\epsilon,\delta),\frac 12 - \epsilon - (1-\delta)^k)$-hard
with respect to the uniform distribution.
\end{theorem}

Let $H$ be a set as in Lemma \ref{lm:hardcore}.
The main idea in the proof is that if we are a small circuit, then our chances of
computing $f(x)$ for $x\sim H$ are about the same as our chances of guessing the
value of a random coin flip. Now, we are given $x_1,\ldots,x_k$ and we need
to compute $f(x_1)\oplus\cdots \xor f(x_k)$; if some $x_j$ is in $H$, then, intuitively,
our chances of correctly doing the computation are about the same as our chances
of computing $f(x_1)\oplus \cdots \oplus f(x_{j-1}) \oplus b \oplus f(x_{j+1}) \cdots \oplus f(x_k)$,
where $b$ is a random bit. A random bit xor-ed with other independent values is
also a random bit, and so, in that case, we will be correct only with probability $1/2$.
So our probability of being correct is at most $1/2$ plus $(1-\delta)^k$ (the probability
that none of the $x_j$ is in $H$) plus $\epsilon$ (to account for the difference between
our ability to guess a random bit and our ability to compute $f(x)$ for $x\sim H$).

Even though this proof sketch may look completely unsound, it leads to a surprisingly simple formal proof, that we present below.

\bigskip

\begin{proof*}[Proof of Theorem \ref{th:xorimp}]
Apply Lemma \ref{lm:hardcore}, and let $H$ be the set of size $\delta 2^n$ such that $f$
is $(S\cdot \poly(\epsilon,\delta),\frac 12 - \epsilon)$-hard
with respect to the uniform distribution over $H$. 

Let $C$ be a circuit of size $S'$ such that
\[ \pr[C(x_1,\ldots,x_k) = f(x_1) \xor \cdots \xor f(x_k) ] > \frac 12 + (1-\delta)^k + \epsilon \]
Let $D$ be the uniform distribution over $k$-tuples $(x_1,\ldots,x_k) \in (\B^n)^k$ conditioned
on at least one $x_j$ being an element of $H$. By conditioning on the event that some $x_j \in H$, we obtain
\[ \pr_{(x_1,\ldots,x_k) \sim D} [C(x_1,\ldots,x_k) = f(x_1) \xor \cdots \xor f(x_k) ] > \frac 12 +  \epsilon \]
We can see the process of picking a $k$-tuple $(x_1,\ldots,x_k)\sim D$ as first 
picking a non-empty subset $S\subseteq [k]$ with an appropriate distribution, then, for each $j\in S$, picking $x_j$ uniformly from $H$, and, for each $j\not\in S$, picking $x_j$ uniformly from $\B^n - H$, so the above expression can be rewritten as
\[ \E_{S \neq \varnothing}\bigl[\pr_{x_j \sim H, j\in S; x_j \sim(\B^n - H), j\not\in S} [C(x_1,\ldots,x_k) = f(x_1) \xor \cdots \xor f(x_k) ]\bigr]
 > \frac 12  + \epsilon \]
Fix the set $S$ that maximizes the outside expectation, and let $i$ be the first element of $S$.  Then we have
\[ \pr_{x_j \sim H, j\in S; x_j \sim(\B^n - H), j\not\in S} [C(x_1,\ldots,x_k) = f(x_1) \xor \cdots \xor f(x_k) ]
 > \frac 12  + \epsilon \]
or equivalently
\[ \E_{x_j \sim H, j\in S - \{i\}; x_j \sim(\B^n - H), j\not\in S} 
  \bigl[\pr_{x_i \sim H}[C(x_1,\ldots,x_k) = 
    f(x_1) \xor \cdots \xor f(x_k) ] \bigr] > \frac 12  + \epsilon \]
Let $a_j$ for $j\neq i$ be the assignment for $x_j$ that maximizes the above expectation. Then we have
\begin{multline*}
\pr_{x_i \sim H}  [C(a_1,\ldots,a_{i-1},x_i,a_{i+1},\ldots,a_k) \\ =
f(a_1) \xor \cdots \xor f(a_{i-1}) \xor f(x_i) \xor f(a_{i+1}) \xor \cdots \xor f(a_k) ] > \frac 12 + \epsilon 
\end{multline*}
which we can rearrange as
\begin{multline*}
\pr_{x \sim H} [C(a_1,\ldots,a_{i-1},x_i,a_{i+1},\ldots,a_k) \\
  \xor f(a_1) \xor \cdots \xor f(a_{i-1}) \xor f(a_{i+1}) \xor
    \cdots \xor f(a_k) = f(x) ] > \frac 12 + \epsilon
\end{multline*}
Note that the left-hand side expression above can be computed by a circuit of size
at most $S'+1$, showing that $f$ is not $(S'+1,\frac 12 - \epsilon)$-hard with respect to the
uniform distribution over $H$. We can choose $S'=S\cdot \poly(\epsilon,\delta)$ in a way that
contradicts our assumption about $f$ being $(S,\delta)$-hard with respect to $U_n$,
and so we conclude that $g$ is indeed $(S\cdot \poly(\epsilon,\delta),\frac 12 - \epsilon - (1-\delta)^k)$-hard
with respect to the uniform distribution.
\end{proof*}

\section{O'Donnell's Approach}

The XOR Lemma does not allow us to prove results of the form ``if there is a midly hard-on-average
distributional problem in NP with respect to the uniform distribution then there is a very hard-on-average
distributional problem in NP with respect to the uniform distribution.'' The difficulty is that
if $L$ is (the characteristic function of) a problem in \np, then, given $x,y$, it is not clear
that the problem of computing $L(x)\xor L(y)$ is still in \np. Indeed, if $L$ is \np-complete, then
computing $L(x)\xor L(y)$ is not in \np\ unless $\np = co\np$.

We note, however, that if $g:\B^k \to \B$ is a {\em monotone} function, and $L$ is in \np,
then computing $g(L(x_1),\ldots,L(x_k))$ given $(x_1,\ldots,x_k)$ is a problem in \np. We
may then ask whether there are monotone functions $g$ such that, if $L$ is mildly hard on average,
then computing  $g(L(x_1),\ldots,L(x_k))$ is very hard on average.

To address this question, we return to the informal proof of the XOR Lemma outlined in
the previous section. Let $f:\B^n \to \B$ be a $(S,\delta)$-hard function, and let
$H$ be a set as in Impagliazzo's Lemma. Define the probabilistic function $F$ such
that $F(x)=f(x)$ for $x\not\in H$ and $F(x)$ is a random bit for $x\in H$. Our informal
proof of the XOR Lemma was that, for a small circuit, computing $F(x_1)\xor \cdots \xor F(x_k)$ given $(x_1,\ldots,x_k)$
is about as hard as computing $f(x_1)\xor \cdots f(x_k)$ given  $(x_1,\ldots,x_k)$; no algorithm,
however, can solve the former problem with probability larger than $\frac 12 + (1-\delta)^k$,
for information-theoretic reasons, and so this is also an approximate upper bound to the
probability that a small circuit correctly solves the latter problem.

O'Donnell \cite{OD02} shows that there are monotone functions $g$ such that
computing $g(F(x_1),\ldots,F(x_k))$ given $(x_1,\ldots,x_k)$ cannot be
done with probability larger than $1/2 + \epsilon$, provided $k$ is at least $\poly(1/\epsilon,1/\delta)$, 
and a similar upper bound holds for the probability that a small circuit
can compute $g(f(x_1),\ldots,f(x_k))$ given $(x_1,\ldots,x_k)$.

Let us start with a formalization of the information-theoretic result.
For a function $f:\B^n \to \B$ and a set $H\subseteq \B^n$, we denote
by $F_H$ a random variable distributed over functions $\B^n \to \B$, 
defined so that $F_H(x)$ is a random bit for $x\in H$ and $F_H(x)=f(x)$ for $x\not\in H$.
We say that a Boolean function is {\em balanced} if $\pr [ f(U_n) = 1] = \frac 12$.

\begin{lemma}[O'Donnell] \label{lm:od-it}
For every $\epsilon>0$, $\delta>0$ there is a $k = \poly(1/\epsilon,1/\delta)$ and a
monotone function $g:\B^k\to \B$, computable by a circuit of size $O(k)$, such that
for every balanced  function $f:\B^n \to \B$, every subset $H\subseteq \B^n$ of size $\delta 2^n$
and every function $A:\B^{kn} \to \B$ we have
\[ \pr_{x_1,\ldots,x_k} [ A(x_1,\ldots,x_k) = g(F_H(x_1),\ldots,F_H(x_k ) )] \leq \frac 12 + \epsilon \]
where different occurrences of $F_H$ in the above expression are sampled independendently.
\end{lemma} 

The proof of the Lemma is not easy, and we refer the reader to \cite{OD02} for more details.
Let us see how to use the Lemma for the sake of hardness amplification. We need to formalize
the notion of $g(F_H(x_1),\ldots,F_H(x_k))$ and $g(f(x_1),\ldots,f(x_k))$ being similarly hard
to compute for a small circuit. Specifically, we prove the following result.

\begin{lemma} \label{lm:od-ind} Let $f:\B^n \to \B$ be a $(S,\delta)$-hard function. Then, for every $\alpha>0$,
there is a set $H$ of size $\delta 2^n$ such that for every $k$, and every function $g:\B^k \to \B$
computable by a circuit of size at most $s$, and for every circuit $A$ of size at 
most $S\cdot \poly(\alpha,\delta) -s$, we have
\[ \pr [ A(x_1,\ldots,x_k ) = g(f(x_1),\ldots,f(x_k)) ] \leq
\pr [ A(x_1,\ldots,x_k ) = g(F_H(x_1),\ldots,F_H(x_k)) ]+ k\cdot \alpha \delta \]
\end{lemma}

In order to skecth the proof Lemma \ref{lm:od-ind}, we first need to introduce the
notion of {\em computational indistinguishability}. We say that two distributions
$X,Y$ ranging over $\B^n$ are $(S,\epsilon)$-indistinguishable if for every circuit
$C$ of size $\leq S$ we have

\[ \bigl| \pr [ C(X)=1] - \pr [ C(Y)=1] \bigr| \leq \epsilon \]

\begin{proof*}[Proof sketch of Lemma \ref{lm:od-ind}]
Given a $(S,\delta)$-hard function $f$, we first find a set $H$ as in Impagliazzo's Lemma,
such that $f$ is $(S',1/2-\alpha)$-hard with respect to the uniform distribution
on $H$, where $S'=S\cdot \poly(\alpha,\delta)$. Then we consider the distributions $(x,f(x))$ and $(x,F_H(x))$, for uniformly
distributed $x$, and we prove that they are $(S'-O(1),\alpha\delta)$-indistinguishable.
From this point, it is not hard to show, using a hybrid argument, that the distributions
\[ (x_1,\ldots,x_k,f(x_1),\ldots,f(x_k))\]
and
\[ (x_1,\ldots,x_k,F_H(x_1),\ldots,F_H(x_k)) \]
are $(S'-O(1),k\alpha\delta)$-indistinguishable.
Suppose now that $g$ is a function computable in size $s$ and that $A$ is a circuit of size $S''$ such that
\[ \pr [ A(x_1,\ldots,x_k ) = g(f(x_1),\ldots,f(x_k)) ] >
\pr [ A(x_1,\ldots,x_k ) = g(F_H(x_1),\ldots,F_H(x_k)) ]+ k\cdot \alpha \delta \]
Define the circuit 
\[ C(x_1,\ldots,x_k,b_1,\ldots,b_k):= A(x_1,\ldots,x_k) \xor g(b_1,\ldots,b_k) \]
of size $S'' + s+O(1)$ showing that the two above distributions
are not $(S''+s+O(1),k\alpha\delta)$-indistinguishable. It is possible
to choose $S'' = S\cdot \poly(\alpha,\delta)$ so that this is a contradiction.
\end{proof*}

Lemma \ref{lm:od-ind}, together with Lemma \ref{lm:od-it},  is
sufficient to provide amplification of hardness within \np\ for problems whose characteristic function is balanced.

\begin{lemma}
Let $f:\B^n \to \B$ be a balanced $(S,\delta)$-hard function. Then for every $\epsilon$
there is a $k=\poly(1/\epsilon,1/\delta)$ and a monotone $g:\B^k \to \B$ computable by
a circuit of size $O(k)$ such that if we define
\[ h(x_1,\ldots,x_k) := g(f(x_1),\ldots,f(x_k)) \]
we have that $h$ is $(S\cdot \poly(\epsilon,\delta),1/2 - \epsilon)$-hard.
\end{lemma}

\begin{proof}
Apply Lemma \ref{lm:od-it}
and find a $k=\poly(1/\epsilon,1/\delta)$ and a function $g:\B^k \to \B$ such that for every set $H$
of size $\delta 2^n$ and every $A$ we have
\[ \pr_{x_1,\ldots,x_k} [ A(x_1,\ldots,x_k) = g(F_H(x_1),\ldots,F_H(x_k))] \leq \frac 12 + \frac \epsilon 2 \]
Apply Lemma  \ref{lm:od-ind} with $\alpha = \epsilon \delta /2k$ to find a set $H$ such that for every circuit $A$ of size 
at most $S\cdot \poly(\alpha,\delta) -s = S\cdot\poly(\epsilon,\delta)$ we have
\[ \pr [ A(x_1,\ldots,x_k ) = g(f(x_1),\ldots,f(x_k)) ] \leq
\pr [ A(x_1,\ldots,x_k ) = g(F_H(x_1),\ldots,F_H(x_k)) ]+ \frac \epsilon {2} \]
Combining the two expressions, we have that for every circuit $A$ of size at most $S\cdot\poly(\epsilon,\delta)$
\[ \pr [ A(x_1,\ldots,x_k ) = g(f(x_1),\ldots,f(x_k)) ] \leq \frac 12 + \epsilon.\hfill\qed\]
\end{proof}

Some extra work is needed to remove the assumption that the funtion be balanced, and to
optimize the constants. O'Donnell final result is the following.

\begin{theorem}[O'Donnell]
Suppose that for every language $L$ in \np\ we have $(L,{\cal U}) \in \heur_{1/2 - 1/{n^{.33}}} \p/\poly$.
Then for every polynomial $p$ and for every language $L$ in $\np$ we have
\[ (L,{\cal U}) \in \heur_{1/p(n)}\p/\poly. \]
\end{theorem}

The result was improved by Healy et al. \cite{HVV04}, but only for balanced languages (that is, for languages whose characteristic function is balanced on every input length). 

\begin{theorem}[Healy et al.]
Suppose that for every balanced 
language $L$  in \np\ there is a polynomial $p$ such that
 $(L,{\cal U}) \in \heur_{1/2 - 1/{p(n)}} \p/\poly$.
Then for every polynomial $p$ and for every balanced language $L$ in $\np$ we have
\[ (L,{\cal U}) \in \heur_{1/{p(n)}} \p/\poly \]
\end{theorem}

Trevisan \cite{T03,T05} proves weaker results for the uniform $\heur\bptime$ classes.
Specifically, Trevisan proves
that there is a constant $c$ such that
if $(\np,{\cal U}) \subseteq \heur_{1/2 - 1/{(\log n)^c} } \bpp$
then, for every polynomial $p$,  $(\np,{\cal U}) \in \heur_{1/{p(n)} } \bpp$.

Indeed, the actual result is slightly stronger. 

\begin{theorem}[Trevisan]
Suppose that  for every language $L$ in \np\ there is a polynomial time
randomized algorithm $A$ such that for every $n$

\[ \pr_{x\sim U_n; {\rm coin\ tosses\ of}\ A}  [A(x)\neq L(x) ] \leq \frac 12 + \frac 1 {(\log n)^c} \]
Then, for every polynomial $p$,  $(\np,{\cal U}) \in \heur_{1/{p(n)} } \bpp$.
\end{theorem}

Note that the assumption in the theorem is (possibly) weaker
than $(\np,{\cal U}) \subseteq \heur_{1/2 - 1/{(\log n)^c} } \bpp$,
which requires 
\[ \pr_{x\sim U_n} \left[ \pr_{{\rm coin\ tosses\ of}\ A} [A(x)\neq L(x)] > \frac 14 \right] \leq \frac 12 + \frac 1 {(\log n)^c} \]

\chapter{Worst-Case versus Average-Case and Cryptography}
\label{sec:ff}

The results on hardness amplification from \secref{sec:amp} indicate that the notion of average-case hardness is very robust with respect to the hardness parameter.  Namely, it is just as hard to solve hard problems in $(\np, \uniform)$ on slightly more than half their inputs as it is to solve them on a $1 - 1/\poly(n)$ fraction of inputs.  It is reasonable to ask if this connection can be pushed to the extreme: Is it the case that solving problems in $(\np, \uniform)$ on slightly more than half their inputs is no easier than solving them on {\em all} inputs?  In other words, are there problems in $(\np, \uniform)$ whose tractability would imply that $\np \subseteq \bpp$?

A related and fundamental question in cryptography is whether the security of
various cryptographic primitives can be reduced to a reasonable
worst-case complexity theoretic assumption, such as $\np \not\subseteq
\bpp$.  This question has not been settled yet, and there is contrasting
evidence about the possibility of such a connection.  In this Section we
review and explain several results related to this topic.  As we shall
see, at the heart of the question of basing cryptography on a worst-case
assumption is the connection between worst-case and average-case
complexity.

Various cryptographic tasks require cryptographic primitives of
seemingly different strength.  Here, we focus on the worst-case
assumptions necessary for the existence of one-way functions
(equivalently, symmetric key cryptography) and public key encryption.

Since under the assumption $\np \subseteq \bpp$ no one-way functions
exist, a worst-case assumption necessary for the existence of one-way
functions must be at least as strong as $\np \not\subseteq \bpp$.  Is
this assumption sufficient for the existence of one-way functions?  And
if it is not, is it possible to base the existence of one-way functions
on a possibly relaxed, but still reasonable worst-case complexity
assumption?

Assuming the worst-case intractability of certain promise problems on
lattices, it is possible to obtain provably secure constructions of
cryptographic one-way functions, as well as seemingly stronger
primitives such as collision resistant hash functions and public-key
encryption schemes.  However, all known worst-case intractable problems
that yield secure cryptographic primitives are both in $\np$ and
$\co\np$, thus are unlikely to be $\np$ hard.\footnote{The worst-case
assumption that statistical zero knowledge contains intractable
problems, which seems to be much stronger than $\np \not\subseteq \bpp$, is
known to imply the existence of infinitely often one-way functions, a
primitive object seemingly weaker than the one-way function \cite{O91}.
This primitive does not appear to have any useful applications.}

At this point, it is an open question whether the average-case tractability of $(\np, \uniform)$ would imply that $\np \subseteq \bpp$, and whether 
any form of cryptography can be based on the assumption $\np \not\subseteq \bpp$.  In this Section we review evidence that points to some difficulties in
establishing such connections.

\section{Worst-Case to Average-Case Reductions}
\label{sec:worstavg}

What do we mean when we say that the existence of one way functions
can be based on the assumption $\np \not\subseteq \bpp$?  The most
general interpretation would be to say that there exists a proof of
the statement ``$\np \not\subseteq \bpp$ implies that one-way
functions exist''.  At this point no such proof is known; however, it
is difficult to rule out the existence of a proof, for that would
imply that either ``$\np \not\subseteq \bpp$'' or ``one-way functions
exist'' would not be provable.  One plausible interpretation of the
claim that the existence of one-way functions requires assumptions
stronger than $\np \subseteq \bpp$ would be to say that any
``plausible'' way to obtain a worst-case algorithm for $\sat$ (or some other
$\np$-complete problem) from an imagined inverter for the universal
one-way function fails, or at least violates some reasonable assumption.

To see what we mean by ``plausible'', let us see how a possible proof
of the claim might go.  Generally such proofs are carried out by
reduction; namely, there is an efficiently computable procedure that
maps candidate inverters for the one-way function to algorithms for
$\sat$.  Moreover, the reductions typically use the one-way function
inverter as a black box only.  Such a reduction can be modeled as an
efficient oracle procedure $R$ that, when given oracle access to an
average case inverter for the one-way function, solves $\sat$ correctly
on almost all instances.  With this in mind, the notion that one-way
functions can be based on the assumption ``$\np \not\subseteq \bpp$''
can be liberally interpreted as the existence of a reduction $R$ of the
form described above.

We would also like to consider the possibility that one-way functions
can be based on stronger assumptions.  This motivates the notion of a
worst-case to average-case reduction.  First, we define the notion of an
``inversion oracle'' for a one-way function.

\begin{definition}[Inversion oracle]
Let $\{f_n: \B^n \to \B^*\}$ be a family of functions.  An inversion oracle
for $\{f_n\}$ with error $\delta(n)$ is a family of (possibly
randomized) functions
$\{I_n: \B^* \to \B^n\}$ such that for all $n$,
$$\pr_{x \sim U_n, I_n}[I_n(f_n(x)) \not\in f_n^{-1}(f_n(x))]
  \leq \delta(n).$$
\end{definition}

Thus, if there is an efficiently computable inversion oracle for
$f$ with inverse polynomial error, then $f$ is not strongly
one-way.

\begin{definition}[Worst-case to average-case reduction]
A {\em worst-case to average-case reduction} from a language $L$ to inverting 
a family of functions $\{f_n\}$ with average-case error $\delta(n)$ is an
oracle procedure $R$ such that for all inversion oracles $I$ with
error $\delta(n)$, all sufficiently large $n$, and all $x$ of length $n$,
$$\pr_{R,I}[R^I(x) \neq L(x)] < 1/3.$$
\end{definition}

The reduction is called {\em non-adaptive} if the reduction makes all
its queries in parallel, that is, each query are independent of answers
to previous queries.

If the function $f$ were not one-way, the inversion oracle could be
implemented by an efficient algorithm, and the reduction would give an
efficient algorithm for $L$.  Thus a worst-case to average-case
reduction can be viewed as a fairly general tool for establishing a
connection between the average-case complexity of inverting $f$ and the
worst-case complexity of $L$.

In a similar fashion, we can define worst-case to average-case
reductions for other primitives in average-case complexity, in
particular distributional decision problems and distributional search
problems (of which one-way functions are a special case).  The only part
of the definition that differs for these primitives is the notion of an
inversion oracle, which we call ``approximate oracle'' in this context.
For illustration we state the definition for deterministic oracles, and
for decision problems only.

\begin{definition}
\label{dfn:worsttoavg}
Let $L$ be a language and $\calD$ an ensemble of distributions.  An
{\em approximate oracle} for $(L, \calD)$ with error $\delta(n)$ is a
function $A: \B^* \to \{0, 1, \bot\}$ such that for all $n$, 
\[\pr_{x \sim D_n}[A(x) \neq L(x)] < \delta(n).\]  
The approximate oracle is {\em errorless} if for all $x$, $A(x) \in \{L(x), \bot\}$.

A {\em worst-case to average-case reduction} with error $\delta(n)$ from
$L$ to $(L', \calD)$ is an efficient oracle procedure $R$ such that for all
approximate oracles $A$ with error $\delta(n)$, all sufficiently large
$n$, and all $x$ of length $n$, $\pr_R[R^A(x) \neq L(x)] < 1/3$.
\end{definition}

\begin{change}
Thus if $(\bh, \calubh)$ has an efficiently computable approximate oracle,
then $(\np, \psamp) \subseteq \heur\bpp$; if the oracle is errorless,
then $(\np, \psamp) \subseteq \avg\zpp$.  Assuming $\np \not\subseteq
\bpp$, the existence of a worst-case to average-case reduction from
$SAT$ to $(\bh, \calubh)$ implies that $(\np, \psamp) \not\subseteq
\heur\bpp$ (or $(\np, \psamp) \not\subseteq \avg\zpp$, if the reduction
only works with respect to errorless oracles).

This definition of ``worst-case to average-case reduction'' models the framework used to establish the amplification of hardness results from \secref{sec:amp}.  Also, in the extreme case $\delta = 0$, the definition becomes the standard notion of reducibility between worst-case problems.
\end{change}

\paragraph{Alternative Definitions.}
The notion of ``worst-case to average-case reduction'' attempts to capture a reasonable class of possible approaches for basing average-case complexity and cryptography on $\np$-hardness.  We wish to stress, however, that the definition is by no means canonical and that it is natural to consider certain variants.  For simplicity we focus on \dfnref{dfn:worsttoavg}.

One alternative to \dfnref{dfn:worsttoavg} is to consider generic procedures that, given oracle access to {\em any} worst-case hard language $L$, produce an average-case hard language $(L', \calD)$.  For such a procedure $A$ to be useful for $\np$-languages it should be the case that $A$ itself is an $\np$ procedure with access to an oracle.  This notion is interesting because such procedures exist in higher complexity classes such as $\pspace$ and $\exptime$, where they are used to establish worst-case to average-case connections.  The amplification results of \secref{sec:amp} are also of this type.  Viola~\cite{V05} (see also \cite{V03}) shows that no such oracle procedure exists in $\np$, and even in the polynomial hierarchy (unless $(\np, \psamp) \not\subseteq \heur\p / \poly$, in which case $A$ exists trivially.)

In summary, Viola's result shows that any worst-case to average-case reduction in $\np$ must use specific properties of the worst-case language it is reducing from.  Indeed, the worst-case to average-case reductions of Ajtai, Micciancio, and Regev heavily exploit properties that are specific to lattices.

A serious limitation of \dfnref{dfn:worsttoavg} is that it does not impose any computational restriction on the average-case oracle.\footnote{In fact, all results presented in this Section hold for $\Sigma_2$ oracles, and in some cases for $\np$ oracles.}  In reality, to base average-case complexity on $\np$-hardness, the reduction need only consider candidate average-case oracles that can be implemented in $\bpp$.  This intriguing type of a reduction is called a ``$\bpp$-class black-box reduction'' by Gutfreund and Ta-Shma~\cite{GT06}:  As in \dfnref{dfn:worsttoavg}, the reduction only obtains oracle (black-box) access to the average-case solver, but is allowed to behave arbitrarily if the oracle cannot be implemented in $\bpp$.  Gutfreund, Shaltiel, and Ta-Shma~\cite{GST05, GT06} show an interesting setting in which $\bpp$-class black-box reductions are provably more powerful than ordinary worst-case to average-case reductions (under reasonable assumptions.)  However, it is not known whether such reductions can be used to base average-case complexity for $\np$ and cryptography on $\np$-hardness.

It is of course possible to further relax the definition and allow the reduction non-black box access to an implementation of the inversion oracle.  Little is known about the power of such a setting.

\section{Permutations and Range-Computable Functions}
\label{sec:permutations}

What is the hardest language $L$ for which we can expect to have a
worst-case to average-case reduction from $L$ to inverting some one-way function? Let us look at some simple cases first.

First, let us consider the case of a reduction $R$ from $L$ to a one-way
permutation $f: \B^n \to \B^n$.  Then it is not difficult to see that
$L$ must be in $\am \cap \co\am$ ($\np \cap \co\np$ if the
reduction is deterministic).  The situation is completely analogous for
$L$ and $\overline{L}$, so it is sufficient to prove that $L \in \am$.
A simple two-round protocol for deciding membership in $L$ works as
follows: In the first round, the verifier sends the coins used by the
reduction to the prover.  In the second round, the prover sends the
verifier a transcript that describes the computation on $R$ when given
access to an oracle that inverts $f$ on all inputs.  When $R$ makes
oracle query $q$, the honest prover answers with the unique $a$ such
that $f(a) = q$.  The verifier can check that all the answers provided
by the prover are consistent with its queries, thus forcing the prover
to perfectly simulate a computation of $R$ when given oracle access to
an inverter for $f$.  At the end of the interaction, the verifier
accepts iff the transcript provided by the prover is an accepting
transcript for $R$.

It follows that the average-case hardness of any one-way permutation can
be based, at best, on the worst-case hardness of some problem in $\am
\cap \co\am$.  Thus there appears to be no hope of basing the hardness
of any cryptosystem that requires one-way permutations on the assumption
$\np \not\subseteq \bpp$.

\subsection{$k$-to-One Functions}

A permutation is a function that is both onto and one-to-one; Akavia et
al.~\cite{AGGM06} consider what happens when the function $f: \B^{n + \log k} \to \B^n$ is $k$-to-one, namely every element in $\B^n$ has exactly $k$ pre-images under $f$.  The crucial difference between the cases $k = 1$ and $k
> 1$ is that when $k = 1$, the function $f$ admits a unique inverting
oracle, while for $k > 1$ there are many such oracles.  To illustrate
the significance of this, let us see what happens when the above
protocol for permutations is applied to a two-to-one function $f$.
Since the number of inverting oracles for $f$ is now doubly exponential
in $n$, it may be the case that for every choice of randomness by the
reduction, there exists some inversion oracle that makes the reduction
output the incorrect answer.  A cheating prover can then force the
verifier to output the incorrect answer by using this inversion oracle
in its simulation.

The solution of Akavia et al. is to force the prover to commit to a
particular oracle that is independent of the randomness used by the
reduction.  Let us first illustrate this with the case $k = 2$.  Then it
is easy to modify the protocol for $L$ so that the prover is always
forced to simulate interaction with the ``smallest'' inverting oracle for $f$:
This is the inverter that, on input $q$, always answers with the
lexicograpically smaller pre-image of $q$ under $f$.  To check
correctness, for every query $q$ the verifier always asks to see {\em
both} preimages of $q$, and always uses the smaller of the two values in
its simulation of the reduction.  It is straightforward that this
argument works for any $k$ up to $\poly(n)$.

For values of $k$ larger than $\poly(n)$, it is infeasible to ask the
prover to provide a complete list of pre-images for each query.
Instead, the prover is forced to provide a {\em random} pre-image, which
is independent of the randomness used by the reduction.  Thus the prover
will simulate the interaction of $R$ with a random inverter.  Let us
outline how such a random pre-image might be obtained.  The random
inverter that the proof system intends to simulate is the following one:
For each possible query $q$, choose a random hash function $h$ mapping
$n$ bits to slightly fewer than $\log_2 (k / s)$ bits, where $s =
\poly(n)$.  With high probability, the size of the set $S = h^{-1}(0)
\cap f^{-1}(q)$ is about $s$.  Out of all the elements of $S$, choose
the lexicographically smallest one (and if $S$ is empty, choose an
arbitrary inverse of $q$).

As a first attempt, consider this proof system for simulating the
inverter on a query $q$: The verifier chooses a random hash function
$h$, asks the prover for a complete list of members of $S$, and chooses
the lexicographically smallest one.  Notice that no prover can include
fictitious members of $S$ in its list, because membership in $S$ is an
efficiently verifiable property.  Therefore, provers can only cheat in a
``one-sided'' manner:  A cheating prover can attempt to omit members of
$S$, but never claim fictitious members of $S$.

\begin{change}
A cheating prover may, of course, fool the verifier by claiming that,
say, $S$ is empty.  The verifier knows that the size of $S$ must be
approximately $s$, so the verifier can protect against such an attack by
rejecting all sets $S$ whose size deviates substantially from $s$.  The
problem is that the cheating prover may fool the verifier even by
omitting a {\em single} entry of $S$, namely the lexicographically
smallest one.  Hence the verifier must ensure that the prover has not
omitted even a single element of $S$.

This appears impossible to achieve in general, as deviation bounds on
the size of $S$ only guarantee that $S$ will have roughly the expected
number of elements.  Instead, Akavia et al. consider what happens when
we fix the randomness used by the reduction and execute this protocol
$t = \poly(n)$ times independently in parallel.  Let $S_i$ denote the 
set $S$ resulting from the $i$th run of the protocol.  

Now suppose that for every potential query $q$, it can be guaranteed that in a $1 - \epsilon$ fraction of the $t$ protocol runs, the prover provides the correct set $S_i$.  Then at least a $1 - \epsilon$ fraction of the protocol runs provide a correct answer to the first query asked by the reduction; out of those, a $1 - \epsilon$ fraction of runs provide a correct answer to the second query, and so on.  If the verifier asks $\ell$ queries, then a $(1 - \epsilon)^\ell$ fraction of runs will have all their queries answered correctly.  By choosing $\epsilon$ small enough, it can be ensured that a random run simulates the reduction correctly with high probability.

Therefore the main task is to design a verifier test that ensures a $1 - \epsilon$ fraction of the $t$ protocol runs yield the correct set $S_i$.
\end{change}
The crucial point is that in order to make the verifier fail with
probability $\epsilon$, a cheating prover must now omit at least
$\epsilon t$ elements from the union of sets $S_1 \cup\dots\cup S_t$.\footnote{By convention we assume that the sets are pairwise disjoint.}  For $t \gg s / \epsilon^2$, $\epsilon t$ becomes a significant
deviation from $st$, the expected size of this union.  Statistically, we 
know that with high probability,
$$\bigabs{\abs{S_1 \cup\dots\cup S_t} - st} < \epsilon t / 2$$
so if the verifier checks that
$$\sum_{i = 1}^t \abs{\text{prover's claim for $S_i$}} \geq st -
  \epsilon t / 2$$
the honest prover will pass this check with high probability.  On
the other hand, this severely limits the power of a cheating prover:  If
any prover omits more than $\epsilon t$ elements from $S_1 \cup\dots\cup S_t$, then
$$\sum_{i = 1}^t \abs{\text{prover's claim for $S_i$}} <
  \abs{S_1 \cup\dots\cup S_t} - \epsilon t <
  (st + \epsilon t / 2) - \epsilon t <
  st - \epsilon t / 2,$$
and the verifier rejects.  Notice that the soundness of this protocol
relies on the fact that the power of a cheating prover is one-sided:  A
cheating prover can only understate, but never overstate the size of the
sets $S_i$.

One additional condition that must be ensured is that the sets $S_i$ are
nonempty for most $i$, for otherwise not even the honest prover can
correctly simulate the inverter for $f$.  This can be achieved by
an appropriate choice of parameters.

\paragraph{Size-Computable, Size-Approximable, and Size-Certifiable functions.}
A family of functions $f_n: \B^n \to \B^*$ is {\em size-computable} if
there is an efficient algorithm that on inputs $n$ and $y$ runs in time
polynomial in $n$ and outputs the number $\abs{f_n^{-1}(y)}$.  The
$k$-to-one functions considered above can be viewed as a special case of
size-computable functions.  If the algorithm outputs an approximation of 
$\abs{f_n^{-1}(y)}$ within an arbitrary factor that is inverse polynomial in $n$, the family
is called {\em size-approximable}.  If the algorithm is nondeterministic, the family
is called {\em size-certifiable}.  The protocol of Akavia et al. naturally
extends to the case of size-computable, size-approximable, and size-certifiable functions.

\begin{theorem}[Akavia et al.]
\label{thm:akavia}
Suppose there exists a worst-case to average-case reduction from
language $L$ to inverting a size-approximable or size-certifiable family of functions $\{f_n\}$.  Then
$L \in \am \cap \co\am$.
\end{theorem}

An example of a size-certifiable family is the family of functions
\[f_n(p, q) = \begin{cases} p \cdot q &\text{if $p$ and $q$ are $\floor{n/2}$-bit primes,} \\
   0 &\text{otherwise.}\end{cases}\] 
It is widely believed that this family of functions is weakly one-way.  However,  \thmref{thm:akavia} shows that the problem of inverting this family is 
unlikely to be $\np$-hard.

\section{General One-Way Functions and Average-Case Hard Languages}
\label{sec:owfhard}

\thmref{thm:akavia} can be interpreted as evidence that it may not be
possible to base the hardness of one-way functions on an \np-complete
problem.  The requirement the family $\{f_n\}$ be range-certifiable
may appear to be a technical one, and it is often the case that the
existence of one-way functions satisfying some additional technical
requirement is equivalent to the existence of general one-way functions.

We will argue that this interpretation of \thmref{thm:akavia} is
mistaken.  Observe that the protocol of Akavia et al. in fact simulates a 
run of the reduction interacting with a {\em worst-case} inversion oracle 
for $f_n$, not an average case one;  thus it shows that even the 
more difficult problem of inverting $y = f_n(x)$ on {\em every} output $y$ 
is unlikely to be \np-hard.

On the other hand, we do know of one-way functions that are \np-hard to
invert in the worst case.  For instance, consider the function $f$ that
maps a CNF $\phi$ and an assignment $a$ for $\phi$ to $(\phi, \phi(a))$.  A
worst-case inversion algorithm for $f$ solves the search
version of \sat.  Naturally, we do not interpret this as saying that
``$f$ is a one-way function that is \np-hard to invert'', because it
may well be the case that even though $f$ is $\np$ hard to invert on all
inputs, it is invertible on most inputs.  (This is in fact true for many natural
choices of distribution on inputs.)

Thus if it is indeed the case that the hardness of inverting one-way
functions cannot be based on an $\np$ complete problem, the argument
must use the fact that the assumed reduction from the $\np$ complete
problem to the inversion oracle works correctly with respect to an
{\em average-case} inversion oracle, not only for a worst-case one.

At this point it is not known whether such reductions exist in general.
The techniques described in the previous Section can be viewed as
partial progress towards a negative result that are obtained by putting
restrictions on the type of one-way function under consideration.  In
this Section we present a different approach which allows for general
one-way functions but places restrictions on the type of reduction used
to establish the worst-case to average-case equivalence.  In contrast to
\thmref{thm:akavia}, some of the results presented below make essential
use of the fact that the one-way function must be hard to invert on
average.

We begin by looking at the connection between worst-case and
average-case hardness for languages, rather than functions.  In
particular, we focus on the relation between the conjectures $\np
\not\subseteq \bpp$ and $(\np, \calU) \not\subseteq \heur\bpp$.

\subsection{The Feigenbaum-Fortnow Approach}
What can a worst-case to average-case reduction from a language $L$ to a
distributional $\np$ problem $(L', \uniform)$ look like?

To begin with, we observe that if the reduction is deterministic, then
$L$ must be in $\p$: For any $x \in \B^*$, the answer produced by the
reduction on input $x$ must be independent of the choice of average-case
oracle for $L'$.  One such average-case oracle is the oracle that agrees
with $L'$ on all the strings that are not queried by the reduction on
input $x$, and answers $\bot$ on all the other queries.  From the point
of view of the reduction, however, this oracle is indistinguishable from
the oracle that answers $\bot$ on every query.  Therefore, an efficient
algorithm for $L$ can be obtained by simulating the reduction on input
$x$ with access to an oracle that always answers $\bot$.

It follows that any nontrivial worst-case to average-case reduction must
make randomized queries to the average-case oracle.  Feigenbaum and
Fortnow~\cite{FF93} consider the case in which the reduction is non-adaptive and the
distribution of every query made by the reduction on input $x$ of length
$n$ is uniform in $\B^{n'}$ for some $n' = \poly(n)$.  Reductions of
these type are called {\em locally random reductions}.  The reason such
reductions are interesting is that is that they provide a natural way of
establishing a worst-case to average-case connection: If the reduction
asks $q$ queries, then any average-case oracle that is $1/4qn'$-close to
$L'$ with respect to the uniform distribution is indistinguishable from
$L'$ itself from the point of view of the reduction with probability
$3/4$.  Thus if there exists a locally random reduction from $L$ to
$L'$, and $L$ is hard in the worst-case, then $L'$ is hard to solve on
more than a $1 - 1/4qn'$-fraction of inputs.  Locally random reductions
have been used to establish worst-case to average-case connections in 
settings other than $\np$.

Feigenbaum and Fortnow essentially rule out locally random reductions as
a tool for establishing worst-case to average-case connection for all of
$\np$.  More precisely, they show that if there exists a locally random
reduction from a language $L$ to a language $L'$ in $\np$, then it must
be that $L$ is in $\np / \poly \cap \co\np / \poly$.  In particular, $L$
is unlikely to be $\np$-hard: If $L$ is $\np$-hard, then $\np$ is
contained in $\co\np / \poly$, and the polynomial hierarchy collapses to
the third level.

To prove this, Feigenbaum and Fortnow give a way to simulate the
reduction (on input $x$) by an $\am$ proof system that uses polynomial length non-uniform advice.  The outcome of the simulation then determines whether $x$ is a ``yes'' or a ``no''
instance of $L$.  Thus the protocol can be used to determine membership
in both $L$ and $\overline{L}$.  An $\am$ proof system with advice can 
be turned into a non-deterministic circuit, giving the conclusion 
$L \in \np / \poly \cap \co\np / \poly$.

\paragraph{The Feigenbaum-Fortnow Protocol.}  Let $R$ be a locally
random reduction from $L$ to $L' \in \np$.  Suppose that on an input of length 
$n$, $R$ makes $k$ queries, each of which is uniformly distributed in $\B^{n'}$.  
Without loss of generality, assume that $R$ is correct with very high 
probability (say $1 - 1/k^3$) over its random coins.  

We show an interactive protocol for membership in $L$.  The protocol for $\overline{L}$ 
is identical except that it inverts the answers given by $R$.

The non-uniform advice used by the protocol will be the value
$p = \pr_{y \sim \B^{n'}}[y \in L']$.

\medskip
\hrule
\noindent{\bf The protocol.} On input $x \in \B^n$,
\begin{enumerate}
\item {\bf Verifier:} Run $R(x)$ independently $m = 64k^2\log k$ times to generate $m$ sets of queries
  $(y_1^1,\dots,y_k^1),\dots,(y_1^m,\dots,y_k^m)$.  Send all queries to the prover.
\item {\bf Prover:} For each $y_i^j$, respond by saying whether $y_i^j \in L'$.  Accompany
  each claim that $y_i^j \in L'$ by an $\np$-certificate for $y_i^j$.
\item {\bf Verifier:} Accept if all of the following conditions hold:
   \begin{enumerate}
   \item $R(x)$ accepts in all $m$ iterations using the answers provided by the prover,
   \item All certificates sent by the prover are valid, and
   \item For every $1 \leq j \leq k$, at least $pm - m/2k$ of the queries $y_j^1,\dots,y_j^m$ 
      are answered ``yes''.
   \end{enumerate}
\end{enumerate}
\hrule
\medskip

If $x \in L$ and the prover follows the protocol, then $R(x)$ accepts in all $m$ 
iterations with high probability, and the verifier accepts provided condition 3(c) 
is satisfied.  Note that for each fixed $j$, the strings $y_j^1,\dots,y_j^m$ are 
independent and uniformly distributed in $\B^{n'}$, and each one has probability 
$p$ of being a yes instance.  By Chernoff bounds, with probability at least $1/4k$ 
at least $pm - 4\sqrt{m \log k} > pm - m/2k$ of them are yes instances.  By a 
union bound with probability $3/4$ this is satisfied for all $j$ and condition 3(c) holds.

If $x \not\in L$, to make the verifier accept, the prover must send an erroneous answer in
every one of the $m$ runs of $R(x)$, so in particular there must be at least $m$ errors
among the prover's answers.  All the erroneous answers of the prover must be yes 
instances on which it answers no (if the prover tries to cheat the other way, it wouldn't
be able to provide certificates.)  In particular, there must be some $j$ such that among
the queries $y_j^1,\dots,y_j^m$ at least $m/k$ are answered no even though they were
yes instances.  By a Chernoff bound as above, it is unlikely that there are more than 
$pm + 4\sqrt{m\log k}$ yes instances among $y_j^1,\dots,y_j^m$, so the prover is
giving at most $pm + 4\sqrt{m\log k} - m/k < pm - m/2k$ ``yes'' answers for $y_j^1,\dots,y_j^m$.  
Then the verifier rejects with high probability in step 3(c).

\subsection{Arbitrary Non-Adaptive Reductions}

For the result of Feigenbaum and Fortnow, it is not necessary that the
distribution of each query made by the reduction be uniform over
$\B^{n'}$, but it is essential that the marginal distribution of queries
made by the reduction be independent of the reduction's input.  This
restriction is quite strong, and in this sense, the result is extremely
sensitive: If one modifies the distribution of queries even by an
exponentially small amount that depends on the input, all statistical
properties of the reduction are preserved, but one can no longer draw
the conclusion that $L \in \np/\poly \cap \co\np /\poly$.

Bogdanov and Trevisan~\cite{BT03} show that the conclusion of Feigenbaum
and Fortnow holds in a more general setting.  They show that the
existence of any non-adaptive worst-case to average-case reduction from
$L$ to an arbitary problem $(L', \calD)$ in $(\np, \psamp)$ implies
that $L$ is in $\np/\poly \cap \co\np /\poly$, with no restriction on
the distribution of queries made by the reduction.  In particular, the
queries made by the reduction are allowed to depend arbitrarily on the
input $x$.  This formulation extends the result of Feigenbaum and
Fortnow in two directions: First, it allows for a more general class of
worst-case to average-case reductions; second, it allows average-case
complexity to be measured with respect to an arbitrary samplable
distribution, not only the uniform distribution.

\begin{theorem}[Bogdanov and Trevisan]
\label{thm:bt05}
Suppose that there exists a non-adaptive worst-case to average-case
reduction from a language $L$ to a decision problem $(L', \calD)$ in
$(\np, \psamp)$.  Then $L \in \np / \poly \cap \co\np / \poly$.
\end{theorem}

The proof of Bogdanov and Trevisan uses essentially the fact that the
reduction is correct when given access to an arbitrary average-case
oracle for $(L', \calD)$.  The idea of the proof is again to simulate the
reduction querying an average-case oracle for $(L', \calD)$ with an $\am$
protocol using advice.  Observe that the Feigenbaum-Fortnow protocol
works for arbitrary non-adaptive reductions whenever it is given as
auxiliary input the probability $p_x$ that a random query made by the
reduction on input $x$ is a ``yes'' instance of $L'$ according to
distribution $\calD$.  For a general reduction, however, the value $p_x$
cannot be provided as advice for the protocol, because it may depend on
the particular input $x$.

The idea of Bogdanov and Trevisan is to use a different protocol to
compute the value $p_x$, then use the Feigenbaum-Fortnow protocol for
membership in $L$ using the value $p_x$ as auxiliary input.  Initially,
a weaker version of the theorem is proved where $\calD$ is the uniform
distribution.  To begin with, let us allow the distribution of queries
made by the reduction to depend on $x$, but restrict it to be
``$\alpha$-smooth'': We assume that every query $y$ is generated with
probability at most $\alpha \cdot 2^{-\abs{y}}$, where $\alpha$ is a
constant.  Suppose that, given a {\em random} query $y$, we could force
the prover to reveal whether or not $y \in L'$.  Then by sampling enough
such queries $y$, we can estimate $p_x$ as the fraction of ``yes''
queries made by the reduction.  But how do we force the prover to reveal
if $y \in L'$?  The idea is to hide the query $y$ among a sequence of
queries $z_1,\dots,z_k$ for which we {\em do} know whether $z_i \in L'$,
in such a way that the prover cannot tell where in the sequence we hid
our query $y$.  In such a case, the prover is forced to give a correct
answer for $y$, for if he were to cheat he wouldn't know where in the
sequence to cheat, thus would likely be caught.

The problem is that we do not know a specific set of queries $z_i$ with
the desired property.  However, the strings $z_i$ were chosen by
sampling independently from $\calD$, then with high probability $pk \pm
O(\sqrt k)$ of these queries will end up in $L'$, where $p$ is the
probability that a string sampled from $\calD$ is in $L'$.  Since $p$
depends only on the length of $x$ but not on $x$ itself, it can be given
to the verifier non-uniformly.  This suggests the following verifier
strategy: Set $k = \omega(\alpha^2)$, generate $k$ uniformly random
queries $z_1,\dots,z_k$ of length $n$, hide $y$ among $z_1,\dots,z_k$ by
inserting it at a random position in the sequence, send all the queries
to the prover and ask for membership in $L'$ together with witnesses
that at least $pk - O(\sqrt k)$ queries belong to $L'$.  Then with high
probability, either the verifier rejects or the answer about membership
of $y$ in $L'$ is likely correct.  Intuitively, a cheating prover can
give at most $O(\sqrt k)$ wrong answers.  The prover wants to use this
power wisely and assign one of these wrong answers to the query $y$.
However, smoothness ensures that no matter how the prover chooses the
set of $O(\sqrt k)$ queries to cheat on, it is very unlikely that the
query $y$ falls into that set.

For a reduction that is not smooth, it is in general impossible to hide
a query $y$ among random queries from $\calD$ using the above approach.
However, suppose that the verifier had the ability to identify queries
$y$ that occur with probability $\geq \alpha \cdot 2^{-\abs{y}}$; let us
call such queries ``heavy'', and the other ones ``light''.  The fraction
of heavy queries in $\calD$ is at most $1 / \alpha$.  Suppose also that
the prover answers all light queries correctly.  The prover can then
certify membership in $L$ as follows: If the query made by the reduction
is heavy, pretend that the average-case oracle answered $\bot$,
otherwise use the answer provided by the prover.  This process simulates
exactly a run of the reduction when given access to an average-case
oracle that agrees with $L'$ on all the light queries, and answers
$\bot$ on all the heavy queries.  In particular, the oracle agrees with
$L'$ on a $1 - 1/\alpha$ fraction of strings, so the reduction is
guaranteed to return the correct answer.

In general, the verifier cannot identify which queries made by the
reduction are heavy and which are light.  The last element of the
construction by Bogdanov and Trevisan is an $\am$ protocol with advice that
accomplishes this task.

The case of a general samplable distribution $\calD$ can be reduced to the case when $\calD$ is the uniform distribution using \thmref{thm:il}, observing that the reduction in the proof is indeed non-adaptive.

\subsection{Distributional Search Problems and One-Way Functions}

Theorem~\ref{thm:bt05} shows that non-adaptive worst-case to
average-case reductions from an $\np$-hard problem to decision problems
in $(\np, \psamp)$ are unlikely to exist.  How about reductions to search problems?  Using the fact that search-to-decision reduction described in Section~\ref{sec:std} is non-adaptive, we can conclude that non-adaptive reductions from $\np$-hard problems to distributional search problems in $\np$ are also unlikely to exist.

A case of special interest is when the distributional search problem is
inverting a one-way function: If there exists a non-adaptive worst-case to
average-case reduction from a language $L$ to a family of functions
$\{f_n\}$, then $L \in \np / \poly \cap \co\np / \poly$.  Using a more
refined argument for the case of one-way functions, Akavia et al. obtain
a simulation of the reduction by an $\am$ protocol without advice:

\begin{theorem}[Akavia et al.]
\label{thm:akavia2}
Suppose that there exists a non-adaptive worst-case to average-case
reduction from language $L$ to inverting a family of functions $\{f_n\}$.  Then $L \in \am \cap \co\am$.
\end{theorem}

\section{Public Key Encryption}

Do there exist public key encryption schemes whose security can be
based on the assumption $\np \not\subseteq \bpp$?  Since public key
encryption schemes are harder to design than one-way functions, we
expect that this question should be only harder to answer in the
affirmative than the question whether one-way functions follow from
the assumption $\np \not\subseteq \bpp$.  Conversely, the lack of
cryptographic primitives based on $\np$ hardness assumptions should be
easier to explain in the public-key setting than in the symmetric-key
setting.

As in the case of one-way functions, we interpret the question whether
public key encryption can be based on the assumption that $\np
\not\subseteq \bpp$ as asking for the existence of an efficiently
computable reduction that converts any adversary that breaks the
encryption scheme into an algorithm for $\sat$.  By an encryption
scheme, we mean a collection consisting of a key generation algorithm
$G$, an encryption algorithm $E$, and a decryption algorithm $D$ (all
randomized) such that
\begin{itemize}
\item Algorithm $G$ takes as input a hardness parameter $n$, runs in
  time polynomial in $n$, and produces a pair of keys:  the public
  key $pk$ and the secret key $sk$.

\item Algorithm $E$ takes as inputs a hardness parameter $n$, a
public key $pk$, and a bit $b$ to be encrypted, runs in time
polynomial in $n$, and satisfies the property that for most public
keys $pk$ (obtained by running $G(n)$), the distributions $E(n, pk,
0)$ and $E(n, pk, 1)$ are computationally indistinguishable (with
respect to the parameter $n$, by an algorithm that takes as auxiliary
input $n$ and $pk$).

\item Algorithm $D$ takes as inputs a hardness parameter $n$, a secret
  key $sk$, and a ciphertext $c$, runs in time polynomial in $n$, and
  satisfies the property that for all $b$, and most pairs $(pk, sk)$
  obtained from $G(n)$, $D(n, sk, E(n, pk, b)) = b$ with probability
  negligible in $n$.
\end{itemize}
The existence of one bit encryption is sufficient to construct public
key encryption schemes for messages of arbitrary length that satisfy
very strong notions of security.

As in the case of one way functions, it is not known in general
whether there exists a reduction from $\sat$ to an adversary for some
one bit encryption scheme.  However, such reductions can be ruled out
under certain restrictions either on the cryptosystem in question or
on the way the reduction works.

Goldreich and Goldwasser~\cite{GG98:np}, building upon previous work by
Brassard~\cite{B79} restrict attention to encryption schemes where
for all $n$ and $pk$, the sets $E(n, pk, 0)$ ane $E(n, pk, 1)$ are
disjoint, and moreover the set
$$S = \{(1^n, pk, c): c \not\in E(n, pk, 0) \cup E(n, pk, 1)\}$$
is in $\np$ (namely, the property that $c$ is a possible ciphertext is
efficiently refutable).  Goldreich and Goldwasser observe that some,
but not all known one bit encryption schemes satisfy these
properties.  They observe that if there is a reduction from a language
$L$ to an adversary for an encryption scheme of this type, then $L \in
\am \cap \co\am$.  The reason is that the reduction can be simulated
by a two-round proof system in which the prover plays the role of a
distinguishing oracle for the sets $E(n, pk, 0)$ and $E(n, pk, 1)$.
In the first round, the verifier chooses the randomness to be used by
the reduction and sends it to the prover.  In the second round, the
prover sends a transcript of the reduction interacting with an
adversary for the encryption scheme.  When the reduction queries the
adversary on input $(n, pk, c)$, there are three possibilities:  Either
$c \in (n, pk, 0)$, or $c \in (n, pk, 1)$, or $(n, pk, c) \in S$.  By
assumption, all three of these cases are efficiently certifiable.
Therefore, a transcript of the reduction augmented by certificates for
the answers made by every query asked by the reduction constitutes a
valid and efficiently checkable simulation of the reduction
interacting with a distinguishing oracle for one-bit encryption.

The requirement that the sets of possible encryptions of $0$ and $1$
are disjoint can be somewhat relaxed, and the requirement that the set
$S$ is in $\np$ can be substituted by a requirement that the reduction
is ``smart''---it never queries invalid ciphertexts.  Thus, the
observation of Goldreich and Goldwasser can be viewed as saying that
the $\np$ hardness of one bit encryption cannot be established via
``non-smart'' reductions.

Should these arguments be viewed as an indication that public key
cryptography cannot be based on $\np$ hard problems?  Observe that the
proof systems of Brassard and Goldreich and Goldwasser do not use the
fact that the reduction outputs the correct answer even if it
interacts with an average-case distinguisher between the encryptions
of $0$ and $1$.  Thus, these are essentially results about the
worst-case complexity of breaking encryption, showing that under
certain restrictions on the encryption scheme or on the reduction, the
hardness of breaking the encryption {\em in the worst case} is a
problem in $\np \cap \co\np$.  However, these restrictions on the
encryption scheme or on the reduction cannot be so easily removed.  As
was shown by Lempel \cite{L79}, there do exist ``encryption schemes'' which are
$\np$ hard to break in the worst case, but are tractable to break on
average:  Namely, the problem ``On input $(n, pk, E(n, pk, b))$, find $b$'' is
$\np$ hard in the worst case, but is tractable on average.
(Lempel's result generalizes the observation that there exist one-way
functions that are $\np$ hard to invert in the worst case but easy to
invert on average to the setting of public-key cryptography.)  Currently, 
there is no known argument that explains why public-key
cryptography appears to require worst-case assumptions stronger than
$\np \not\subseteq \bpp$ beyond what is known for one-way functions,
i.e., symmetric key cryptography.

\section{Perspective: Is Distributional \np\ as Hard as \np?}

So far we have focused on negative results regarding connections
between the worst case and average case complexity of $\np$.  Since
these results do not rule out the possiblity that distributional
$\np$ is as hard as $\np$, the question remains if such a connection
is possible, and if it is, how one should go about establishing it.

The problem of basing cryptography on $\np$ hardness has played a
central role since the beginnings of cryptography, and much research
effort has been put into answering this question in the affirmative.
A breakthrough was made in work by Ajtai~\cite{A96}, who showed that
the existence of intractable problems in distributional $\np$ follows
from the assumption that there is no efficient algorithm that
approximates the length of the shortest vector on a lattice in the
worst case (within a factor of $n^{O(1)}$, where $n$ is the dimension
of the lattice).  This is the first example of a problem in
distributional $\np$ whose hardness follows from a reasonable
worst-case intractability assumption.  In later works, Ajtai, Dwork,
Micciancio, and Regev substantially extended Ajtai's original result,
showing that (1) The existence of useful cryptographic objects,
including one-way functions and public key encryption schemes, also
follows from reasonable worst-case intractability assumptions and (2)
The worst-case intractability assumption used by Ajtai can be
substantially weakened, giving the hope that further improvements
could replace Ajtai's assumption with the strongest possible
worst-case intractability assumption, namely $\np \not\subseteq \bpp$.

All known worst case to average case connections for $\np$ are
established by reductions, and all known reductions start from a
problem that is known to reside inside $\np \cap \co\np$.  One view of
this situation is that membership in $\np \cap \co\np$ does not reveal
anything fundamental about the relation between worst case and average
case complexity for $\np$, but is merely an artifact of the current
reductions; improved reductions could go beyond this barrier, and
eventually yield an equivalence between worst case and average case
hardness for $\np$.

On the other hand, the results presented in this section, if
liberally interpreted, seem to indicate the opposite:  The mere
existence of a worst-case to average-case reduction for $\np$ often
implies that the problem one is reducing from is in $\np \cap \co\np$
(or $\am \cap \co\am$, or $\np / \poly \cap \co\np / \poly$.)  Moreover, the
reason for this connection appears to be fairly universal:  A
worst-case to average-case reduction can be viewed as a proof system
in which the verifier runs the reduction, and the prover simulates the
average-case oracle.  The difficulty is in forcing even a cheating
prover to simulate the average-case oracle correctly;  currently, it
is known how to do this only under restrictive assumptions on the
reduction (Theorems~\ref{thm:bt05} and~\ref{thm:akavia2}).  However,
further improvements may lead to the conclusion that this connection
between worst-case to average-case reduction and constant-round proof
systems is a universal one, and thus there is no hope of basing
average-case complexity for $\np$ on $\np$ hardness assumptions by
means of a reduction.

\chapter{Other Topics}

The theory of average-case complexity for $\np$ lacks the wealth of natural complete problems encountered in worst-case complexity.  Yet, there are many natural distributional problems that are believed to be intractable on average.  

One such problem is random $k$SAT, whose instances are generated by choosing clauses independently at random.  In \secref{sec:rksat} we survey some of the known results about random $k$SAT, especially for $k = 3$.  While random $3$SAT 
is not known to be average-case complete, some versions of it are not known to have efficient errorless heuristics.  An unusual result of Feige shows that the intractability of random $3$SAT would have some interesting consequences in approximation complexity. 

Another class of problems that are believed to be intractable on average is derived from lattice based cryptography.  The importance of these problems stems from the fact that they are the only known examples of problems in distributional $\np$ that are hard according to a worst-case notion of hardness:  If these problems were easy on average, then the corresponding problems on lattices, long believed to be hard, could be solved in the worst case.  We survey some key results in \secref{sec:lattices}.

\section{The Complexity of Random $k$SAT}
\label{sec:rksat}

A widely investigated question in both statistics and the theory of computing is the
tractability of random $k$CNF instances with respect to natural distributions.  The
most widely studied distribution on $k$CNF instances is the following:  Given 
parameters $n > 0$ and $m_k(n) > 0$, choose at random $m_k(n)$ out of the 
$2^k\binom{n}{k}$ possible clauses of a $k$CNF on $n$ boolean variables.  
An essentially equivalent model is to choose each of the possible $2^k\binom{n}{k}$
clauses independently with probability $m_k(n) / 2^k\binom{n}{k}$.

By a counting argument, it follows that when $m_k(n)/n \geq 2^k \ln 2$, a random
$k$CNF is almost always unsatisfiable as $n$ grows large.  Better analysis improves this
upper bound by a small additive constant.  Achlioptas and Peres \cite{AP04}, following Achlioptas and Moore~\cite{AM02}, prove
that when $m_k(n) < 2^k \ln 2 -k \ln 2 / 2 -c$ (for a constant $c$), then a random $k$CNF is
almost always satisfiable. Their result is non-constructive, that is, they do not
provide an efficient algorithm that finds satisfying assignments for a large
fraction of such formulas.

For specific values of $k$, better lower and upper bounds are known.
All known such lower bounds, except for the Achlioptas-Peres and Achlioptas-Moore results, are algorithmic.  In particular, it is known that $3.51 < m_3(n)/n < 4.51$.  

Friedgut~\cite{Fr99} showed that for every $k \geq 2$, satisfiability of random $k$CNF exhibits
a (possibly) non-uniform threshold.  More precisely, for every $\epsilon > 0$ and sufficiently large $n$
there exists a value $c_k(n)$ such that a random $k$CNF is satisfiable with probability 
$1 - \epsilon$ when $m_k(n) / n \leq (1 - \epsilon) c_k(n)$, and with probability at most 
$\epsilon$ when $m_k(n) / n \geq (1 + \epsilon) c_k(n)$.  It is conjectured that the sequence 
$c_k(n)$ converges to a value $c_k$, known as the $k$SAT threshold, as 
$n \to \infty$.  Experiments indicate for instance that $c_3(n) \to c_3 \approx 4.26$.

Assuming the existence of a threshold for $k$SAT, the existence of heuristic
algorithms for random $k$SAT with respect to this family of distributions becomes trivial 
everywhere except possibly at the threshold.\footnote{In the literature on random $k$SAT, 
usually the error parameter of the average-case algorithm is implicitly fixed to 
$o(1)$ or $n^{-c}$ for some fixed $c$.  Not much is known for the case of algorithms 
with negligible error or heuristic schemes.}  However, the situation is different with respect to errorless 
algorithms.  Below the threshold, where
most of the formulas are satisfiable, an errorless algorithm must certify most
satisfiable formulas efficiently.  In fact, since the lower bounds for $m_k(n)$ are 
algorithmic, we know that for every $k$ there is an errorless algorithm for $k$SAT
when $m_k(n) / n < a_k 2^k / k$, where the sequence $a_k$ converges to some positive value.  It is conjectured that algorithms
for finding satisfying assignments on most $k$CNF instances exist all the way up to
the $k$SAT threshold.

\subsection{Refuting Random CNF Instances}

Above the $k$SAT threshold, where most of the formulas are unsatisfiable, an errorless algorithm 
is required to refute most $k$CNF instances efficiently.  A useful way
of thinking of such a refutation algorithm is the following:  The algorithm is given a $k$CNF
instance $\phi$ and wants to distinguish between the case when $\phi$ is satisfiable and
when $\phi$ is ``typical'' for the distribution on inputs.  The algorithm can subject $\phi$ to 
any efficiently computable test that a random $\phi$ passes with high probability.
If the instance $\phi$ does not pass these tests, the algorithm can output $\bot$.
The challenge is to design a set of tests such that every $\phi$ that passes all the tests
must be unsatisfiable, in which case the algorithm rejects $\phi$.

When $m_k(n) > \Omega_k(n^{k-1})$, the following naive refutation algorithm works:
Take a variable, say $x_1$, and consider all the clauses that contain it.  Fixing $x_1$
to true yields a $(k-1)$CNF consisting of those $\Omega_k(n^{k-2})$ clauses that 
contain the literal $\overline{x}_1$, and this formula can be refuted recursively 
(the base case being a $2$CNF, for which an efficient refutation algorithm exists.)  
Repeat by fixing $x_1$ to false.  (For an improved version of this approach, see~\cite{BKPS98}.)

A more sophisticated approach for refuting random $k$CNF that handles smaller values
of $m_k(n)$ was introduced by Goerdt and Krivelevich~\cite{GK01}.  Their idea is to 
reduce $k$CNF instances to graphs (using a variant of Karp's reduction from $3$SAT to 
maximum independent set) so that satisfiable formulas map to graphs with large independent
sets, while the image of a random $k$CNF instance is unlikely to have a large independent 
set.  Moreover, they show that for most graphs derived from random $k$CNF, it is possible to
efficiently certify that the graph does not have a large independent set via eigenvalue 
computations.  Subsequent improvements of this argument 
yield refutation algorithms for random $k$CNF with $m_k(n) = \omega(n^{\ceil{k/2}})$
\cite{CGLS03}.  For the case $k = 3$ there are better refutation algorithms, and the best known
works for $m_3(n) = \omega(n^{3/2})$~\cite{FO04}.  This algorithm departs from previous
work in that it does not reduce $3$SAT to maximum independent set but uses a different 
reduction by Feige~\cite{F02}, which we describe in the next Section.

Do refutation algorithms for random $k$CNF exist when $m_k(n)$ is above the satisfiability
threshold $c_k n$, but below $n^{k/2}$?  For the case of $3$CNF, there is evidence 
suggesting that refuting random formulas may be hard for $m_3(n) < n^{3/2 - \epsilon}$ for every 
$\epsilon > 0$.  Ben-Sasson and Wigderson~\cite{BW01} (following~\cite{CS88})
show that for this range of 
parameters, most formulas require refutations {\em by resolution} of size 
$2^{\Omega(n^{\epsilon/(1-\epsilon)})}$.  (The naive refutation algorithm above can be 
viewed as implementing a simple proof by resolution.)  Recently, Feige and 
Ofek~\cite{FO06} showed that a different approach based on semi-definite programming 
that subsumes the algorithm of~\cite{FO04} also fails to certify unsatisfiability when 
$m_3(n) < n^{3/2}/\poly\log(n)$. 

A very recent breakthrough of Feige, Kim, and Ofek~\cite{FKO06} gives a {\em non-deterministic} refutation algorithm for $m_3(n) = \omega(n^{7/5})$, thus showing that random \trisat\ with respect to this distribution is in $\avg_{o(1)}\co\np$.\footnote{This class is defined in a way analogous to $\avg_\delta\p$; see \secref{sec:definitions}).}

\subsection{Connection to Hardness of Approximation}

Feige~\cite{F02} conjectures that for every constant $c$, unsatisfiability of random $3$CNF
is hard to certify (within negligible error) whenever $m_3(n) < c n$.  In particular, Feige's 
conjecture implies that $(\np, \psamp) \not\subseteq \avg_{\rm neg}\p$, but there is no evidence as to
whether random \trisat\ with parameter $m_3(n) < c n$ is complete for the class 
$(\np, \psamp)$.

Instead of pursuing connections with average-case complexity, 
Feige views his conjecture as a strengthening of the famous result by H\aa stad~\cite{H97} 
about the inapproximability of $3$SAT in the worst case.  Indeed, H\aa stad shows that assuming 
$\p \neq \np$, it is hard to distinguish between satisfiable $3$CNF instances and $3$CNF instances 
where no more than a $7/8 + \epsilon$ fraction of the clauses can be satisfied.  The class of 
instances on which no more than $7/8 + \epsilon$ fraction of the clauses can be satisfied in
particular includes most random $3$CNF instances with $cn$ clauses for sufficiently large 
$c$.  Feige's conjecture says that even if we restrict ourselves to these random instances, the
distinguishing problem remains intractable.  As several inapproximability results assuming
$\p \neq \np$ follow by reduction from the hardness of approximating $3$SAT, it can be hoped 
that Feige's stronger conjecture may yield new or stronger conclusions.

The main technical result of Feige is the following theorem. For notation purposes, given a
$3$CNF $\phi$ and an assignment $a$, let $\mu_i(\phi, a)$ denote the fraction of clauses in
$\phi$ where $a$ satisfies exactly $i$ literals, for $0 \leq i \leq 3$.

\begin{theorem}[Feige]
For every $\epsilon > 0$ there exists an algorithm $A$ that for all sufficiently large $c$ has the
following properties:
\begin{itemize}
\item $A$ accepts all but a negligible fraction of random $3$CNF on $n$ variables and $cn$ clauses.
\item For sufficiently large $n$, if $\phi$ is a satisfiable $3$CNF with $n$ variables and $cn$ clauses 
  and $A$ accepts $\phi$, then for every satisfying assignment $a$ of $\phi$, it holds that
  $\mu_1(\phi, a) = 3/4 \pm \epsilon$, $\mu_2(\phi, a) < \epsilon$, and 
  $\mu_3(\phi, a) = 1/4 \pm \epsilon$.
\end{itemize}
\end{theorem}

Observe that, in contrast, for most random $3$CNF $\phi$ and every assignment $a$, we have that 
$\mu_1(\phi, a) = \mu_2(\phi, a) = 3/8 \pm \epsilon$ and $\mu_0(\phi, a) = \mu_3(\phi, a) = 1/8 \pm 
\epsilon$.  

Assuming the conjecture, the theorem for instance implies the following:  For a $3$CNF
$\phi$ with $n$ variables and $cn$ clauses, it is hard to distinguish between the following cases:
\begin{itemize}
\item There exists an assignment for $\phi$ that satisfies {\em all} literals in a $1/4 - \epsilon$ fraction of 
  clauses
\item No assignment for $\phi$ satisfies all literals in more than a $1/8 + \epsilon$ fraction of clauses.
\end{itemize}

This hardness of approximation result is not known to follow from $\p \neq \np$.  Feige shows that hardness of approximation results for balanced bipartite clique, min bisection, dense subgraph, and the $2$-catalog problem follow from it\footnote{To be precise, Feige proves and needs a slightly more general result.} via combinatorial reductions.

\iffalse
Another natural distribution of $\trisat$ instances is the ``planted assignment'' model, where for given parameters $n > 0$ and $m(n) > 0$, an random instance is generating by first picking a random assignment $a \in \B^n$, then generating a formula that is consistent with the assignment by choosing
at random $m(n)$ of the possible $7\binom{n}{3}$ clauses that are not violated by the assignment.  A formula from this distribution will always be satisfiable.  The {\em search} problem of finding a satisfying assignment for such a formula may be of interest in cryptography:  If this problem is hard on average, then the function that given a random formula and a planted assignment outputs the formula is one-way.  It is known that there exist constants $c_1$ and $c_2$ ($c_1 < c_2$) such that finding a satisfying assignment is easy for most formulas when $m(n) < c_1 n$ and when $m(n) > c_2 n$.  There are no known results that indicate the problem may be hard in the intermediate range.
\fi

\section{The Complexity of Lattice Problems}
\label{sec:lattices}

Discrete lattices in $\R^n$ provide examples of problems in $\np$ that are believed to be 
intractable in the worst case and which worst-case to average-case reduce to certain 
distributional problems in $(\np, \psamp)$.  Some of these reductions yield stronger objects
such as one-way functions, collision resistant hash functions, and public-key cryptosystems.

The lattice problems in question are all {\em promise} problems~\cite{ESY84,G05}. Instead of attempting to list all their variants and the connections between them, for illustration we focus on the shortest vector problem.  (Other lattice problems exhibit similar behavior.  For a more general
treatment, see \cite{GM:book} and \cite{MR04}.)  A lattice $\mathcal{L}$ in $\R^n$ is represented by specifying a basis of $n$ vectors for it (all vectors have $\poly(n)$ size descriptions.)

\noindent{\bf The shortest vector problem $\svp_{\gamma(n)}$.} The instances are pairs 
  $(\mathcal{L}, d)$, where $\mathcal{L}$ is a lattice in $\R^n$ and $d$ is a number.  
  In yes instances, there exists a vector $\mathbf{v}$ in $\mathcal{L}$ of length at most 
  $d$.\footnote{To be specific we measure length in the $\ell_2$ norm. The problem is no easier 
  for other $\ell_p$ norms, see~\cite{RR06}.}  In no instances, every vector in $\mathcal{L}$ has 
  length at least $\gamma(n) d$.

This problem is in $\np$ (for $\gamma(n) \geq 1$.) The following seemingly easier variant also turns out to be useful.
  
\noindent{\bf The unique shortest vector problem $\usvp_{\gamma(n)}$.} This is the same as 
  $\svp_{\gamma(n)}$, except that in yes instances we require that every vector in $\mathcal{L}$ 
  whose length is at most $\gamma(n) d$ be parallel to the shortest vector $v$.

We stress that we are interested in the {\em worst-case} hardness of these problems as the
dimension of the lattice $n$ grows.  The best known polynomial time approximation algorithm 
for the shortest vector problem, due to Ajtai, Kumar, and Sivakumar~\cite{AKS01}, solves $\svp_{\gamma(n)}$ for $\gamma(n) =  2^{\Theta(n \log \log n / \log n)}$ (previous algorithms
of Lenstra, Lenstra, and Lov\'{a}sz~\cite{LLL82} and Schnorr~\cite{S87} achieve somewhat worse
approximation factors.)  For polynomial approximation factors $\gamma(n) = \poly(n)$, the best 
known algorithms run in time $2^{\Theta(n)}$~\cite{AKS01, KS03}.  

In a seminal paper Ajtai~\cite{A96} showed that assuming $\svp_{O(n^c)}$ is intractable for some 
fixed $c > 0$ there exist one-way functions.  He constructs a family of functions $\{f_n\}$ for 
which there exists a worst-case to average-case reduction from $\svp_{O(n^c)}$ to inverting 
$\{f_n\}$.  Later, Ajtai and Dwork~\cite{AD97} showed that public key encryption exists assuming
$\usvp_{O(n^c)}$ is intractable for some fixed $c > 0$.  The 
parameter $c$ has been improved since the original constructions, and it is known that
\begin{itemize}
\item One-way functions and collision resistant hash functions exist assuming 
$\svp_{\tilde{O}(n)}$ is intractable~\cite{MR04}.
\item Public key encryption exists assuming $\usvp_{\tilde{O}(n^{1.5})}$ is intractable~\cite{R03}.
\item Public key encryption exists assuming $\svp_{\tilde{O}(n^{1.5})}$ is intractable by quantum
  algorithms~\cite{Regev05}.
\end{itemize}

A short, self-contained outline of a basic worst-case to average-case reduction from $\usvp$ can be found in a tutorial of Regev~\cite{Regev06}.

These results greatly motivate the study of hardness of lattice problems:  For instance, if it were true
that $\svp_{n^{1.5 + \epsilon}}$ is $\np$-hard for some $\epsilon > 0$, it would follow that one-way functions exist (and in particular $(\np, \psamp) \not\subseteq \heur\bpp$) assuming only 
$\np \not\subseteq \bpp$.  

However, the best hardness results known for the shortest vector problem fall short of what is 
necessary for the current worst-case to average-case reductions.  Micciancio~\cite{M01} (following
Ajtai~\cite{Ajtai98}) showed that $\svp_{\gamma(n)}$ where $\gamma(n) = \sqrt{2} - \epsilon$ is 
$\np$-hard under randomized polynomial-time reductions for every $\epsilon > 0$.  More 
recently, Khot~\cite{K04} improved the hardness to $\gamma(n) = 2^{(\log n)^{1/2 - \epsilon}}$ 
for every $\epsilon > 0$, but his reduction runs in randomized quasipolynomial time.

On the other hand, Goldreich and Goldwasser~\cite{GG98} showed that 
$\svp_{\gamma(n)} \in \co\am$ for $\gamma(n) = \Omega(\sqrt{n / \log n})$ and Aharonov and 
Regev~\cite{AR05} showed that $\svp_{\gamma(n)} \in \co\np$ for $\gamma(n) = \Omega(\sqrt{n})$.
This can be taken as evidence that $\svp_{\gamma(n)}$ is not $\np$-hard when $\gamma(n)$ 
exceeds $\sqrt{n}$, but one must be careful because $\svp_{\gamma(n)}$ is a promise problem,
not a language.  While it is true that assuming $\np \neq \co\np$, languages in $\np \cap \co\np$
cannot be $\np$-hard, this conclusion fails in general for promise problems: Even, Selman, and
Yacobi~\cite{ESY84} give an example of a promise problem that is $\np$-hard yet resides in 
$\np \cap \co\np$.

It is interesting to observe that the one-way functions constructed by Ajtai~\cite{A96} and 
Micciancio and Regev~\cite{MR04} are size-approximable (in fact, almost regular), so by 
\thmref{thm:akavia} in the best case the hardness of these functions can be based on problems 
in $\am \cap \co\am$.  

\section*{Acknowledgements}

We thanks Scott Aaronson, Jonathan Katz, Chris Moore and the anonymous referee for their
helpful comments. 

%\bibliography{macros,luca}

\end{document}